\documentclass[11pt,a4paper]{article}

\usepackage[a4paper,margin=1in]{geometry}
\usepackage{amsmath,amssymb,amsthm,mathtools}
\usepackage{graphicx}
\usepackage{booktabs}
\usepackage{array}
\usepackage{xcolor}
\usepackage[colorlinks=true,linkcolor=blue,citecolor=blue,urlcolor=blue]{hyperref}
\usepackage{cite}
\usepackage{enumitem}
\usepackage{microtype}
\usepackage{tikz}
\usetikzlibrary{arrows.meta,positioning,shapes.geometric,fit,backgrounds}
\usepackage{caption}
\usepackage{tabularx}

\newtheorem{proposition}{Proposition}

\newtheorem{remark}{Remark}

\newcommand{\derived}{\textcolor{black!70!green}{\textbf{[D]}}}
\newcommand{\conditional}{\textcolor{black!50!orange}{\textbf{[C]}}}
\newcommand{\matching}{\textcolor{black!50!red}{\textbf{[M]}}}

\newcommand{\Tr}{\mathrm{Tr}\,}

\newcommand{\eg}{e.g.\ }

\newcommand{\Ssym}{S^{\mathrm{sym}}}
\newcommand{\Hub}{H_0}

\title{\textbf{Candidate collapse-noise correlators from\\
Generalized Trace Dynamics: a Hubble-scale\\
spectral line under structural assumptions}}

\author{Tejinder P. Singh\\[2pt]
\small Tata Institute of Fundamental Research, Homi Bhabha Road, Mumbai 400005, India\\
\small \texttt{tpsingh@tifr.res.in}}

\date{May 28, 2026}

\begin{document}

\maketitle

\begin{abstract}
\noindent
We present a conditional construction of candidate CSL-type collapse-noise correlators inspired by Generalized Trace Dynamics (GTD). The construction is not a parameter-free derivation from the minimal GTD Grassmann algebra. It rests on a chain of explicit structural postulates, listed in Section~\ref{sec:ledger}; within that auxiliary structure the spectral form and amplitude follow by computation rather than by phenomenological fitting. The resulting narrow-band spectrum at the Hubble scale lies outside the bands of current CSL bounds, so the framework is not in tension with existing high-frequency data.

We compute the two-point function of a candidate collapse-noise operator associated with the GTD aikyon decomposition $q_i = q_B + a_0\beta_i q_F$. In the minimal Grassmann algebra, $q_F$ appears only multiplied by Grassmann generators $\beta_i$, the reduction of $\mathrm{Tr}(q_F^\dagger\Gamma^\mu q_F)$ to ghost-mode operators is obstructed by the nilpotent $\delta\beta = \beta_2 - \beta_1$, and the pure-fermion coefficient $\beta_1\beta_2$ has no ordinary sign, modulus, or inverse. We therefore introduce an auxiliary canonical fermionic Fock-space sector for $q_F$, equivalently replacing the nilpotent pure-fermion coefficient by an ordinary effective scalar body parameter. This replacement is an independent structural postulate, not a consequence of the original minimal action.

Under this auxiliary postulate, together with a scalar bilinear $J = \mathrm{Tr}(q_F^\dagger q_F)$ as bath operator, positive-norm canonical quantization, and an effective sign choice $\sigma = \pm1$ for the scalarized pure-fermion sector, elementary Wick contraction gives a Wightman line at $|\omega| = 2\omega_0$ with amplitude $\mathcal A_J = (\hbar/2m_R\omega_0 L_{\mathrm{aik}}^2)^2\cdot\mathcal N\cdot D$. The cosmological identification $\omega_0 \sim \Hub$ places the line at twice the Hubble scale. Every laboratory experiment, however, sits in the quasi-static regime $T \ll \gamma^{-1}\sim \Hub^{-1}$, where dephasing depends on the equal-time variance $\mathcal A_J$ alone and not on $\omega_0$; the line position is therefore empirically inert. A holographic per-mode mass $m_R^{\mathrm{hol}} = m_{\mathrm{Pl}}/\sqrt{N_{\mathrm{dS}}}\simeq 1.4\times10^{-69}\,\mathrm{kg}$ gives $m_R^{\mathrm{hol}}c^2 \sim \hbar\Hub$ and a natural rate $\lambda_{\mathrm{natural}}\sim \Hub\,\mathcal C_{\mathrm{match}}$; the benchmark $\lambda_{\mathrm{bench}}\sim\alpha_{\mathrm{em}}^2\Hub\sim10^{-22}\,\mathrm{s}^{-1}$ is conditional on an undetermined prefactor.

The contribution is structural rather than predictive: GTD supplies a candidate microscopic source for collapse-model noise within a fundamental theory, but converting this into a parameter-free objective-collapse theory still requires further matching hypotheses --- the system--bath coupling, spatial kernel, Adler--Millard vertex/propagator scaling, stochastic unravelling, and the microscopic origin of $m_R^{\mathrm{hol}}$ --- none of which is closed here. The paper makes these explicit and catalogues the open problems.
\end{abstract}

\tableofcontents

\section{Introduction}
\label{sec:intro}

The dynamical reduction programme \cite{GRW1986,Pearle1989,GhirardiPearleRimini1990,Bassi2013} proposes that the linear superposition principle is modified by stochastic, non-unitary, norm-preserving terms in the Schr\"odinger equation. In its standard nonrelativistic mass-proportional form (Continuous Spontaneous Localization, CSL), it is governed by two phenomenological parameters: a rate $\lambda$ and a spatial correlation length $r_C$. The model is in tension with a growing set of laboratory tests --- X-ray spontaneous emission from atoms \cite{Donadi2021}, the XENONnT bound \cite{XENON2026}, the Majorana Demonstrator search \cite{Majorana2022}, ultracold cantilevers \cite{Vinante2020}, cold-atom experiments \cite{Bilardello2016}, and gravitational-wave detectors \cite{Carlesso2016} --- but the noise field, its spectrum, and the localization length remain phenomenological inputs rather than derived quantities. The general framework for coloured (non-white) collapse noise was developed by Adler and Bassi \cite{AdlerBassi2007,AdlerBassiDonadi2013}; coloured-noise CSL is the natural setting for the present paper. Gravitationally motivated alternatives, in particular the Di\'osi--Penrose family of models \cite{Diosi1987,Diosi1989,Penrose1996}, replace the postulated CSL noise by a noise field of gravitational origin, but the underlying microscopic source of the noise is still postulated rather than derived from a deeper dynamics.

Generalized Trace Dynamics (GTD), built on the pre-quantum trace-dynamics programme of Adler \cite{Adler2004,Adler2007} and developed in the pre-quantum, pre-spacetime  formulation by Singh and collaborators \cite{Palemkota2020,Roy2021,Kakade2023,Singh2026Fermions,Finster2026}, aims to derive standard quantum mechanics as an emergent thermodynamic limit, with deviations driven by Adler--Millard fluctuations. The natural question is whether, as part of this emergence, GTD can supply a microscopic candidate for the noise field of collapse models, including its colour and absolute amplitude. The answer developed below is conditional: it requires an auxiliary fermionic oscillator sector that is motivated by GTD but not derived from the minimal Grassmann algebra.

\paragraph{Aim of this paper.} The aim of the present paper is to test whether GTD can supply, after explicit auxiliary structural assumptions, a microscopic source for the colour and amplitude of collapse-driving noise. To our knowledge, no prior collapse model produces a noise spectrum from an underlying candidate dynamics: in standard CSL and its coloured-noise extensions \cite{AdlerBassi2007,AdlerBassiDonadi2013}, the spectrum is chosen to fit phenomenology, and even the gravitationally motivated Di\'osi--Penrose family postulates rather than derives the noise field. The proposal of this paper is that GTD, equipped with the structural postulates catalogued in the ledger below, supplies a candidate microscopic source whose two-point function can be computed by Wick contraction. The result is a narrow-band correlator at the Hubble scale, with an explicit amplitude computed as a combination of GTD parameters inside the auxiliary fermionic sector. Whether this constitutes a viable collapse model --- as opposed to a candidate spectral mechanism within a broader matching framework --- depends on closing the further matching hypotheses we identify; we are explicit throughout about which calculations are direct consequences of the written action, which are conditional on auxiliary postulates, and which are phenomenological matching choices.

\paragraph{A calibration about laboratory bounds.} A point of calibration about the laboratory phenomenology should be stated up front, since the paper's claim to ``consistency with laboratory bounds'' rests on a structural rather than a parametric mechanism. Existing high-frequency CSL bounds (XENONnT \cite{XENON2026}, Majorana \cite{Majorana2022}, X-ray spontaneous emission \cite{Donadi2021}) constrain the noise spectral density at the relevant photon or phonon frequency, typically in the keV or mechanical-Hz range. The GTD-motivated spectrum sits at $2\Hub \sim 10^{-18}\,\mathrm{s}^{-1}$, far below any laboratory band; the off-resonance suppression at X-ray frequencies is of order $10^{-72}$ (Table~\ref{tab:suppression}). Existing bounds therefore do not constrain the GTD line, but this is because the line lies in an empirically inaccessible band, not because the framework satisfies a non-trivial constraint. The laboratory content of the proposal is the equal-time variance $\mathcal A_J$ and the corresponding $T^2$ Gaussian dephasing law, both of which apply at all frequencies and which the existing bounds do not directly address.

In this paper we show that the answer is conditionally yes for the colour (the spectral location of the noise) within the auxiliary fermionic oscillator construction, and, under further matching hypotheses, only as a placeholder for the amplitude. We adopt the following convention throughout: each non-trivial equation is tagged \derived\ if it is derived directly from the written GTD action and its specified quantization, \conditional\ if it follows from a clearly stated additional assumption or auxiliary effective construction, and \matching\ if it represents an identification of GTD quantities with phenomenological CSL quantities. Statements without a tag are standard.

\paragraph{Scope of the present paper.} The main calculations are:
\begin{enumerate}
\item[(R)] The principal result, conditional on the postulates enumerated below: in an auxiliary canonical fermionic Fock-space quantization of the matrix variable $q_F$ from the GTD aikyon decomposition, with the nilpotent pure-fermion coefficient replaced by an ordinary scalar parameter, the connected Wightman two-point function of the bilinear $J = \mathrm{Tr}(q_F^\dagger q_F)$ is $\langle 0|J(\tau)J(0)|0\rangle = \mathcal A_J\,e^{-2i\sigma\omega_0\tau}$ with computed amplitude $\mathcal A_J = (\hbar/2 m_R\,\omega_0\,L_{\mathrm{aik}}^2)^2\cdot\mathcal N\cdot D$. The Fourier spectrum has support at $|\omega| = 2\omega_0$. (Sections \ref{sec:gtd}, \ref{sec:spectrum}.)
\item[(M1)] A Born--Markov master equation derivation for a system bilinearly coupled to this bath, in both the broadened-Markovian and quasi-static regimes, with complete positivity verified in each. The bilinear coupling is itself a matching hypothesis. (Section \ref{sec:master}.)
\item[(M2)] A spatial-kernel analysis identifying the precise mechanism by which a homogeneous bath fails to localize translated branches, and three concrete routes (sub-leaf locality, internal-state coupling, matter-proportional clustering) that could generate a localization length. We do \emph{not} derive a CSL-scale $r_C$ from GTD. (Section \ref{sec:spatial}.)
\item[(M3)] An enumeration of bath-size scalings consistent with the Adler--Millard relative fluctuation $N_1^{-1/2}$, showing exactly which microscopic structural assumption produces a finite collapse rate in the thermodynamic limit. (Section \ref{sec:AM}.)
\item[(M4)] A dimensional analysis of the bath amplitude with cosmological matching, and identification of why the cosmological energy budget does not by itself fix the collapse-rate normalization. (Section \ref{sec:norm}.)
\item[(M5)] Energy balance, sign-convention discussion, complete positivity, and the distinction between decoherence and collapse. (Section \ref{sec:energy}.)
\end{enumerate}
Section \ref{sec:rate} converts these into a benchmark estimate for the CSL rate; Section \ref{sec:newphysics} identifies the genuinely new physics; Section \ref{sec:tests} lists falsifiable experimental signatures; Section \ref{sec:conclusions} concludes.

\paragraph{Ledger of load-bearing structural postulates.}\label{sec:ledger} Before proceeding it is important to make explicit the structural postulates on which the result (R) depends. None of these is derived from the GTD trace action in its current formulation; each is either an independent postulate or a matching choice. Reading the paper in light of this ledger is essential for calibrating the strength of every downstream claim.

\begin{itemize}
\item[(P1)] \textbf{Canonical fermionic Fock-space quantization of $q_F$.} In the GTD aikyon decomposition $q_i = q_B + a_0\beta_i q_F$, the matrix variable $q_F$ appears only as a coefficient of the Grassmann generators $\beta_i$. Treating $q_F$ as an independent fermionic matrix field with its own canonical anti-commutation relations $\{b^{(a)}, b^{(c)\dagger}\} = \delta^{ac}$ is a postulate, not a consequence of the original minimal Lagrangian. It is adopted because the natural reduction of the fermionic-current ansatz $J^\mu \propto \mathrm{Tr}(q_F^\dagger\Gamma^\mu q_F)$ to ghost-mode operators would require inverting the nilpotent $\delta\beta = \beta_2 - \beta_1$, which is ill-defined in the minimal $\{\beta_1, \beta_2\}$ Grassmann algebra (Section~\ref{sec:Jansatz}). All ``derivations'' of the current correlator in this paper should be read as derivations conditional on (P1) and on the scalarization postulate (P5) below.''
\item[(P2)] \textbf{Positive-norm canonical quantization.} The fermionic ghost Lagrangian arising under (P5) below has an unbounded-below Hamiltonian in the positive-norm Fock representation, with known dynamical pathologies. Alternative quantizations (Krein-space, PT-symmetric, BRST-constrained) are expected to preserve the spectral support at $|\omega| = 2\omega_0$ but may modify the explicit form of $\mathcal A_J$ or its sign. Explicit verification across these alternatives is left to future work.
\item[(P3)] \textbf{Scalar bilinear bath operator.} The choice $J = \mathrm{Tr}(q_F^\dagger q_F)$ (or its vector generalization) over other admissible bilinears of $q_F$ is not derived. Different bilinears would generally give different spectra.
\item[(P4)] \textbf{Cosmological identification $\omega_0 = \Hub$.} The GTD oscillator frequency is identified with the Hubble scale on the basis of cosmological matching with the de~Sitter mode count (Section~\ref{sec:norm}); this is order-of-magnitude reasoning, not a parameter-free derivation.
\item[(P5)] \textbf{Effective scalar body for the pure-fermion coefficient.} In the minimal Grassmann algebra the product $\beta_1\beta_2$ is Grassmann-even but nilpotent; it has no ordinary sign, no ordinary modulus, and no inverse. Consequently the equation $\beta_1\beta_2(\ddot q_F+\omega_0^2 q_F)=0$ does not imply an ordinary oscillator equation for $q_F$. We therefore replace the pure-fermion coefficient $a_0^2\beta_1\beta_2$ by an auxiliary ordinary scalar coefficient $\varepsilon_F=\sigma\xi^2$ with $\xi>0$ and $\sigma=\pm1$, and define a canonically normalized variable $\tilde q_F=\xi q_F$ (absorbing $a_0$ into $\xi$). This is a regularizing/effective-body postulate, not a derivation from the minimal algebra. We adopt $\sigma=+1$ as the working branch throughout, giving a Wightman line at $\omega=-2\omega_0$; the alternative $\sigma=-1$ is recorded in Section~\ref{sec:FockJ}.
\item[(P6)] \textbf{Mass-matching $m_F = m_R$.} The cancellation $H_{\mathrm{as}}|0\rangle = 0$ on the fermionic Fock vacuum (Appendix~\ref{app:hasvac}) requires the fermionic and bosonic oscillator masses to coincide. This is a matching condition not derived from the action.
\item[(P7)] \textbf{Bilinear matter-bath coupling.} The local interaction $g_{\mathrm{int}}\hat M\,\Xi$ between matter and the bath source (Section~\ref{sec:master}) is a phenomenological matching choice, not derived from the GTD action.
\item[(P8)] \textbf{Spatial localization kernel.} GTD does not derive a CSL-scale spatial kernel; routes (S1)--(S3) of Section~\ref{sec:spatial} identify the candidate mechanisms.
\item[(P9)] \textbf{Adler--Millard $N_1^{-1/4}$ propagator suppression.} The thermodynamic limit requires both vertex and propagator $N_1^{-1/4}$ scalings (Section~\ref{sec:AM}); the propagator side is not motivated by the central-limit argument and remains a postulate.
\item[(P10)] \textbf{Stochastic unravelling.} The mapping from linear master equation to nonlinear Born-rule-respecting collapse equation is not derived in this paper.
\item[(P11)] \textbf{Holographic identification of the aikyon oscillator mass.} The Lagrangian parameter $m_R \equiv a_1 a_0/2$ from \eqref{eq:LagAik} is identified with the de-Sitter per-mode mass scale $m_R^{\mathrm{hol}} = m_{\mathrm{Pl}}/\sqrt{N_{\mathrm{dS}}} \simeq 1.4\times10^{-69}\,\mathrm{kg}$ (Section~\ref{sec:norm}, Eq.~\eqref{eq:mholo}). This identification is not derived from the GTD action: the microscopic constants $a_0, a_1$ are not computed within GTD as currently formulated, and the holographic per-mode mass is an order-of-magnitude matching with the cosmological-constant scale rather than an equation of motion. With this choice, the dimensionless ratio $\hbar\Hub/(m_R c^2)$ is $O(1)$ and the natural rate is $\lambda_{\mathrm{natural}}\sim \Hub\,\mathcal C_{\mathrm{match}}$; under an alternative Planck-mass assignment $m_R = m_{\mathrm{Pl}}$ the same formula would give $\lambda_{\mathrm{natural}}\sim 10^{-122}\,\Hub\sim 10^{-140}\,\mathrm{s}^{-1}$. The viability of the benchmark rate is therefore conditional on (P11).
\end{itemize}
Postulates (P1)--(P5) underlie the spectrum calculation in Section~\ref{sec:FockJ}; in particular, (P5) is the step that turns the degenerate nilpotent pure-fermion coefficient of the minimal algebra into a non-degenerate oscillator coefficient. (P6)--(P10) enter at later stages of the matching model; (P11) enters the absolute normalization in Sections~\ref{sec:norm} and~\ref{sec:rate}. This is the structural cost of the present construction; the paper's contribution is not to remove these postulates but to make them explicit and to compute the consequences of each.

\paragraph{Empirical inertness of the Hubble-scale line position.} It is also important to flag, before the technical sections, a quantitative fact that the casual reader could otherwise miss. Every conceivable laboratory experiment sits deep in the quasi-static regime $T \ll \gamma^{-1} \sim \Hub^{-1} \sim 10^{18}\,\mathrm{s}$. In this regime, the dephasing functional depends only on the equal-time variance $\mathcal A_J$ of the symmetrized correlator (Section~\ref{sec:howitworks}); the line frequency $2\omega_0$ does not enter the laboratory rate. The much-advertised ``bandwidth-suppression argument'' against existing high-frequency CSL bounds therefore reduces, on inspection, to the observation that the predicted spectral support lies far below any laboratory band --- which is to say, the predicted spectrum is at an unobservable frequency. The line position re-enters only at cosmological time scales no experiment reaches. The substantive laboratory content of the proposal is the amplitude $\mathcal A_J$ and the implied $T^2$ Gaussian dephasing, not the line position per se.

\paragraph{Holographic natural scale and the benchmark.} Finally, the holographic mass estimate used below is
\begin{equation}
 m_R^{\mathrm{hol}} = \frac{m_{\mathrm{Pl}}}{\sqrt{N_{\mathrm{dS}}}} \simeq 1.4\times10^{-69}\,\mathrm{kg},
\end{equation}
so that $m_R^{\mathrm{hol}}c^2 \simeq 1.3\times10^{-52}\,\mathrm{J}$, of order $\hbar\Hub \simeq 2.3\times10^{-52}\,\mathrm{J}$. In the holographic per-mode normalization, the natural Markovian-surrogate rate has the form
\begin{equation}
\lambda_{\mathrm{natural}} \;\sim\; \Hub\,\mathcal C_{\mathrm{match}},
\end{equation}
where $\mathcal C_{\mathrm{match}}$ collects the still-undetermined system--bath coupling, spatial-kernel normalization, matrix trace factor, Dirac trace factor, and Adler--Millard matching factors. If $\mathcal C_{\mathrm{match}}\sim1$, the one-second amplification threshold is $N\sim\Hub^{-1/2}\sim7\times10^8$ nucleons. If $\mathcal C_{\mathrm{match}}=\alpha_{\mathrm{em}}^2$, the benchmark becomes $\lambda_{\mathrm{bench}}=\alpha_{\mathrm{em}}^2\Hub\simeq1.2\times10^{-22}\,\mathrm{s}^{-1}$ and the usual $N\sim10^{11}$ threshold follows. The benchmark rate should still be read \emph{not} as a parameter-free prediction of GTD, but as a placeholder for the outcome of an as-yet-unperformed microscopic derivation of $\mathcal C_{\mathrm{match}}$.

\subsection{The derivation at a glance}
\label{sec:roadmap}

For the reader's convenience we summarise the structure of the argument in nine steps before proceeding to the technical sections.

\begin{enumerate}
\item \textbf{The setup.} The single-aikyon GTD Lagrangian has the cross-kinetic form $L = m_R\Tr(\dot q_1 \dot q_2 - \omega_0^2 q_1 q_2)$. Varying $q_1$ produces the equation of motion for $q_2$, and vice versa: the two matrix variables are dynamically coupled at the free level. (Section~\ref{sec:gtd}.)

\item \textbf{Bateman diagonalisation.} The normal-mode combinations $Q_\pm = (q_1\pm q_2)/\sqrt 2$ diagonalise the Hamiltonian into a \emph{Bateman pair}: one ordinary oscillator and one ghost-sign oscillator at the same frequency $\omega_0$. The minus sign appears in the energy contribution, not in the Fock-space norm. (Section~\ref{sec:gtd}.)

\item \textbf{Splitting the Hamiltonian.} Adler's self-adjoint/anti-self-adjoint decomposition of trace-dynamics Hamiltonians places ordinary quantum mechanics in the self-adjoint sector and locates collapse-driving dynamics in the anti-self-adjoint corner.

\item \textbf{The candidate noise operator.} The natural ansatz for a noise operator built from the anti-self-adjoint sector is the fermionic current $J^\mu = (1/L_{\mathrm{aik}}^2)\,\mathrm{Tr}(q_F^\dagger \Gamma^\mu q_F)$, or its scalar version $J = (1/L_{\mathrm{aik}}^2)\,\mathrm{Tr}(q_F^\dagger q_F)$, built from the matrix variable $q_F$. The naive reduction of $q_F$ to a function of the bosonic Bateman ghost mode requires inverting the Grassmann-odd combination $\delta\beta = \beta_2 - \beta_1$, which is not well-defined in the minimal $\{\beta_1, \beta_2\}$ algebra. We therefore work with $J$ directly in an auxiliary canonical fermionic Fock-space quantization of $q_F$. (Section~\ref{sec:spectrum}.)

\item \textbf{The effective scalar coefficient.} The pure-fermionic part of the minimal Lagrangian is $L_F = m_R\,a_0^2\,\beta_1\beta_2\,\mathrm{Tr}(\dot q_F^2 - \omega_0^2 q_F^2)$. Because $\beta_1\beta_2$ is nilpotent in the minimal algebra, it cannot be assigned an ordinary sign or modulus and cannot by itself define a canonical oscillator normalization. Postulate (P5) replaces this nilpotent coefficient by an auxiliary ordinary scalar coefficient $\varepsilon_F=\sigma\xi^2$. We adopt $\sigma = +1$ throughout the body of the paper (branch (-) of Section~\ref{sec:FockJ}): a standard fermionic Lagrangian with Heisenberg evolution $b(\tau) = b\,e^{-i\omega_0\tau}$, yielding a Wightman line at $\omega = -2\omega_0$ in the standard energy-flow direction. The alternative $\sigma = -1$ (branch (+), fermionic ghost Lagrangian with reversed Heisenberg phase, line at $\omega = +2\omega_0$, anti-KMS direction) is recorded as a remark; its phenomenological predictions through the symmetrized correlator are identical to those of branch (-), so the choice does not affect the laboratory observables.

\item \textbf{The spectral result.} The connected two-point function of $J$ on the fermionic Fock vacuum, computed by Wick contraction in the working branch $\sigma = +1$, is
\begin{equation*}
\langle 0|J(\tau)J(0)|0\rangle \;=\; \mathcal A_J\,e^{-2i\omega_0\tau}, \qquad \mathcal A_J \;=\; \left(\frac{\hbar}{2 m_R\,\omega_0\,L_{\mathrm{aik}}^2}\right)^{\!2}\!\!\cdot \mathcal N \cdot D,
\end{equation*}
where $\mathcal N$ is a matrix-trace counting factor (scaling as $N^2$ for an $N\times N$ matrix variable $q_F$) and $D$ is the Dirac-trace factor for the chosen current. The Fourier transform $S^{>}(\omega) = 2\pi\mathcal A_J\,\delta(\omega + 2\omega_0)$ is supported at a single frequency $\omega = -2\omega_0$, the bath at zero effective temperature emitting into the system at $\omega = 2\omega_0$ (standard energy-flow direction). The amplitude $\mathcal A_J$ is computed by Wick contraction in the auxiliary sector; as the ledger above notes, $m_R$, $\alpha_{\mathrm{GTD}}$, $\mathcal N$, and $D$ are themselves either free parameters of the theory or matching choices, so the result is not a parameter-free prediction. The corresponding symmetrized correlator is $C_{\mathrm{sym}}(\tau) = \mathcal A_J\cos(2\omega_0\tau)$. A surrogate bosonic ghost-mode operator $J_{\mathrm{eff}} = \kappa\,{:}X^2:$ is recovered as a consistency check (Section~\ref{sec:Jeff_consistency}) under the alternative $\sigma = -1$ choice; the surrogate's free parameter $\kappa$ is identified with the corresponding combination of GTD microscopic parameters. \emph{Remark on branch} (+). With $\sigma = -1$ the line moves to $\omega = +2\omega_0$ (anti-KMS direction) and $\mathcal A_J$ is unchanged. Because the symmetrized correlator $C_{\mathrm{sym}}$ is the same in both branches, all laboratory phenomenology developed below is independent of the choice. (Section~\ref{sec:spectrum}.)

\item \textbf{Cosmological identification.} If $\omega_0\sim\Hub$, motivated by the cosmological matching between aikyon energies and the de Sitter degree-of-freedom count, the angular-frequency line sits at $\omega\sim 10^{-18}\,\mathrm{s}^{-1}$ (cyclic frequency $f\sim 10^{-19}\,\mathrm{Hz}$), far below all laboratory bands. (Sections~\ref{sec:norm}, \ref{sec:rate}.)

\item \textbf{Phenomenological consequence.} High-frequency emission tests --- X-ray, mechanical, interferometric --- lie in the off-resonant Lorentzian tail of this line and are bandwidth-suppressed by factors as small as $10^{-72}$. The white-noise CSL bounds (XENONnT included) therefore do not directly constrain a narrow GTD line without a model-dependent spectral-broadening calculation. (Sections~\ref{sec:invisible}, \ref{sec:tests}.)

\item \textbf{What remains.} A complete collapse model needs four further matching hypotheses: a system--bath coupling, a spatial localisation kernel, the Adler--Millard vertex scaling (contingent on an unsupplied propagator-suppression argument), and a stochastic unravelling beyond the linear reduced dynamics derived here. We treat each explicitly and identify the calculational gap at each step. The benchmark rate of Section~\ref{sec:rate} rests on a choice of suppression factor that the present theory does not derive. The diagrammatic framework within which these calculations could be addressed is outlined in Appendix~\ref{app:future}.
\end{enumerate}

\paragraph{Note on additional matching ingredients.} The four matching hypotheses listed above are the headline ingredients connecting the spectral mechanism to a CSL phenomenology. Three further matching ingredients are required at specific points in the analysis: a two-kernel ansatz (Section~\ref{sec:spatial}) separating the short-range localization kernel of individual-apparatus dephasing from any long-range coherence kernel responsible for cross-sensor correlations; a fermionic--bosonic mass-ratio condition $m_F = m_R$ (postulate (P6); Appendix~\ref{app:hasvac}, Eq.~\eqref{eq:mFmatching}) required for the cancellation $H_{\mathrm{as}}|0\rangle = 0$ on the Fock vacuum; and the holographic identification of the aikyon oscillator mass $m_R$ with the de-Sitter per-mode scale $m_R^{\mathrm{hol}} = m_{\mathrm{Pl}}/\sqrt{N_{\mathrm{dS}}}$ (postulate (P11); Section~\ref{sec:norm}, Eq.~\eqref{eq:mholo}) required for the natural rate of Section~\ref{sec:rate} to lie within reach of the cosmological-arithmetic window. We tag each [M] where it arises rather than introducing them here, but the reader should be aware that the catalogue of matching hypotheses is in this sense broader than the headline four.

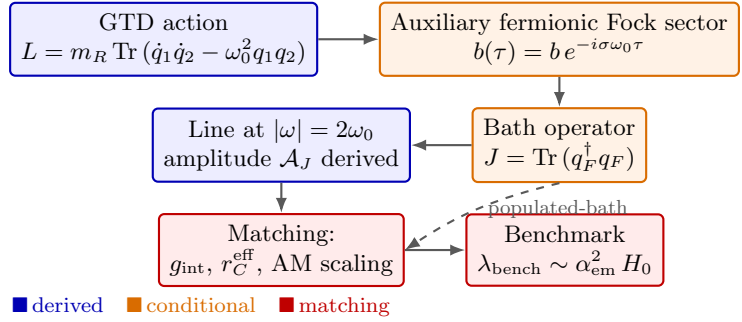
\begin{figure}[ht]
\centering
\begin{tikzpicture}[
  node distance=4mm and 8mm,
  every node/.style={font=\footnotesize},
  derived/.style={draw=blue!70!black,thick,rectangle,rounded corners=2pt,inner sep=4pt,fill=blue!7,align=center,minimum height=8mm},
  conditional/.style={draw=orange!85!black,thick,rectangle,rounded corners=2pt,inner sep=4pt,fill=orange!10,align=center,minimum height=8mm},
  matching/.style={draw=red!75!black,thick,rectangle,rounded corners=2pt,inner sep=4pt,fill=red!8,align=center,minimum height=8mm},
  arr/.style={-{Latex[length=2mm]},thick,gray!70!black}
]
\node[derived] (lag) {GTD action \\ $L = m_R\,\Tr(\dot q_1\dot q_2 - \omega_0^2 q_1 q_2)$};
\node[conditional,right=of lag] (bat) {Auxiliary fermionic Fock sector \\ $b(\tau) = b\,e^{-i\sigma\omega_0\tau}$};
\node[conditional,below=of bat] (jeff) {Bath operator \\ $J = \Tr(q_F^\dagger q_F)$};
\node[derived,left=of jeff] (line) {Line at $|\omega| = 2\omega_0$ \\ amplitude $\mathcal A_J$ derived};
\node[matching,below=of line] (matchg) {Matching: \\ $g_{\mathrm{int}}$, $r_C^{\mathrm{eff}}$, AM scaling};
\node[matching,right=of matchg] (bench) {Benchmark \\ $\lambda_{\mathrm{bench}} \sim \alpha_{\mathrm{em}}^2\,\Hub$};
\draw[arr] (lag) -- (bat);
\draw[arr] (bat) -- (jeff);
\draw[arr] (jeff) -- (line);
\draw[arr] (line) -- (matchg);
\draw[arr] (matchg) -- (bench);
\draw[arr,dashed] (jeff.south) to[bend right=10] node[midway,right,scale=0.85]{populated-bath} (matchg.east);
\node[anchor=north west,xshift=-1mm] at (current bounding box.south west)
  {\scriptsize{\textcolor{blue!70!black}{$\blacksquare$\,derived}\quad\textcolor{orange!85!black}{$\blacksquare$\,conditional}\quad\textcolor{red!75!black}{$\blacksquare$\,matching}}};
\end{tikzpicture}
\caption{Dependency tree of the central result. The GTD action (top-left, derived) motivates, but does not by itself derive, the auxiliary fermionic Fock sector (top-right, conditional), in which the nilpotent pure-fermion coefficient is replaced by an ordinary scalar parameter. Within that sector the bath operator $J = \mathrm{Tr}(q_F^\dagger q_F)$ is adopted as a candidate matter-coupling current. The connected two-point function gives a single Wightman line at $\omega = -2\omega_0$ with computed amplitude $\mathcal A_J$ (in the working branch $\sigma = +1$ adopted throughout the paper). Three matching hypotheses (red) enter the linear master equation and benchmark rate: the system--bath coupling $g_{\mathrm{int}}$, the spatial kernel $r_C^{\mathrm{eff}}$, and the Adler--Millard vertex scaling. A fourth matching hypothesis --- the stochastic state-vector unravelling beyond the linear reduced dynamics --- is orthogonal to the dependency chain shown here and is not visualised. The dashed arrow indicates the populated-bath correction (Section~\ref{sec:vacpop}), which attenuates the spectral asymmetry but preserves the line location.}
\label{fig:dependency}
\end{figure}

\subsection{Why the colour is unavoidable}
\label{sec:unavoidability}

A natural reaction to step~6 is to ask whether the line-spectrum result is an artefact of an inessential algebraic detail. Within the auxiliary fermionic oscillator construction adopted here, it is not. Given the algebraic inputs of that construction, the colour cannot be removed without changing the inputs themselves. The spectrum is forced to be a delta function at $|\omega| = 2\omega_0$ (with the sign of $\omega$ set by the effective scalar sign $\sigma$) by three structural inputs:
\begin{enumerate}
\item[(i)] the GTD aikyon Lagrangian with $q_i = q_B + a_0\beta_i q_F$, together with postulate (P5), which replaces the degenerate nilpotent pure-fermion coefficient by an ordinary quadratic oscillator coefficient at frequency $\omega_0$;
\item[(ii)] the Fock vacuum $b|0\rangle = d|0\rangle = 0$ on which expectation values are taken;
\item[(iii)] the choice of a bilinear bath operator $J = \mathrm{Tr}(q_F^\dagger q_F)$ (or its vector generalization) from the fermionic sector.
\end{enumerate}
The following modifications can change the sign of the line or its amplitude but cannot turn it into a white-noise spectrum without an additional and presently unmotivated structural ingredient:
\begin{itemize}
\item Alternative fermionic-ghost quantisations (PT-symmetric, Krein-space, BRST-constrained) in branch (+) should preserve the line at $|2\omega_0|$ with possibly different sign or amplitude conventions; explicit verification is left for future work.
\item Many-aikyon couplings broaden the delta function into a finite-width Lorentzian, and indeed are one mechanism by which the line could acquire a non-zero $\gamma$; but they cannot produce a frequency-independent noise floor unless the inter-aikyon dynamics introduces a hierarchy of frequency scales spanning many orders of magnitude, which is not present in the simplest extensions.
\item Alternative noise-driving operators from the fermionic sector (higher-order in $q_F$, or with derivatives) could produce different lines or multi-line spectra but, by the same argument, line spectra rather than noise floors.
\end{itemize}
The qualitative prediction --- that the present GTD-motivated auxiliary construction produces a narrow-band rather than a white-noise spectrum --- is therefore structural rather than an artefact of fitting. This is the principal way in which the proposal departs from the prior collapse-model literature, and it is the source of the experimental discriminants developed in Section~\ref{sec:tests}.

\subsection{How a single spectral line drives decoherence}
\label{sec:howitworks}

A natural conceptual question about the result of step~6 is how a delta-function spectrum --- support on a single frequency $|2\omega_0|$ --- can drive decoherence of laboratory superpositions whose Bohr frequencies $\omega_S$ are far from $2\omega_0$. The answer involves a subtlety that distinguishes the GTD bath from white-noise CSL and is worth stating before the technical sections, since it pre-empts a natural objection.

The decoherence rate produced by a bath depends on the regime in which the system samples it. Two regimes are relevant.

\paragraph{Markovian regime ($T \gg \gamma^{-1}$).} When the experimental duration exceeds the bath correlation time, the system samples the full bath spectrum and the decoherence rate is the spectral density evaluated at the system Bohr frequency, $\Gamma \propto S(\omega_S)$. For a delta-function spectrum at $\omega = 2\omega_0$ and system Bohr frequencies $\omega_S \neq 2\omega_0$, this rate is zero: the bath is off-resonant and decouples from the system. This is the regime in which a single line would, naively, be inert.

\paragraph{Quasi-static regime ($T \ll \gamma^{-1}$).} For a Hubble-width line, $\gamma^{-1} \sim \Hub^{-1} \sim 10^{18}\,\mathrm{s}$, and every laboratory experiment sits firmly in $T \ll \gamma^{-1}$. Here the system does not resolve the bath as a frequency-dependent spectrum at all. Each experimental run samples a single realisation of the bath, which is effectively static over the run, and the decoherence rate is governed by the equal-time correlator $C_{\mathrm{sym}}(0) = \mathcal A_J$ of the \emph{symmetrized} bath noise (route (A) of Section~\ref{sec:qvsc}) via the Gaussian-dephasing formula \eqref{eq:Gaussiandephasing}. The off-diagonal density-matrix elements decay as $\exp(-\Gamma_{\mathrm{qs}} T^2)$ with $\Gamma_{\mathrm{qs}} = (a-b)^2 \mathcal A_J/2\hbar^2$. The line frequency $2\omega_0$ does not appear in this rate. What matters is the amplitude $\mathcal A_J$. The line drives decoherence through its equal-time variance, not through resonant transitions.

The reconciliation is that $C_{\mathrm{sym}}(0) = \mathcal A_J = \int (d\omega/2\pi)\,S_{\mathrm{sym}}(\omega)$ regardless of where the spectrum is supported, and the route (B) quantum-bath calculation yields the same leading-order $T^2$ behaviour in the regime $\omega_0 T \ll 1$ (since the leading double-time integral of $e^{-2i\omega_0(t-s)}$ in the working branch, or its complex conjugate in the alternative branch, over $[0,T]^2$ reduces to $\mathcal A_J T^2$ to leading order when $\omega_0 T \ll 1$). A bath whose entire spectral weight sits at a single frequency still has a well-defined and non-zero variance, and that variance is what drives dephasing in the quasi-static regime. The location of the line determines \emph{when} the system crosses from quasi-static to Markovian (i.e.\ whether $T^2$ or $T$ scaling applies); the area under the line determines the \emph{magnitude} of the rate. For GTD with $\omega_0 = \Hub$, all foreseeable laboratory experiments sit deep in the $T^2$-scaling quasi-static regime, and the line position is irrelevant to the rate while being decisive for the bandwidth-suppression argument of Section~\ref{sec:invisible}.

This is how the proposal evades the naive objection that a sharp line at $\omega = 2\Hub$ should be inert at laboratory frequencies. It does not need to drive resonant absorption at $\omega_S \neq 2\omega_0$ in order to cause decoherence. It drives non-resonant Gaussian dephasing through the equal-time variance of the symmetrized correlator, and only the variance --- not the line position --- enters the laboratory rate. The line position re-enters only at cosmological time scales, which no experiment reaches.

\subsection{The measurement problem and macroscopic superpositions}
\label{sec:measurement}

Since the paper develops a microscopic source for collapse-model noise, the natural question is what it claims about the measurement problem and the absence of macroscopic superpositions. The honest answer separates into three points, recorded here at the outset to avoid the paper being read as claiming more than it delivers.

\paragraph{What is decoherence, not collapse.} The mechanism developed in this paper is decoherence in the standard open-systems sense: the linear master equation derived in Section~\ref{sec:master} drives the off-diagonal elements of the reduced density matrix to zero, but it does not select a definite outcome. The mapping from the linear master equation to a Born-rule-respecting nonlinear stochastic Schr\"odinger equation is non-unique \cite{Bassi2013}, and a GTD-specific unravelling, if it exists, must be derived from the structural content of the Adler--Millard fluctuations. This is postulate (P10) of the ledger and problem (F7) in Appendix~\ref{app:future}; the present paper does not solve the measurement problem in the strict sense.

\paragraph{Amplification is standard CSL, not a GTD-specific result.} The conventional collapse-model amplification $\Gamma_N \sim N^2 \lambda$ for rigid objects of $N$ nucleons with branch separation greater than the spatial kernel is a feature of the mass-proportional CSL coupling adopted in Section~\ref{sec:master}, postulate (P7) of the ledger. It is not specific to GTD; the same $N^2$ scaling appears in any mass-proportional collapse model with the same coupling form. The framework inherits this amplification because it adopts the same coupling, not because GTD derives it. The corresponding amplification threshold $N \sim 10^{11}$ nucleons quoted below is a phenomenological consequence of plugging the benchmark rate into the standard CSL amplification arithmetic; the GTD-specific content is the candidate microscopic source of the noise, not the amplification.

\paragraph{The benchmark threshold is conditional.} Under the benchmark $\lambda_{\mathrm{bench}} \sim \alpha_{\mathrm{em}}^2\Hub \sim 10^{-22}\,\mathrm{s}^{-1}$, the one-second amplification threshold is $N \sim 10^{11}$ nucleons or $m \sim 10^{-16}\,\mathrm{kg}$. Above this mass the off-diagonals of the reduced density matrix decay on sub-second time scales; below it, quantum coherence is preserved on laboratory time scales. The benchmark is conditional because the $\alpha_{\mathrm{em}}^2$ prefactor is not derived from the GTD action. In the holographic mass normalization $m_R^{\mathrm{hol}} = m_{\mathrm{Pl}}/\sqrt{N_{\mathrm{dS}}} \simeq 10^{-69}\,\mathrm{kg}$, the natural scale of the Markovian-surrogate rate is $\lambda_{\mathrm{natural}}\sim\Hub\,\mathcal C_{\mathrm{match}}$. Thus the microscopic-to-macroscopic crossover that collapse models are designed to produce depends on the dimensionless matching factor $\mathcal C_{\mathrm{match}}$, the spatial kernel, and the stochastic unravelling.

\paragraph{The position adopted.} This paper takes the position that GTD identifies a candidate microscopic source for collapse-model noise within a fundamental theory, computes the structural content of the resulting spectrum and master equation, and makes explicit the chain of postulates and matching factors needed to convert this content into a parameter-free objective-collapse theory. It does not claim to solve the strict measurement problem, to derive a parameter-free collapse rate, or to predict laboratory-detectable signatures at current sensitivities. The holographic normalization places the natural rate at $\Hub$ times a dimensionless matching factor; deriving that factor, the spatial kernel, and the Born-rule stochastic dynamics from GTD remains essential.

\section{The GTD framework and the Bateman structure}
\label{sec:gtd}

\subsection{Lagrangian and equations of motion}

In the Roy--Sahu--Singh variable choice \cite{Roy2021}, the single-aikyon Lagrangian is
\begin{equation}
L \;=\; m_R\,\Tr\!\left(\dot q_1 \dot q_2 - \omega_0^2\,q_1 q_2\right),
\qquad m_R \equiv \frac{a_1 a_0}{2},\quad \omega_0 \equiv \frac{\alpha_{\mathrm{GTD}}\, c}{L_{\mathrm{aik}}}.
\label{eq:LagAik}
\end{equation}
The dimensionless parameter $\alpha_{\mathrm{GTD}}$ is the GTD-internal coupling; it is \emph{not} a priori equal to the QED fine-structure constant $\alpha_{\mathrm{em}}$, and we keep the distinction explicit to avoid the conflation that has appeared in earlier literature. \derived\ The variables $q_1, q_2$ are matrix-valued and split into bosonic and fermionic parts,
\begin{equation}
q_1 = q_B + a_0 \beta_1 q_F, \qquad q_2 = q_B + a_0 \beta_2 q_F, \qquad \beta_1\neq\beta_2,
\label{eq:qdecomp}
\end{equation}
with $\beta_1,\beta_2$ fixed Grassmann-odd constants. Variation gives the cross-coupled equations $\ddot q_1 + \omega_0^2 q_1 = 0$ and $\ddot q_2 + \omega_0^2 q_2 = 0$, but the momenta are crossed: $p_1 = m_R \dot q_2$, $p_2 = m_R \dot q_1$. The Hamiltonian is
\begin{equation}
H \;=\; \Tr\!\left(\frac{p_1 p_2}{m_R} + m_R \omega_0^2\, q_1 q_2 \right).
\label{eq:Hamcross}
\end{equation}

\subsection{Bateman normal modes}

Introduce
\begin{equation}
Q_\pm = \frac{q_1\pm q_2}{\sqrt 2}, \qquad P_\pm = \frac{p_1\pm p_2}{\sqrt 2}.
\end{equation}
The Hamiltonian becomes
\begin{equation}
H \;=\; \Tr\!\left[\frac{1}{2 m_R}(P_+^2 - P_-^2) + \frac{m_R\omega_0^2}{2}(Q_+^2 - Q_-^2)\right],
\label{eq:HBateman}
\end{equation}
which is the Bateman pair: one positive-sign oscillator and one negative-sign (ghost-sign) oscillator at the same frequency \cite{Bateman1931}. \derived\ Standard creation/annihilation operators are introduced as
\begin{align}
Q_\pm &= \sqrt{\tfrac{\hbar}{2 m_R \omega_0}}\,(a_\pm + a_\pm^\dagger), \\
P_\pm &= -i\sqrt{\tfrac{\hbar m_R \omega_0}{2}}\,(a_\pm - a_\pm^\dagger),
\end{align}
giving
\begin{equation}
H = \hbar\omega_0\!\left(a_+^\dagger a_+ - a_-^\dagger a_-\right).
\label{eq:Hbateman_diag}
\end{equation}
We adopt $[a_\pm,a_\pm^\dagger]=1$, so the Fock space carries a positive-definite norm; what is negative in the $-$ sector is the contribution to the energy, not the norm. We define the Fock vacuum by $a_\pm|0\rangle = 0$.

\conditional\ This is not a ground state of \eqref{eq:Hbateman_diag}: on the Fock basis used here the eigenvalues of \eqref{eq:Hbateman_diag} are real but unbounded below in the $-$ sector. It is nevertheless the natural Fock vacuum for the Bateman algebra under the standard positive-norm canonical-commutation choice. The known pathologies of this choice are a substantive issue. The unbounded spectrum, and in more complete Bateman treatments the appearance of unstable modes or complex resonant poles, have been known since the original Bateman dual-oscillator analyses \cite{Bateman1931,FeshbachTikochinsky1977}. They have motivated a body of work developing alternative quantizations --- in particular PT-symmetric and imaginary-scaling treatments, and Krein-space inner-product constructions \cite{BlasoneJizba2002,Chruscinski2006,Deguchi2019} --- designed to cure these instabilities. Under those alternative quantizations, the structure of the spectral asymmetry derived below may change in significant ways: the line at $|\omega| = 2\omega_0$ could persist with a different amplitude or sign, could be replaced by a real symmetric two-line structure at $\pm 2\omega_0$, or could acquire a different physical interpretation as bath absorption versus emission. The simplest positive-norm choice we adopt here is therefore a load-bearing structural choice; we record it as one of the open structural points (Section~\ref{sec:conclusions}, item~9) and proceed with the simplest choice, with the understanding that the entire spectral-asymmetry phenomenology of (N3) is conditional on this choice.

All subsequent calculations are conditional on this quantization.

The Heisenberg evolution is
\begin{equation}
a_+(\tau) = a_+ e^{-i\omega_0\tau},\qquad a_-(\tau) = a_- e^{+i\omega_0\tau}.
\label{eq:heisenberg}
\end{equation}
The ghost-sign Bateman mode carries the time-reversed phase: this is the only place in the calculation where the ghost character of the construction enters, and it is what drives the entire spectral asymmetry.

\subsection{The anti-self-adjoint Hamiltonian on the Fock vacuum}
\label{sec:Hasvac}

Adler's trace-dynamics formulation splits the Hamiltonian \eqref{eq:Hamcross} of any single-aikyon configuration into a self-adjoint and an anti-self-adjoint part with respect to the matrix trace:
\begin{equation}
H \;=\; H_{\mathrm{sa}} + H_{\mathrm{as}}.
\end{equation}
$H_{\mathrm{sa}}$ generates ordinary unitary quantum dynamics in the thermodynamic limit, while $H_{\mathrm{as}}$ is the residue identified in Adler's programme as the natural seed of collapse-driving non-unitarity. It is therefore the first operator to inspect as a candidate noise source.

\paragraph{Complementary GTD-internal route via $H_{\mathrm{FF}}$.} A companion note by the present author~\cite{Singh2026Fermions} addresses the same decomposition from a complementary angle and is directly relevant to the obstruction encountered in Section~\ref{sec:Jansatz}. Starting from the STM-atom trace Lagrangian written in terms of two inequivalent matrix velocities $\dot Q_1, \dot Q_2$, and computing the trace Hamiltonian via trace-derivative canonical momenta with bosonic and fermionic variations treated separately, that construction isolates a purely fermionic contribution
\[
H_{\mathrm{FF}} \;=\; c\,\mathrm{Tr}\!\bigl[\lambda^3\,\beta_1\,\dot Q_F^\dagger\,\beta_2\,\dot Q_F\bigr]
\]
and shows, using graded cyclicity together with the natural adjoint properties $\beta_a^\dagger = \eta_a\beta_a$ with $\eta_1\eta_2 = +1$, that $H_{\mathrm{FF}}^\dagger = -H_{\mathrm{FF}}$. In other words, $H_{\mathrm{FF}}$ is purely anti-self-adjoint, and is non-trivial precisely because $\beta_1 \neq \beta_2$: if $\beta_1 = \beta_2 \equiv \beta$, the Grassmann-odd property $\beta^2 = 0$ would force $H_{\mathrm{FF}}$ to vanish identically. The same Grassmann-even product $\beta_1\beta_2$ that obstructs the inversion in Section~\ref{sec:Jansatz} enters here as a non-trivial coefficient: in $H_{\mathrm{FF}}$ it sits between two fermionic factors and is acted on by graded cyclicity, neither of which requires its sign or modulus to be defined as an ordinary scalar. The construction in~\cite{Singh2026Fermions} therefore extracts the anti-self-adjoint content of the GTD trace Hamiltonian without inverting $\delta\beta$ and without taking $\mathrm{sgn}(\beta_1\beta_2)$. This is consistent with our finding below that $H_{\mathrm{as}}|0\rangle=0$ on the free-aikyon Fock vacuum (under the mass-ratio matching of Appendix~\ref{app:hasvac}), with collapse-channel content emerging instead from the higher-order fermionic current of Section~\ref{sec:spectrum}. The two constructions are complementary, not redundant: the present paper computes the two-point function of a fermionic-current bilinear in an auxiliary Fock model, while~\cite{Singh2026Fermions} identifies the linear anti-self-adjoint Hamiltonian directly from the GTD trace structure. Bridging the two --- deriving the spectral content of $H_{\mathrm{FF}}$ correlation functions, or showing that $H_{\mathrm{FF}}$ acts trivially on the Fock vacuum of the present model so that bilinear correlators become the operative noise --- is an explicit open problem (item~12 of Section~\ref{sec:conclusions}).

To exhibit $H_{\mathrm{as}}$ explicitly, substitute the bosonic--fermionic decomposition \eqref{eq:qdecomp} into the crossed Hamiltonian \eqref{eq:Hamcross}. With $q_i = q_B + a_0\beta_i q_F$ and the corresponding momentum decomposition $p_i = p_B + a_0\beta_i p_F$ (obtained from $p_1 = m_R \dot q_2$, $p_2 = m_R \dot q_1$ and \eqref{eq:qdecomp}), the products $q_1 q_2$ and $p_1 p_2$ each generate three structurally distinct kinds of terms:
\begin{itemize}
\item Pure bosonic bilinears in $(q_B, p_B)$, with no Grassmann factors;
\item Pure fermionic bilinears in $(q_F, p_F)$, weighted by $\beta_1\beta_2$;
\item Boson--fermion cross bilinears such as $q_B q_F$, $q_F q_B$, $p_B p_F$, $p_F p_B$, weighted by a single power $a_0\beta_1$ or $a_0\beta_2$.
\end{itemize}
Under the matrix trace, the pure-bosonic and pure-fermionic bilinears (and combinations thereof) are self-adjoint and constitute $H_{\mathrm{sa}}$. The cross bilinears carry a single Grassmann-odd factor that makes them anti-self-adjoint under the trace; collected together, they constitute $H_{\mathrm{as}}$:
\begin{equation}
H_{\mathrm{as}} \;=\; a_0(\beta_1+\beta_2)\,\Tr\!\left[\frac{1}{m_R}\bigl(p_B\,p_F + p_F\,p_B\bigr) + m_R\omega_0^2\bigl(q_B\,q_F + q_F\,q_B\bigr)\right].
\label{eq:Hasdef}
\end{equation}
This is the operator we will evaluate on the Fock vacuum.

The bosonic operators $(q_B, p_B)$ can be re-expressed in the Bateman ladder basis $(a_\pm,\,a_\pm^\dagger)$ introduced in \eqref{eq:HBateman}--\eqref{eq:Hbateman_diag}; the bosonic Fock vacuum is defined by $a_\pm |0\rangle_B = 0$. The fermionic operators $(q_F, p_F)$ inherit canonical anticommutation relations on a separate fermionic Fock space with vacuum $|0\rangle_F$, and the combined Fock vacuum is $|0\rangle = |0\rangle_B \otimes |0\rangle_F$. Each cross bilinear in \eqref{eq:Hasdef} is then a product of a bosonic-sector factor (linear in the Bateman ladder operators) and a fermionic-sector factor (linear in the fermionic ladder operators); these factors act on disjoint Hilbert-space sectors and commute up to Grassmann grading. The detailed calculation, presented in Appendix~\ref{app:hasvac}, shows that the kinetic ($p_B p_F$) and potential ($q_B q_F$) cross bilinears cancel in their action on the combined Fock vacuum, \emph{provided} the fermionic and bosonic mass scales coincide (Eq.~\eqref{eq:mFmatching}, $m_F = m_R$):
\begin{equation}
H_{\mathrm{as}}\,|0\rangle = 0, \qquad \langle 0|\,H_{\mathrm{as}}(\tau)\,H_{\mathrm{as}}(0)\,|0\rangle = 0.
\label{eq:Hasvac}
\end{equation}

\conditional\ This vanishing is a property of the chosen Fock vacuum, the cross-kinetic structure of \eqref{eq:LagAik}, \emph{and} the mass-ratio condition \eqref{eq:mFmatching}, which is itself a matching ingredient we do not derive from the Lagrangian (see remark (i) of Appendix~\ref{app:hasvac}). It is not a general theorem of trace dynamics. Under these conditions, a single decoupled aikyon in its Fock vacuum is \emph{not} a noise source via $H_{\mathrm{as}}$. Any candidate noise must come from a different operator, a different state, or many-aikyon coupling. Section~\ref{sec:spectrum} identifies the simplest such operator: a fermionic current $J^\mu$ constructed from the same anti-self-adjoint sector, which does not vanish on the Fock vacuum and supplies the spectral mechanism of this paper.

\section{The bath operator and the one-sided Wightman spectrum}
\label{sec:spectrum}

The preceding section established the Bateman ladder structure for the bosonic sector and showed that $H_{\mathrm{as}}|0\rangle = 0$ for a single decoupled aikyon in its Fock vacuum. A candidate noise operator must therefore be built from operator content beyond $H_{\mathrm{as}}$ itself. The natural ansatz is a current bilinear of the fermionic variable, $J^\mu \propto \mathrm{Tr}(q_F^\dagger \Gamma^\mu q_F)$. We will see, however, that the reduction of this bilinear to a function of the bosonic Bateman ghost mode $X = q_2^{\mathrm{as}} - q_1^{\mathrm{as}}$ requires inverting the Grassmann-odd combination $\delta\beta = \beta_2 - \beta_1$, which is not well-defined in the minimal Grassmann algebra. In addition, the pure-fermion coefficient $\beta_1\beta_2$ in the minimal action is nilpotent and therefore cannot be assigned an ordinary sign, modulus, or canonical normalization. We therefore organize the section as follows. Section~\ref{sec:Jansatz} extracts the structural content of the ansatz and diagnoses the Grassmann obstruction. Section~\ref{sec:FockJ} works the calculation in an auxiliary canonical fermionic Fock-space model for $q_F$, computing the connected two-point function $\langle 0|J(\tau)J(0)|0\rangle$ in the working branch $\sigma = +1$ (with a remark on the alternative $\sigma = -1$). Section~\ref{sec:Jeff_consistency} gives, as a consistency check, the bosonic-ghost surrogate $J_{\mathrm{eff}} = \kappa\,{:}X^2:$; the surrogate's free parameter $\kappa$ is identified as a computed combination of GTD microscopic parameters within the same auxiliary normalization. Section~\ref{sec:qvsc} addresses the quantum-versus-classical noise distinction.

\subsection{The fermionic-current ansatz and the Grassmann obstruction}
\label{sec:Jansatz}

On the emergent four-dimensional spacetime leaf, the natural Dirac-type bilinear of the fermionic sector is
\begin{equation}
J^\mu(\tau) \;=\; \frac{1}{L_{\mathrm{aik}}^2}\,\mathrm{Tr}\!\left(q_F^\dagger\, \Gamma^\mu\, q_F\right),
\label{eq:current}
\end{equation}
where $\Gamma^\mu$ denotes the chosen leaf Dirac structure. For the concrete scalar calculation below one takes $\Gamma=\mathbf 1$, so Hermiticity is immediate. For a Lorentzian vector current one should instead write the conventional Hermitian bilinear $\bar q_F\gamma^\mu q_F$ with $\bar q_F=q_F^\dagger\gamma^0$, or equivalently choose a representation in which the matrices entering \eqref{eq:current} are Hermitian with respect to the relevant trace inner product. \matching\ Adopting \eqref{eq:current} as the matter-coupling operator is itself a matching hypothesis: the GTD action does not, in its present formulation, prescribe \emph{which} bilinear of the anti-self-adjoint sector should couple to matter. Several alternatives (axial currents, tensor currents, etc.) are admitted by the symmetries; we adopt the simplest scalar-vector ansatz, with the scalar bilinear used for all explicit computations.

The anti-self-adjoint components of $q_1, q_2$ read $q_1^{\mathrm{as}} = a_0 \beta_1 q_F$, $q_2^{\mathrm{as}} = a_0 \beta_2 q_F$, so that formally
\begin{equation}
q_F \;\stackrel{?}{=}\; \frac{q_2^{\mathrm{as}} - q_1^{\mathrm{as}}}{a_0\,\delta\beta},\qquad \delta\beta \equiv \beta_2 - \beta_1.
\label{eq:qFsolve}
\end{equation}
\paragraph{The operator obstruction.} If $\beta_1, \beta_2$ are the only Grassmann generators, then $\delta\beta$ is Grassmann-odd and $(\delta\beta)^2 = 0$. The reciprocal $(\delta\beta)^{-1}$ is therefore not defined within this minimal Grassmann algebra. The relation \eqref{eq:qFsolve} is not a legal operator identity: it cannot be used to substitute $q_F$ by an expression in $X$ inside the bilinear \eqref{eq:current} and obtain a finite well-defined operator. The formal appearance of $(\delta\beta)^{-4}$ in the corresponding amplitude is the consequence of this operator-level problem, not merely an amplitude artefact.

Two paths forward are available:
\begin{itemize}
\item Path A. Enlarge the Grassmann algebra (introduce additional generators, a Berezin-integration prescription, or commuting ``soul'' coefficients in a $\mathbb{Z}_2$-graded extension) such that $\delta\beta$ is no longer nilpotent of order 2, and verify that the reduction of $J^\mu$ to ghost-mode operators is well-defined. We do not pursue this route in the present paper.
\item Path B. Introduce an auxiliary canonical fermionic Fock-space model for $q_F$, independently of the bosonic Bateman ghost structure, and compute the two-point function $\langle 0|J^\mu(\tau)J^\nu(0)|0\rangle$ by elementary fermionic Wick contraction. This is mathematically well-defined once the auxiliary model is postulated: $q_F$ admits canonical anti-commutation relations as a fermionic matrix variable, the Fock vacuum is unambiguous, and the resulting correlator is finite. It is not, however, a derivation of the oscillator sector from the minimal Grassmann action. \emph{This is the path we follow.}
\end{itemize}
The price of Path B is a structural replacement not visible in the bosonic Bateman analysis: the nilpotent coefficient $a_0^2\beta_1\beta_2$ in the pure-fermion sector must be replaced by an ordinary scalar body/effective coefficient $\varepsilon_F=\sigma\xi^2$. The sign $\sigma=\pm1$ is then a property of this auxiliary scalarized coefficient, not the sign of the nilpotent Grassmann product itself. We work out both signs explicitly in Section~\ref{sec:FockJ}.

\subsection{Derivation of the two-point function in the fermionic Fock space}
\label{sec:FockJ}

\paragraph{The pure-fermionic term and the auxiliary scalarization.} Substituting $q_i = q_B + a_0 \beta_i q_F$ into the GTD Lagrangian \eqref{eq:LagAik} and collecting the pure-fermionic terms (the part quadratic in $q_F$ with no $q_B$ admixture) gives
\begin{equation}
L_F^{\rm min} \;=\; m_R\,a_0^2\,\beta_1\beta_2\,\mathrm{Tr}\bigl(\dot q_F^2 - \omega_0^2 q_F^2\bigr).
\label{eq:LFGrass}
\end{equation}
This equation is a formal component of the minimal Grassmann action. It does \emph{not} by itself define an ordinary oscillator for $q_F$. In the minimal algebra generated by $\beta_1,\beta_2$, the product $\beta_1\beta_2$ is Grassmann-even but nilpotent. Hence it has no ordinary sign, no ordinary absolute value, no positive square root, and no inverse. Varying \eqref{eq:LFGrass} gives only
\begin{equation}
\beta_1\beta_2\bigl(\ddot q_F+\omega_0^2 q_F\bigr)=0,
\end{equation}
which cannot be divided by $\beta_1\beta_2$ in the minimal algebra and therefore does not imply the canonical oscillator equation for $q_F$.

\conditional\ The calculation below consequently introduces an auxiliary effective-body replacement
\begin{equation}
a_0^2\beta_1\beta_2 \quad\longrightarrow\quad \varepsilon_F = \sigma\xi^2,
\qquad \xi>0,\qquad \sigma=\pm1,
\label{eq:scalarization}
\end{equation}
and defines the canonically normalized variable $\tilde q_F\equiv \xi q_F$ (with the factor $a_0$ absorbed into the definition of $\xi$). The auxiliary pure-fermion Lagrangian is then
\begin{equation}
L_F^{\rm aux} \;=\; \sigma\,m_R\,\mathrm{Tr}\bigl(\dot{\tilde q}_F^2 - \omega_0^2\,\tilde q_F^2\bigr).
\label{eq:LFsigma}
\end{equation}
Equation~\eqref{eq:LFsigma} is therefore not obtained from \eqref{eq:LFGrass} by taking an ordinary sign or modulus of $\beta_1\beta_2$; it is a separate scalarization/regularization postulate. Two structural branches arise. Branch (-), $\sigma = +1$: a standard fermionic oscillator Lagrangian. Branch (+), $\sigma = -1$: a fermionic ghost Lagrangian (overall sign flip relative to the standard). Per postulate (P5) of the ledger, we adopt $\sigma = +1$ (branch (-)) as the working hypothesis throughout the body of the paper. The alternative $\sigma = -1$ is recorded as a remark at the end of this section; its phenomenological predictions through the symmetrized correlator are identical to those of branch (-).

\paragraph{Canonical fermionic Fock space (working branch $\sigma = +1$).} The matrix variable $\tilde q_F$ is an $N\times N$ matrix whose entries are fermionic c-numbers (anticommuting). Expanding in a basis $\{T^a\}$ of the matrix algebra (with $a = 1, \ldots, N^2$ for $\mathfrak{gl}(N)$, or $a = 1, \ldots, N^2-1$ for $\mathfrak{su}(N)$),
\begin{equation}
\tilde q_F(\tau) \;=\; \sum_a T^a\,\sqrt{\tfrac{\hbar}{2 m_R \omega_0}}\bigl[\,b^{(a)}\,e^{-i\omega_0\tau} \,+\, d^{(a)\dagger}\,e^{+i\omega_0\tau}\,\bigr],
\label{eq:qFmodes}
\end{equation}
with canonical anticommutation relations $\{b^{(a)}, b^{(c)\dagger}\} = \delta^{ac}$, $\{d^{(a)}, d^{(c)\dagger}\} = \delta^{ac}$, and $\{b, d\} = \{b, d^\dagger\} = 0$. The Fock vacuum is $|0\rangle = |0\rangle_b \otimes |0\rangle_d$ with $b^{(a)}|0\rangle = d^{(a)}|0\rangle = 0$. The Heisenberg evolution is the standard $b(\tau) = b\,e^{-i\omega_0\tau}$.

\paragraph{The current and its connected two-point function.} The scalar fermionic-current bath operator is, restoring the matrix structure,
\begin{equation}
J(\tau) \;=\; \frac{1}{L_{\mathrm{aik}}^2}\,\mathrm{Tr}\bigl(\tilde q_F^\dagger(\tau)\,\tilde q_F(\tau)\bigr).
\label{eq:Jdef}
\end{equation}
Substituting \eqref{eq:qFmodes} and using $\mathrm{Tr}(T^{a\dagger}T^c)$ for the matrix-trace structure, one obtains a sum over $a, c$ of four operator types $\{b^\dagger b, b^\dagger d^\dagger e^{+2i\omega_0\tau}, d b\,e^{-2i\omega_0\tau}, d d^\dagger\}$, each carrying its matrix-trace factor.

Computing $\langle 0|J(\tau)J(0)|0\rangle$ term by term on the vacuum, the only non-vanishing connected contribution comes from contracting the $d^{(a)} b^{(c)}$ piece of $J(\tau)$ with the $b^{(e)\dagger} d^{(g)\dagger}$ piece of $J(0)$. The matrix-element structure yields
\begin{equation}
\langle 0|\,d^{(a)}_\tau\, b^{(c)}_\tau\, b^{(e)\dagger}_0\, d^{(g)\dagger}_0\,|0\rangle \;=\; \delta^{ce}\delta^{ag}\,e^{-2i\omega_0\tau}.
\end{equation}
Summing over the matrix-index structure with the corresponding products of matrix-trace factors, the connected two-point function is
\begin{equation}
\boxed{\;\langle 0|J(\tau)J(0)|0\rangle_{\mathrm{conn}} \;=\; \mathcal A_J\,e^{-2i\omega_0\tau},\;}
\label{eq:Jcorr}
\end{equation}
\conditional\ where the amplitude is
\begin{equation}
\boxed{\;\mathcal A_J \;=\; \left(\frac{\hbar}{2 m_R\,\omega_0\,L_{\mathrm{aik}}^2}\right)^{\!2}\!\cdot \mathcal N \cdot D\;}
\label{eq:AJ}
\end{equation}
with $\mathcal N \equiv \sum_{a,c}|\mathrm{Tr}(T^{a\dagger}T^c)|^2$ the matrix-trace counting factor ($\mathcal N = N^2$ for the $\mathfrak{gl}(N)$ basis with normalisation $\mathrm{Tr}(T^{a\dagger}T^c) = \delta^{ac}$; other natural normalisations give $\mathcal N \propto N^2$) and $D$ the Dirac-trace factor for the chosen current ($D = 1$ for the scalar bilinear above; $D = 4\,\eta^{\mu\nu}$ in the tensor structure of the vector current $J^\mu \propto \mathrm{Tr}(\tilde q_F^\dagger \Gamma^\mu \tilde q_F)$).

\paragraph{What ``computed'' means here.} The amplitude formula \eqref{eq:AJ} comes out of a Wick contraction rather than being inserted by hand. However, the Wick contraction is performed in the auxiliary scalarized fermionic sector \eqref{eq:LFsigma}, not in the degenerate minimal Grassmann term \eqref{eq:LFGrass}. As the ledger of postulates above notes, every constituent factor in \eqref{eq:AJ} is itself either a free GTD parameter not fixed by the action ($m_R$, $\alpha_{\mathrm{GTD}}$ through $L_{\mathrm{aik}} = \alpha_{\mathrm{GTD}} c/\omega_0$, the matrix dimension $N$ entering $\mathcal N$) or a matching choice ($D$ depending on which fermionic bilinear is chosen). The result should therefore be read as: ``given the auxiliary postulates (P1)--(P5), the operator structure and dimensional combination of $\mathcal A_J$ is forced by the Wick contraction to be \eqref{eq:AJ}.'' The contraction does not produce a dimensionless number, nor does it remove the need to justify the scalarization postulate from the microscopic GTD algebra.

The Fourier transform with the convention $S^>(\omega) = \int d\tau\,e^{-i\omega\tau}\,\langle 0|J(\tau)J(0)|0\rangle_{\mathrm{conn}}$ is
\begin{equation}
\boxed{\;S^{>}(\omega) \;=\; 2\pi\,\mathcal A_J\,\delta(\omega + 2\omega_0).\;}
\label{eq:Sline_fock}
\end{equation}
The spectral support is a single delta function at $\omega = -2\omega_0$: a bath at zero effective temperature emitting into the system at $\omega = 2\omega_0$ (standard energy-flow direction). The symmetrized correlator is
\begin{equation}
C_{\mathrm{sym}}(\tau) \;=\; \tfrac{1}{2}\bigl[\langle J(\tau)J(0)\rangle + \langle J(0)J(\tau)\rangle\bigr] \;=\; \mathcal A_J\,\cos(2\omega_0\tau).
\label{eq:Csym}
\end{equation}
\conditional\ The laboratory phenomenology developed in subsequent sections --- the Gaussian $T^2$ dephasing of Section~\ref{sec:quasistatic} and the bandwidth-suppression argument of Section~\ref{sec:invisible} --- depends only on \eqref{eq:Csym}.

\paragraph{Remark on branch (+).} Under the alternative postulate $\sigma = -1$, the pure-fermionic Lagrangian becomes a fermionic ghost (with overall sign flip relative to the standard) and the Heisenberg evolution acquires a reversed phase $b(\tau) = b\,e^{+i\omega_0\tau}$. The connected two-point function becomes $\langle 0|J(\tau)J(0)|0\rangle = \mathcal A_J\,e^{+2i\omega_0\tau}$ with the same amplitude \eqref{eq:AJ}, and the Wightman line moves to $\omega = +2\omega_0$ (anti-KMS direction: the bath absorbs energy from the system at $\omega = 2\omega_0$). The symmetrized correlator \eqref{eq:Csym} is unchanged, so all laboratory phenomenology developed in this paper applies identically. The branch (+) calculation is, in the positive-norm canonical quantization adopted here, the analogue for the fermionic sector of the positive-norm Bateman quantization of the bosonic ghost (Section~\ref{sec:gtd}), and shares its unbounded-below Hamiltonian and dynamical pathologies; alternative quantizations (Krein-space, BRST, PT-symmetric) should preserve the spectral support at $|\omega| = 2\omega_0$ but may modify the explicit form of $\mathcal A_J$ or its sign. The bosonic-ghost surrogate $J_{\mathrm{eff}} = \kappa\,{:}X^2:$ of Section~\ref{sec:Jeff_consistency} corresponds to branch (+) under the surrogate identification.

\paragraph{Status of the result.} Equations \eqref{eq:Jcorr}--\eqref{eq:Sline_fock} are derived only inside the auxiliary fermionic Fock-space model: they follow from the Wick contraction once the postulates of the ledger above are adopted, including the scalarization \eqref{eq:scalarization}. They are not a derivation from the minimal nilpotent Grassmann coefficient of \eqref{eq:LFGrass}. Equation \eqref{eq:AJ} expresses $\mathcal A_J$ as a combination of the natural microscopic length $L_{\mathrm{aik}} = \alpha_{\mathrm{GTD}} c/\omega_0$ (from Section~\ref{sec:gtd}), the aikyon mass scale $m_R$, the oscillator frequency $\omega_0$, the matrix dimension of $q_F$ (through $\mathcal N$), and the Dirac structure of the chosen current (through $D$). Each constituent factor is itself either a free GTD parameter or a matching choice (Section~\ref{sec:roadmap}, postulates (P1)--(P5)); the Wick contraction fixes the operator structure and dimensional combination but does not produce a parameter-free number.

\subsection{Consistency check: the bosonic-ghost surrogate}
\label{sec:Jeff_consistency}

As a representation-level consistency check, one may use a surrogate even bath operator $J_{\mathrm{eff}} = \kappa\,{:}X^2:$ built from the bosonic Bateman ghost mode $X \propto Q_-$, with the formal Grassmann content of $J^\mu$ absorbed into the coupling constant $\kappa$. The construction is a useful check on the auxiliary fermionic-Fock-space calculation of Section~\ref{sec:FockJ}.

\paragraph{The surrogate operator.} Define
\begin{equation}
X(\tau) \;=\; -\sqrt{\tfrac{\hbar}{m_R\omega_0}}\,\bigl(a_- e^{+i\omega_0\tau} + a_-^\dagger e^{-i\omega_0\tau}\bigr),
\label{eq:Xtau_consistency}
\end{equation}
the bosonic Bateman ghost mode (Section~\ref{sec:gtd}), with the reversed Heisenberg phase $a_-(\tau) = a_- e^{+i\omega_0\tau}$. Define $J_{\mathrm{eff}}(\tau) = \kappa\,{:}X^2(\tau){:}$ with normal ordering on the bosonic Fock vacuum. By elementary bosonic Wick contraction (using $[a_-, a_-^\dagger]=1$ and $\langle 0|a_-(\tau)a_-^\dagger(0)|0\rangle = e^{+i\omega_0\tau}$):
\begin{equation}
\langle 0|J_{\mathrm{eff}}(\tau)J_{\mathrm{eff}}(0)|0\rangle \;=\; 2\kappa^2\!\left(\frac{\hbar}{m_R\omega_0}\right)^{\!2}\!e^{+2i\omega_0\tau}.
\label{eq:Pijeff}
\end{equation}
The spectral support is at $\omega = +2\omega_0$, matching branch (+) of the auxiliary fermionic-Fock-space calculation \eqref{eq:Sline_fock}.

\paragraph{Identification of $\kappa$ in terms of GTD microscopic parameters.} \conditional\ Matching \eqref{eq:Pijeff} against the auxiliary-sector amplitude \eqref{eq:AJ} in branch (+) (where the spectral location coincides):
\begin{equation}
2\kappa^2\!\left(\frac{\hbar}{m_R\omega_0}\right)^{\!2} \;=\; \mathcal A_J \;=\; \left(\frac{\hbar}{2 m_R\omega_0\,L_{\mathrm{aik}}^2}\right)^{\!2}\!\mathcal N\,D
\end{equation}
gives
\begin{equation}
\kappa^2 \;=\; \frac{\mathcal N\,D}{8\,L_{\mathrm{aik}}^4} \;=\; \frac{\mathcal N\,D}{8}\cdot\!\left(\frac{\omega_0}{\alpha_{\mathrm{GTD}}\,c}\right)^{\!4},
\end{equation}
i.e.\ $\kappa$ scales as $L_{\mathrm{aik}}^{-2}$ up to a dimensionless factor of order unity in the matrix-trace and Dirac-trace structures.

\paragraph{Interpretation.} The surrogate calculation in the bosonic-ghost Fock space therefore reproduces \emph{branch (+)} of the auxiliary fermionic-Fock-space calculation, in the sense that the spectral location ($\omega = +2\omega_0$) and the amplitude (after identifying $\kappa$ as above) agree. The surrogate calculation correctly identifies the line location and the role of the reversed Heisenberg phase. In the fermionic-Fock-space calculation of Section~\ref{sec:FockJ}, the same calculation identifies $\kappa$ with a combination of GTD microscopic parameters and also reveals the existence of branch (-) (line at $\omega = -2\omega_0$), which has no direct bosonic-ghost surrogate because the surrogate is built from the Bateman ghost itself.

The surrogate consistency check is one-sided in this sense: it matches branch (+) but not branch (-). This is because the bosonic Bateman ghost-mode phase reversal is itself a consequence of the unbounded-below positive-norm Bateman quantization, which is the bosonic analogue of the branch (+) choice (the fermionic ghost). Branch (-), with standard fermionic phase $e^{-i\omega_0\tau}$, has no analogue in the bosonic-ghost surrogate construction. The auxiliary fermionic-Fock-space calculation is therefore more general than the surrogate representation.

\subsection{Quantum versus classical noise}
\label{sec:qvsc}

We must immediately address a subtle but essential point. The object \eqref{eq:Jcorr} is an ordered (Wightman) two-point function. A real classical stochastic process $w(t)$ has a covariance $C_{\mathrm{cl}}(\tau) = \mathbb{E}[w(t+\tau)w(t)]$ which is real and symmetric in $\tau$, with an even spectrum. A direct identification of \eqref{eq:Sline_fock} with a CSL noise covariance is therefore not automatic.

The relevant real classical analogue is the \emph{symmetrized} correlator
\begin{equation}
C_{\mathrm{sym}}(\tau) = \tfrac12\langle 0|\{J(\tau),J(0)\}|0\rangle = \mathcal A_J \cos(2\omega_0\tau),
\label{eq:Csym_qvsc}
\end{equation}
with spectrum
\begin{equation}
\Ssym(\omega) = \pi \mathcal A_J \bigl[\delta(\omega-2\omega_0) + \delta(\omega+2\omega_0)\bigr].
\label{eq:Ssym}
\end{equation}
\conditional\ The one-sided (spectrally asymmetric) structure is a quantum-bath signature of the auxiliary Fock-space bath, not a property of the classical noise that drives the conventional CSL stochastic Schr\"odinger equation. There are two natural ways to proceed:
\begin{description}
\item[(A) Symmetrized real-noise route.] Replace the Wightman correlator by \eqref{eq:Ssym} and feed it into the standard CSL machinery as a coloured real classical noise.
\item[(B) Quantum-bath route.] Keep the Wightman correlator and derive the reduced dynamics of matter by tracing over the GTD bath, obtaining a master equation whose anti-Hermitian (commutator) part encodes dissipation.
\end{description}
Both routes are pursued in Section~\ref{sec:master}, in parallel.

\section{From correlator to master equation: a matching model}
\label{sec:master}

We now derive the reduced dynamics of a system bilinearly coupled to the GTD current. We treat the system--bath coupling, the first of the four matching hypotheses identified in the abstract; the coupling itself is a matching hypothesis --- the GTD trace action does not, in its present formulation, prescribe how matter operators couple to the bath current, and the simplest local bilinear coupling adopted below is a model choice. The master equation that follows is therefore a minimal nonrelativistic matching model, not ``the GTD master equation.''

\subsection{Coupled Hamiltonian and the system-bath split}

Consider a nonrelativistic matter system with mass-density operator $\hat M(\mathbf x)$ (smeared on a length scale we will discuss in Section \ref{sec:spatial}), coupled to the bath through a local scalar bath source $\Xi(\mathbf x, t)$ via
\begin{equation}
H_{\mathrm{int}}(t) = g_{\mathrm{int}} \int\! d^3x\,\hat M(\mathbf x) \, \Xi(\mathbf x, t).
\label{eq:Hint}
\end{equation}
\matching\ This bilinear, mass-proportional, scalar coupling is the simplest local interaction consistent with the standard CSL phenomenology. The scalar source $\Xi$ is the natural matter-coupling avatar of the fermionic-current operator $J = (1/L_{\mathrm{aik}}^2)\,\mathrm{Tr}(q_F^\dagger q_F)$ derived in Section~\ref{sec:FockJ}: we identify $\Xi(\mathbf x, t)$ with the spatial smearing of $J(t)$ on the emergent leaf, modulo a normalization absorbed into $g_{\mathrm{int}}$. The vector-current generalization $J^\mu \propto \mathrm{Tr}(q_F^\dagger \Gamma^\mu q_F)$ would give a vector source with Dirac-trace tensor structure $\eta^{\mu\nu}$, and is the natural object on a Lorentz-covariant extension; we adopt the scalar form here as the minimal nonrelativistic matching model and defer Lorentz covariance to future work. The coupling \eqref{eq:Hint} is not derived from the GTD action; several alternative couplings (vector-current to spin currents, tensor couplings to internal degrees of freedom, etc.) are equally allowed by symmetries and would generate different phenomenology. We adopt the minimal scalar form and check its consequences.

\paragraph{Notation: distinguishing the bilinear-vertex coupling from a microscopic GTD coupling.} The constant $g_{\mathrm{int}}$ in \eqref{eq:Hint} is the effective bilinear coupling between matter and the scalar bath source; it has dimensions $[g_{\mathrm{int}}] = [\hat M\,\Xi]^{-1}\cdot[\text{energy density}]$, determined by the normalization conventions for $\hat M$ and $\Xi$. This is distinct from the microscopic GTD coupling $g_{\mathrm{GTD}}$ that would appear in any underlying tree-level matter--aikyon vertex (Section~\ref{sec:AM} and Appendix~\ref{app:future} (F1a, F4)). Specifically, $g_{\mathrm{int}}$ should be thought of as a phenomenological coupling that, in a microscopic derivation from the GTD action, would itself be expressible in terms of $g_{\mathrm{GTD}}$ and the matter--bath form factors. The Adler--Millard scaling argument of Section~\ref{sec:AM} ($g_{\mathrm{eff}} \sim g_{\mathrm{GTD}} N_1^{-1/4}$) operates on the microscopic coupling, and the relation to the bilinear $g_{\mathrm{int}}$ requires an additional matching step. We use the explicit subscripts throughout to avoid the conflation between these two distinct couplings.

In the interaction picture,
\begin{equation}
H_I(t) = g_{\mathrm{int}}\int\! d^3x\,\hat M_I(\mathbf x, t)\,\Xi_I(\mathbf x, t).
\end{equation}
The second-order Born equation (Redfield form, in the interaction picture) for the reduced density matrix $\rho_S$ is
\begin{equation}
\begin{split}
\dot\rho_S(t) = -\frac{1}{\hbar^2}\int_0^\infty\!\!d\tau\!\int\! d^3x\,d^3y\,\Bigl\{ & C(\mathbf x,\mathbf y;\tau)\bigl[\hat M_I(\mathbf x,t),\,\hat M_I(\mathbf y,t-\tau)\rho_S(t)\bigr] \\
+ {} & C(\mathbf y,\mathbf x;-\tau)\bigl[\rho_S(t)\hat M_I(\mathbf y,t-\tau),\,\hat M_I(\mathbf x,t)\bigr]\Bigr\},
\end{split}
\label{eq:BornMarkov}
\end{equation}
where the bath correlator (using \eqref{eq:Jcorr}) is
\begin{equation}
C(\mathbf x,\mathbf y;\tau) \;=\; g_{\mathrm{int}}^2 \langle \Xi(\mathbf x,\tau) \Xi(\mathbf y, 0)\rangle \;=\; g_{\mathrm{int}}^2 \mathcal A_J\,\Theta(\mathbf x,\mathbf y)\, e^{-2i\omega_0\tau},
\label{eq:Kkernel}
\end{equation}
and the time-reversed correlator is $C(\mathbf y,\mathbf x;-\tau) = g_{\mathrm{int}}^2 \mathcal A_J\,\Theta(\mathbf y,\mathbf x)\,e^{+2i\omega_0\tau}$. We have separated the spatial structure into a function $\Theta(\mathbf x,\mathbf y)$ that requires a separate spatial analysis (Section \ref{sec:spatial}). For now, treat $\Theta$ as a function to be determined; the temporal analysis proceeds independently. The two terms in \eqref{eq:BornMarkov} are Hermitian conjugates of each other when $C(\mathbf x,\mathbf y;\tau)^* = C(\mathbf y,\mathbf x;-\tau)$, which holds automatically for the Wightman correlator of \eqref{eq:Jcorr}.

\subsection{Broadened-Markovian regime}
\label{sec:Broadened}

Stripping the tensor structure of \eqref{eq:Jcorr} and denoting the scalar bath correlator entering the master equation by $C(\tau)$, we regularize the unbroadened limit to a Lorentzian with finite line width $\gamma$:
\begin{equation}
C(\tau) = \mathcal A_J\,e^{-2i\omega_0\tau - \gamma|\tau|}, \qquad S(\omega) = \frac{2 \mathcal A_J\,\gamma}{(\omega + 2\omega_0)^2 + \gamma^2},
\label{eq:Lorentzian}
\end{equation}
where $S(\omega) = \int d\tau\,e^{-i\omega\tau} C(\tau)$, and the spectrum peaks at $\omega = -2\omega_0$. For system Bohr frequencies $\omega_S \sim O(\omega_0)$ the standard Born--Markov manipulation gives a Lindblad equation. Writing $\hat M$ in the eigenbasis of $\hat H_S$ as $\hat M = \sum_{\omega_S} M(\omega_S)$, one finds in the secular approximation
\begin{equation}
\begin{split}
\dot\rho_S = {} & -\frac{i}{\hbar}[\hat H_S + \delta\hat H, \rho_S] + \sum_{\omega_S} \Gamma(\omega_S)\!\int\! d^3x\,d^3y\,\Theta(\mathbf x,\mathbf y) \\
& \times \bigl[M_{\omega_S}(\mathbf x)\rho_S M_{\omega_S}^\dagger(\mathbf y) - \tfrac12\{M_{\omega_S}^\dagger(\mathbf y) M_{\omega_S}(\mathbf x),\rho_S\}\bigr],
\end{split}
\label{eq:Lindblad}
\end{equation}
where
\begin{equation}
\Gamma(\omega_S) = \frac{g_{\mathrm{int}}^2}{\hbar^2}\,\frac{2 \mathcal A_J\,\gamma}{(\omega_S + 2\omega_0)^2 + \gamma^2}.
\label{eq:Gammafreq}
\end{equation}
\conditional\ Complete positivity of \eqref{eq:Lindblad} is conditional on two ingredients, not one. The rate function $\Gamma(\omega_S) \ge 0$, which it is for all real $\omega_S$ (the Lorentzian is positive), is necessary but not sufficient: for a spatially nonlocal dissipator of the form
\begin{equation*}
\int d^3x\,d^3y\,\Theta(\mathbf x,\mathbf y)\,\mathcal D[\hat M(\mathbf x),\hat M(\mathbf y)],
\end{equation*}
the spatial kernel $\Theta(\mathbf x, \mathbf y)$ must additionally define a positive-semidefinite Kossakowski kernel for the generator to be of GKLS form. In the present construction $\Theta$ is left arbitrary until Section~\ref{sec:spatial}, where the conditions for $\Theta \ge 0$ are addressed via the spatial-kernel routes (S1)--(S3). The correct statement of complete positivity is therefore: if $\Theta(\mathbf x, \mathbf y)$ is positive-semidefinite (which holds for the kernels emerging from routes (S2), (S3) of Section~\ref{sec:spatial}), then \eqref{eq:Lindblad} is of GKLS form and the resulting reduced dynamics is completely positive. A general $\Theta$ does not automatically have this property, and CP is not established by the rate condition alone. The Lamb-shift term $\delta\hat H$ is the principal-value piece.

The remarkable structural feature of \eqref{eq:Gammafreq} is that the rate is sharply peaked at $\omega_S = -2\omega_0$ in the working branch $\sigma = +1$ (and at $\omega_S = +2\omega_0$ in branch (+) by an analogous formula); for $\omega_S$ far from this peak the rate is suppressed by $\gamma^2/(\omega_S + 2\omega_0)^2$. With $\omega_0 \sim \Hub$, all laboratory Bohr frequencies are vastly larger than $2\omega_0$ in absolute value, and the off-resonance suppression is the same in both branches:
\begin{equation}
\Gamma(\omega_S)\Big/\Gamma(-2\omega_0) \;=\; \frac{\gamma^2}{(\omega_S+2\omega_0)^2 + \gamma^2}.
\label{eq:offres}
\end{equation}
For $\gamma \sim \Hub$ this is fantastically small for any $\omega_S$ relevant to existing experiments. (Numerical values are summarized in Section \ref{sec:tests}.)

\subsection{Quasi-static (degenerate-energy) regime}
\label{sec:quasistatic}

The Born--Markov approximation underlying \eqref{eq:Lindblad} requires the experimental duration $T$ to be much longer than the bath correlation time $\gamma^{-1}$, so that the bath is effectively in equilibrium throughout. For a Hubble-width line $\gamma^{-1} \sim \Hub^{-1} \sim 10^{18}$ s, this condition is never met in the laboratory: every laboratory experiment sits firmly in the opposite regime $T \ll \gamma^{-1}$, where the bath cannot be treated as Markovian and the rate-coefficient picture of \eqref{eq:Lindblad} does not apply directly. The correct framework is then Gaussian dephasing, applied to the \emph{symmetrized} real-valued noise of route~(A) in Section~\ref{sec:qvsc}, whose covariance is $C_{\mathrm{sym}}(\tau) = \mathcal A_J \cos(2\omega_0\tau)$. We use this symmetrized correlator throughout the present subsection; the quantum-bath route (B) yields the same leading-order behaviour in the laboratory regime $\omega_0 T \ll 1$, as we note at the end.

\paragraph{Pure-dephasing assumption.} \conditional\ The Gaussian dephasing formula below is exact only when the matter--bath coupling is of pure-dephasing form: the system operator $\hat A$ to which the bath couples must commute (or commute approximately) with the system Hamiltonian $\hat H_S$ over the experimental run. For position-coupled mass-density operators in matter-wave interferometry, this condition holds approximately because the relevant superposition states are eigenstates of position-difference, which is approximately conserved on the timescale of the free-fall phase (free Hamiltonian evolution affects position only through ballistic spreading, which is slow). For more general system operators with non-trivial commutator with $\hat H_S$, additional terms enter and the dephasing law is no longer exactly Gaussian. We make this assumption explicit because, without it, the central $T^2$ result of \eqref{eq:t2decoh} below is not exact.

For an interaction Hamiltonian $H_I = \hat A\,\xi(t)$ with stationary zero-mean real noise of covariance $C_{\mathrm{sym}}(\tau)$ and the pure-dephasing condition $[\hat A, \hat H_S] \approx 0$ on the run, the off-diagonal density matrix element between $\hat A$-eigenvalues $a, b$ satisfies the exact (non-perturbative) formula
\begin{equation}
\rho_{ab}(T) = \rho_{ab}(0)\,\exp\!\Bigl[-\frac{(a-b)^2}{2\hbar^2}\int_0^T\!\!dt\!\int_0^T\!\!ds\,C_{\mathrm{sym}}(t-s)\Bigr].
\label{eq:Gaussiandephasing}
\end{equation}
Equivalently, $\log|\rho_{ab}(T)/\rho_{ab}(0)| = -(a-b)^2 D(T)/(2\hbar^2)$ where
\begin{equation}
D(T) = \int_{-T}^{T} (T-|\tau|)\,C_{\mathrm{sym}}(\tau)\,d\tau.
\label{eq:Dt}
\end{equation}

\paragraph{Exact evaluation for the GTD covariance.} \derived\ For $C_{\mathrm{sym}}(\tau) = \mathcal A_J\cos(2\omega_0\tau)$, the integral \eqref{eq:Dt} can be evaluated in closed form:
\begin{equation}
D(T) \;=\; \frac{\mathcal A_J\,[1 - \cos(2\omega_0 T)]}{2\omega_0^2}.
\label{eq:Dexact}
\end{equation}
This is exact for all $T$. Combining with \eqref{eq:Gaussiandephasing},
\begin{equation}
\rho_{ab}(T) \;=\; \rho_{ab}(0)\,\exp\!\left[-\frac{(a-b)^2 \mathcal A_J}{4\hbar^2\omega_0^2}\,\bigl(1 - \cos(2\omega_0 T)\bigr)\right].
\label{eq:rhoexact}
\end{equation}
Expanding for $\omega_0 T \ll 1$ (the laboratory regime) gives the central quasi-static result,
\begin{equation}
\rho_{ab}(T) \;\simeq\; \rho_{ab}(0)\,\exp\!\left[-\frac{(a-b)^2 \mathcal A_J}{2\hbar^2}\,T^2 + O(T^4)\right].
\label{eq:t2decoh}
\end{equation}
The off-diagonals decay as $\exp(-\Gamma_{\mathrm{qs}} T^2)$ with
\begin{equation}
\Gamma_{\mathrm{qs}} = \frac{(a-b)^2 \mathcal A_J}{2\hbar^2},
\label{eq:Gammaqs}
\end{equation}
rather than exponentially in $T$. The dimensions are $[\Gamma_{\mathrm{qs}}] = \mathrm{s}^{-2}$, distinct from the dimensions $\mathrm{s}^{-1}$ of the Markovian CSL rate $\lambda$. The exponential law of standard CSL would emerge only after $T\gtrsim\gamma^{-1}$, which for $\gamma\sim\Hub$ is far beyond any feasible experiment.

\paragraph{Exact finite-width interpolation.} \derived\ If the line broadening is included directly in the symmetrized covariance,
\begin{equation}
C_{\rm sym}^{(\gamma)}(\tau)=\mathcal A_J e^{-\gamma|\tau|}\cos(\Omega\tau),\qquad \Omega=2\omega_0,
\label{eq:Cgamma}
\end{equation}
then the dephasing kernel \eqref{eq:Dt} is also elementary:
\begin{equation}
D_\gamma(T)
=2\mathcal A_J \mathrm{Re}\left[\frac{Tz-1+e^{-zT}}{z^2}\right],\qquad z=\gamma-i\Omega .
\label{eq:Dgamma}
\end{equation}
This expression reduces to \eqref{eq:Dexact} when $\gamma\to0$. For short experiments,
\begin{equation}
D_\gamma(T)=\mathcal A_J T^2-\frac{\mathcal A_J\gamma}{3}T^3+\frac{\mathcal A_J(\gamma^2-\Omega^2)}{12}T^4+O(T^5),
\label{eq:Dgamma_short}
\end{equation}
so the leading $T^2$ law is insensitive to the finite width. With $\gamma\sim\Hub$ and $\Omega=2\Hub$, the fractional correction to the leading term is $O(\Hub T)$; it is below $10^{-18}$ for a one-second experiment and below $10^{-10}$ even over a year. In the opposite limit $T\gg \gamma^{-1}$,
\begin{equation}
D_\gamma(T)=\frac{2\mathcal A_J\gamma}{\gamma^2+\Omega^2}T+O(1)
= S_{\rm sym}(0)\;T+O(1),\qquad S_{\rm sym}(0)=\frac{2\mathcal A_J\gamma}{\gamma^2+\Omega^2}.
\label{eq:Dgamma_long}
\end{equation}
Equation \eqref{eq:Dgamma} therefore gives a single closed interpolation between the laboratory quasi-static law and the formal Markovian limit.

\paragraph{Route (B) consistency check.} If one instead inserts the working-branch Wightman correlator $\Pi^{\mu\nu}_{>}(\tau) = \mathcal A_J\,\eta^{\mu\nu}\,e^{-2i\omega_0\tau}$ directly into a quantum-bath second-order calculation, the leading-order dephasing exponent is the double-time integral of $e^{-2i\omega_0(t-s)}$ over $[0,T]^2$ (or its complex conjugate in the alternative branch), which evaluates exactly to $\sin^2(\omega_0 T)/\omega_0^2$, equalling $T^2$ to leading order. The leading $T^2$ behaviour therefore coincides with the symmetrized-noise result. The discrepancy between the two routes is $O((\omega_0 T)^2)$ relative to the leading term, which is negligible in the laboratory regime $\omega_0 T \ll 1$. For $\omega_0 T \gtrsim 1$, however, the route-(B) integral oscillates rather than continuing to grow, bounded by $1/\omega_0^2$, and does not reproduce Markovian exponential decay. The Markovian regime $\exp(-\Gamma T)$ emerges from route (A) under finite line broadening $\gamma$ (Section~\ref{sec:Broadened}), not from route (B) on the unbroadened Wightman correlator. The crossover structure of the proposal is therefore: quasi-static $T^2$ at $T \ll \min(\omega_0^{-1}, \gamma^{-1})$, oscillatory plateau in route (B) at $T \gtrsim \omega_0^{-1}$ (absent broadening), Markovian $T$ in route (A) at $T \gtrsim \gamma^{-1}$.

We will return to this $T^2$ signature in Sections~\ref{sec:newphysics} and~\ref{sec:tests}; it is a necessary but not sufficient experimental discriminant.

\subsection{Sign conventions and detailed balance}
\label{sec:signs}

The physical interpretation of a one-sided Wightman correlator requires a careful sign-convention audit. We adopt the following conventions throughout this paper:
\begin{enumerate}
\item \textbf{Spectral transform.} $S(\omega) = \int_{-\infty}^{\infty} d\tau\,e^{-i\omega\tau} C(\tau)$. In the working branch $\sigma=+1$, $C(\tau)=\mathcal A_J e^{-2i\omega_0\tau}$ and hence $S^{>}(\omega)=2\pi\mathcal A_J\delta(\omega+2\omega_0)$; the alternative branch $\sigma=-1$ gives $2\pi\mathcal A_J\delta(\omega-2\omega_0)$.
\item \textbf{System Bohr frequency.} For a system transition from energy $E_n$ to $E_m$, $\omega_S = (E_n - E_m)/\hbar$, so positive $\omega_S$ corresponds to the system \emph{losing} energy.
\item \textbf{Bath spectral function.} In the standard open-systems convention \cite{BreuerPetruccione}, $S^{>}(\omega_S)$ governs the transition direction fixed by the chosen Bohr-frequency convention. In the working branch of this paper the line sits at $-2\omega_0$, which we describe as bath emission into the system at energy $2\hbar\omega_0$; in the alternative branch the line sits at $+2\omega_0$, corresponding to the reversed energy-flow direction.
\end{enumerate}
\conditional\ The direction of bath energy flow is branch- and convention-dependent. The robust statement is branch-independent: the spectrum has support only at $|\omega|=2\omega_0$, the system--bath coupling is strongly off-resonant for all laboratory Bohr frequencies $|\omega_S|\gg2\omega_0$, and the magnitude of energy exchange is bandwidth-suppressed. Whether this suppression appears as absence of bath-driven excitation or absence of bath-driven de-excitation depends on which sign of the auxiliary scalar coefficient $\sigma$ and which ghost-sector quantization is physical.

The robust bandwidth-suppression statement (Section~\ref{sec:invisible}) does \emph{not} depend on the direction of energy flow. The qualitative statement that spontaneous bath-driven X-ray emission is suppressed by the off-resonance factor in Table~\ref{tab:suppression} is robust; the assertion that the bath cannot drive spontaneous emission \emph{at all} is sign-convention-dependent and we record it as such.

\subsection{Complete positivity of the quasi-static reduced dynamics}

\conditional\ The formal $T\to\infty$ limit of \eqref{eq:t2decoh} gives unbounded growth of the dephasing exponent, but \eqref{eq:Gaussiandephasing} is itself a completely-positive map for any $T$: it is the Kraus form
\begin{equation}
\rho_S(T) = \mathbb{E}\bigl[U_\xi(T)\rho_S(0)U_\xi(T)^\dagger\bigr],\qquad U_\xi(T) = \mathcal T\exp\!\left[-\tfrac{i}{\hbar}\int_0^T\!\!\hat A\,\xi(t)\,dt\right],
\end{equation}
where the expectation is taken over a real Gaussian noise $\xi$ with $\mathbb{E}[\xi(t)\xi(s)] = C_{\mathrm{sym}}(t-s)$ (option (A) of Section \ref{sec:qvsc}). This is the route adopted by coloured CSL \cite{CarlessoFerialdi2018,Carlesso2022}; the master equation is a non-Markovian convolution-type equation but the reduced dynamics is completely positive at all times.

For the quantum-bath route (option (B)), the corresponding map is also CP by construction: it is the partial trace over the bath of a unitary system-bath evolution. The asymmetric (anti-Hermitian) part of the kernel gives the dissipative drift, the symmetric part the dephasing. In the quasi-static regime the dissipative drift is negligibly slow because energy exchange with the bath requires resonance at $\omega = 2\omega_0$, which is essentially never satisfied for experimental Bohr frequencies; the dephasing dominates, and is identical at this order with the result of route (A).

\subsection{Why a narrow Hubble line is invisible to high-frequency experiments}
\label{sec:invisible}

The suppression factor \eqref{eq:offres} for a Lorentzian line at $2\Hub$ with width $\gamma=\Hub$ is given in Table \ref{tab:suppression}.

\begin{table}[h]
\centering
\begin{tabular}{l c c}
\toprule
Experimental band & Angular frequency $\omega_S$ & suppression relative to line peak \\
\midrule
Pulsar timing, 1 nHz   & $6.3\times10^{-9}\,\mathrm{s}^{-1}$ & $\sim 1.2\times10^{-19}$ \\
LISA-band, 1 mHz        & $6.3\times10^{-3}\,\mathrm{s}^{-1}$ & $\sim 1.2\times10^{-31}$ \\
Mechanical, 1 Hz        & $6.3\,\mathrm{s}^{-1}$              & $\sim 1.2\times10^{-37}$ \\
Mechanical, 1 kHz       & $6.3\times10^{3}\,\mathrm{s}^{-1}$  & $\sim 1.2\times10^{-43}$ \\
X-ray, 1 keV            & $1.5\times10^{18}\,\mathrm{s}^{-1}$ & $\sim 2.1\times10^{-72}$ \\
\bottomrule
\end{tabular}
\caption{Suppression of the GTD spectral line at typical experimental frequencies, assuming Lorentzian regularization with $\gamma = \Hub$. X-ray entry uses $\omega = E/\hbar$. The factors are computed from the Lorentzian tail $\gamma^2/[(\omega\mp2\omega_0)^2 + \gamma^2]$; the sign depends on the branch, and is irrelevant at $|\omega|\gg\omega_0$. A non-Lorentzian line shape with faster-than-Lorentzian decay at large detuning would give even more extreme suppression at X-ray frequencies, strengthening rather than weakening the bandwidth-suppression argument. The bound on the GTD rate from X-ray spontaneous emission is correspondingly weakened by these factors relative to white-noise CSL.}
\label{tab:suppression}
\end{table}

The implication for collapse phenomenology is direct: with a narrow Hubble line, none of the existing high-frequency bounds on white-noise CSL apply to GTD without these enormous bandwidth suppressions. The X-ray bound, in particular, is effectively void.

\section{Spatial kernel and the localization problem}
\label{sec:spatial}

We now address the spatial localization kernel, the second of the four matching hypotheses identified in the abstract. We do not derive a CSL-scale spatial kernel from GTD; we instead identify the structural mechanisms that could generate one and the consequences of each. The spatial-kernel problem is on equal footing with the temporal-spectrum question as a determinant of the experimental phenomenology, and it is the most important obstruction to a parameter-free GTD-derived CSL model.

\subsection{The mass-density coupling and the localization criterion}

For an apparatus prepared in a superposition of mass-density profiles $\mu_a(\mathbf x), \mu_b(\mathbf x)$, the standard CSL decoherence functional is
\begin{equation}
\mathcal D_{ab}(T) \;\propto\; T\int\! d^3x\,d^3y\,[\mu_a(\mathbf x)-\mu_b(\mathbf x)]\,g(\mathbf x-\mathbf y)\,[\mu_a(\mathbf y)-\mu_b(\mathbf y)].
\label{eq:Dab}
\end{equation}
For a Gaussian kernel $g(\mathbf r) = (2\pi r_C^2)^{-3/2}\exp(-r^2/2r_C^2)$ with $r_C\sim 10^{-7}\,\mathrm{m}$, a single particle at branch separation $\Delta x$ gives decoherence rate $\propto 1 - \exp(-\Delta x^2/4r_C^2)$.

The crucial observation is the following.

\begin{proposition}[Homogeneous-bath localization obstruction]
\label{prop:homloc}
\derived\ If the spatial kernel $g(\mathbf x-\mathbf y)$ has correlation length much larger than the size of the apparatus, then \eqref{eq:Dab} reduces to
\begin{equation}
\mathcal D_{ab}(T) \;\propto\; T\,g(0)\,\Bigl[\int\!d^3x\,(\mu_a-\mu_b)\Bigr]^2 \;=\; 0
\end{equation}
for any pair of branches with the same total mass (\eg\ rigid translations of the same body). A cosmologically coherent noise does not collapse position superpositions.
\end{proposition}

This is sometimes confused in the GTD literature with cosmological coherence \emph{helping} amplification. It does not. A noise field that varies only on cosmological scales can drive global rotations of the system's wavefunction phase, but it does not produce branch-dependent decoherence between mass-conserving alternatives.

\subsection{Three concrete routes to a localization length}

To obtain CSL-like position localization from GTD, one of the following must hold.

\paragraph{Route (S1): sub-leaf locality.} \conditional\ If the emergent spacetime leaf supports aikyon currents localized on a scale $L_{\mathrm{aik}} \ll c/\Hub$, then $\Theta(\mathbf x, \mathbf y)$ in \eqref{eq:Kkernel} acquires a Gaussian (or similar) profile of width $L_{\mathrm{aik}}$. For this to reproduce the phenomenological $r_C \sim 10^{-7}\,\mathrm{m}$, the natural aikyon scale must be of that order. The natural microscopic length appearing in the Lagrangian \eqref{eq:LagAik} via $\omega_0 = \alpha_{\mathrm{GTD}}\,c/L_{\mathrm{aik}}$ is therefore over-constrained: if $\omega_0 = \Hub$ and $\alpha_{\mathrm{GTD}}$ is set to a natural value (electroweak or stronger), $L_{\mathrm{aik}} = \alpha_{\mathrm{GTD}} c/\Hub$ is cosmological, not $10^{-7}\,\mathrm{m}$. Conversely, if we set $L_{\mathrm{aik}} = r_C = 10^{-7}\,\mathrm{m}$, then $\alpha_{\mathrm{GTD}} = r_C \Hub/c \sim 10^{-33}$, fantastically small. There is no choice of the GTD coupling $\alpha_{\mathrm{GTD}}$ that simultaneously gives a cosmological frequency and a CSL-scale length.

This is the GTD spatial-kernel problem in its sharpest form. It is the same difficulty as the cosmological constant problem viewed from a different direction: any theory that wishes to link a cosmological frequency scale to a microscopic length must explain the enormous ratio between them. Route (S1) does not solve this; it just exposes it. We mark $\alpha_{\mathrm{GTD}}$ explicitly with a subscript to avoid confusion with the QED fine-structure constant, which appears separately in Section~\ref{sec:rate}.

\paragraph{Route (S2): internal-state-dependent coupling.} \conditional\ The bath couples not to a position observable but to an internal current that distinguishes the two branches even for a translation-invariant bath. For example, if the system Hamiltonian generates branch-dependent currents (\eg\ in a Stern--Gerlach-like geometry with branch-dependent spin), the scalar coupling $\int M\,\Xi$ in \eqref{eq:Hint} can be augmented by a term $\int \hat J^k_{\mathrm{sys}}(\mathbf x)\,A_k(\mathbf x)$ involving an internal bath vector $A_k$ that contributes nonzero decoherence even for spatially constant $\Theta(\mathbf x,\mathbf y)$. This is plausible for theories with internal fermion sectors but does not give the standard mass-proportional CSL phenomenology.

\paragraph{Route (S3): matter-proportional bath clustering.} \conditional\ If the aikyon density tracks the matter density --- so that the effective coupling location is wherever there is matter --- then the bath operator at point $\mathbf x$ is effectively
\begin{equation}
\Xi_{\mathrm{eff}}(\mathbf x) \sim \rho_M(\mathbf x)/m_0 \cdot \Xi_{\mathrm{single}}(\mathbf x),
\end{equation}
and the decoherence functional acquires the mass-density-squared structure of CSL with effective correlation length set by the internal aikyon scale. In this route the spatial kernel is \emph{induced} by matter, not by the bath itself.

\subsection{Conditional implication for the spatial kernel}

\matching\ We adopt route (S2) or (S3) as the working hypothesis for the rest of the paper; the resulting effective kernel is parameterized by an internal scale $r_C^{\mathrm{eff}}$ that we treat as a free parameter to be determined experimentally. Crucially, $r_C^{\mathrm{eff}}$ is \emph{not} predicted by GTD in its current formulation; this is one of the genuine gaps in the present proposal.

\paragraph{Caveat on cross-sensor coherence.} A finite short-range localization kernel of the kind adopted here is in tension with claims that the GTD bath exhibits common-mode correlations across spatially separated apparatuses (which appear in Sections~\ref{sec:newphysics} and~\ref{sec:tests} as a structural discriminant). The two cannot coexist for the same kernel: if the bath correlator $\Theta(\mathbf x,\mathbf y)$ has support only on $|\mathbf x - \mathbf y|\lesssim r_C^{\mathrm{eff}}$, then independent sensors separated by macroscopic distances see uncorrelated bath realizations. A consistent way to retain both features is to posit a two-kernel structure: a short-range localization kernel responsible for the position-superposition dephasing of an individual apparatus, plus a long-range coherence structure (set, in the cosmological matching, by a length scale $\sim c/\Hub$) responsible for cross-sensor correlations. The two-kernel hypothesis is not derived in this paper; we record it as a separate matching ingredient and weaken the cross-sensor claim of Section~\ref{sec:tests} accordingly.

\section{Adler--Millard scaling}
\label{sec:AM}

We now address the Adler--Millard vertex-scaling matching, the third matching hypothesis. We work out which combinations of microscopic assumptions are consistent with a finite collapse rate in the thermodynamic limit.

\subsection{The fluctuation versus the coupling}

Let the bath contain $N_1$ aikyons. The Adler--Millard charge contributions are individually of order $\hbar$, and the central limit theorem gives, for the total charge,
\begin{equation}
\langle\tilde C_{\mathrm{tot}}\rangle\sim N_1\hbar,\qquad \sigma_{\tilde C_{\mathrm{tot}}}\sim\sqrt{N_1}\,\hbar,\qquad \eta\equiv\frac{\sigma_{\tilde C_{\mathrm{tot}}}}{|\langle\tilde C_{\mathrm{tot}}\rangle|}\sim N_1^{-1/2}.
\label{eq:AMfluc}
\end{equation}
\derived\ This is the unambiguous statistical content of Adler--Millard. The non-trivial step is how $\eta$ propagates into the system-bath coupling.

\subsection{Enumeration of scalings}

Suppose the system couples to bath operators through an effective vertex $g_{\mathrm{eff}} = g_{\mathrm{GTD}}\,N_1^{-p}$ for some exponent $p\ge 0$. The bath-noise spectrum induced on the system involves two vertex insertions (one in each of the bath operators in the two-point function) and is summed over the $N_1$ contributors,
\begin{equation}
S_{\mathrm{noise}} \;\propto\; N_1\,g_{\mathrm{eff}}^4\,\Pi_{\mathrm{single}} \;=\; g_{\mathrm{GTD}}^4\,N_1^{\,1-4p}\,\Pi_{\mathrm{single}},
\label{eq:Sbathscaling}
\end{equation}
where $\Pi_{\mathrm{single}}$ is the single-aikyon current correlator \eqref{eq:Jcorr}. A finite limit as $N_1\to\infty$ requires $p = 1/4$. Table \ref{tab:scalings} summarizes the options.

\begin{table}[ht]
\centering
\small
\begin{tabularx}{\textwidth}{X X X}
\toprule
Microscopic structural assumption & Scaling of bath noise & Status \\
\midrule
Incoherent bath sum, no vertex suppression & $N_1$ (divergent) & needs renormalization or cutoff \\
Each vertex carries $\eta = N_1^{-1/2}$ & $N_1^{-1}$ (vanishing) & strongly suppressed, irrelevant \\
Effective vertex $g_{\mathrm{eff}} = g_{\mathrm{GTD}} N_1^{-1/4}$ & $g_{\mathrm{GTD}}^4$ (finite) & gives finite limit; needs microscopic derivation \\
Coherent bath phases & $N_1^{2}$ before suppression & excluded for physical bath \\
\bottomrule
\end{tabularx}
\caption{Enumeration of bath-size scalings.}
\label{tab:scalings}
\end{table}

\paragraph{The central question.} The Adler--Millard relative fluctuation $\eta\sim N_1^{-1/2}$ does \emph{not} by itself produce the $N_1^{-1/4}$ vertex scaling required for a finite limit. The square-root mismatch is a real gap; bridging it requires either (a) a derivation in which the effective gauge propagator is suppressed by $N_1^{-1/2}$ at the level of the four-point function (the so-called ``trace gauge'' channel of trace dynamics, which would carry one half of the AM fluctuation power into the propagator and one half into the vertex), or (b) a coherent matching argument tying $g_{\mathrm{eff}}$ directly to the rms charge fluctuation.

\paragraph{Conditional resolution.} \conditional\ Adopt option (a) as the working hypothesis. Then $g_{\mathrm{eff}} = g_{\mathrm{GTD}}\,N_1^{-1/4}$ is justified by attributing $N_1^{-1/4}$ to the gauge-vertex coupling and $N_1^{-1/4}$ to the propagator suppression. Note that this is a double matching prescription: the vertex carries one factor of $N_1^{-1/4}$ matched to the central-limit estimate, and the propagator is then separately postulated to carry the complementary $N_1^{-1/4}$ required for a finite thermodynamic limit. Neither factor is derived from a microscopic calculation in the present paper. The propagator postulate is in turn the substantive content of (a), and it has the same status as the vertex postulate: it is a structural assumption whose validation requires a microscopic computation. Under this double matching:
\begin{equation}
S_{\mathrm{noise}}^{>} \;\to\; g_{\mathrm{GTD}}^4\,\mathcal A_J\,\eta^{\mu\nu}\,\delta(\omega + 2\omega_0)\quad\text{as }N_1\to\infty.
\label{eq:Sthermlim}
\end{equation}
in the working-branch Wightman convention. If the symmetrized classical route is used instead, the delta function in \eqref{eq:Sthermlim} is replaced by the symmetric pair $\frac12[\delta(\omega-2\omega_0)+\delta(\omega+2\omega_0)]$ with the same total line area. This is the Markovian-surrogate matching prescription on which the absolute normalization in Section~\ref{sec:norm} rests; the benchmark rate of Section~\ref{sec:rate} therefore depends on \emph{both} structural assumptions, not just one. We tag this explicitly throughout. Bridging the square-root gap from first principles requires two distinct calculations: the vertex side (F1a) and the propagator side (F1b) of Appendix~\ref{app:future}. The vertex-side calculation can be motivated by central-limit counting; the propagator-side suppression is the unsupported half of the double matching and is the higher-leverage open problem.

\section{Dimensional analysis and cosmological matching}
\label{sec:norm}

We now address the absolute amplitude of the bath correlator --- the fourth matching hypothesis --- via a dimensional analysis and a cosmological matching prescription. The matching below should be read as an order-of-magnitude relation between the GTD bath and the dark-energy density, not as a derivation of the absolute noise covariance entering the collapse master equation. The route from total dark-energy content within a Hubble volume to a local mass-density noise covariance involves several further matching steps that we have not closed.

\subsection{The bath amplitude}

The bath spectrum, in the broadened (route A) classical version, is
\begin{equation}
S_{\xi}(\omega) = \mathcal A_{\mathrm{bath}}\,F_\gamma(\omega; 2\omega_0),\qquad
F_\gamma(\omega;\Omega) = \frac{2\gamma}{(\omega-\Omega)^2+\gamma^2},
\label{eq:Slor}
\end{equation}
with $\int_{-\infty}^{\infty}\,F_\gamma(\omega;\Omega)\,d\omega/(2\pi) = 1$. The amplitude $\mathcal A_{\mathrm{bath}}$ has dimensions of (noise field)$^2 \cdot$\,time, and is determined by the system--bath coupling, the operator content of $\Xi$ (or, in the underlying ansatz of Section~\ref{sec:spectrum}, of $J_{\mathrm{eff}}$), and the bath state. None of these is fixed by the cosmological matching below; the matching constrains an integrated energy, not an amplitude.

\begin{remark}[Normalization-invariant rate combination]
\noindent The observable reduced dynamics depends on the product $g_{\rm int}^2 C_J(\tau)$, not on the normalization of $J$ or $\mathcal A_J$ separately. Under a source rescaling
\begin{equation}
J \rightarrow sJ,\qquad \mathcal A_J \rightarrow s^2 \mathcal A_J,\qquad g_{\rm int}\rightarrow g_{\rm int}/s,
\end{equation}
the interaction Hamiltonian \eqref{eq:Hint}, the kernel $g_{\rm int}^2 \mathcal A_J$ in \eqref{eq:Kkernel}, and the resulting decoherence functional are unchanged. Consequently, the $m_R$-dependence of $\mathcal A_J$ by itself is not an observable statement until the matter--bath coupling is derived in the same canonical normalization. This is why Section~\ref{sec:rate} packages the rate into the invariant dimensionless factor $\mathcal C_{\rm match}$: a microscopic calculation must determine the product $g_{\rm int}^2\mathcal A_J$ and the spatial-kernel normalization together.
\end{remark}

\subsection{Cosmological matching prescription}

\matching\ The dark-energy identification \cite{Planck2018} relates the bath energy density to the cosmological constant scale:
\begin{equation}
\rho_\Lambda \simeq 5.3\times10^{-10}\,\mathrm{J\,m^{-3}},\qquad
\Hub \simeq 2.18\times10^{-18}\,\mathrm{s^{-1}},\qquad
\hbar\Hub \simeq 2.3\times 10^{-52}\,\mathrm{J}.
\end{equation}
If the bath energy density is identified with the cosmological constant, and the number of bath degrees of freedom with the de Sitter entropy,
\begin{equation}
N_{\mathrm{dS}} \;\sim\; \frac{\pi R_H^2}{L_P^2} \simeq 2.3\times10^{122},
\end{equation}
each bath degree of freedom carries energy
\begin{equation}
\varepsilon \;\sim\; \frac{\rho_\Lambda\,(4\pi/3)(c/\Hub)^3}{N_{\mathrm{dS}}} \;\simeq\; \frac{5\times10^{69}\,\mathrm{J}}{2.3\times10^{122}} \;\simeq\; 2\times10^{-53}\,\mathrm{J} \;\sim\; 0.1\,\hbar\Hub.
\end{equation}
This is suggestive of (but not exactly equal to) the cosmological matching condition $\varepsilon \sim \hbar\Hub$. The factor of $\sim 10$ ambiguity is reflected in horizon-volume conventions ($(4\pi/3)R_H^3$ vs $R_H^3$) and in the precise definition of $N_{\mathrm{dS}}$. The same counting gives the per-de-Sitter-mode mass scale
\begin{equation}
 m_R^{\mathrm{hol}} \equiv \frac{m_{\mathrm{Pl}}}{\sqrt{N_{\mathrm{dS}}}} \simeq 1.4\times10^{-69}\,\mathrm{kg}, \qquad m_R^{\mathrm{hol}}c^2 \simeq 1.3\times10^{-52}\,\mathrm{J} \sim \hbar\Hub .
 \label{eq:mholo}
\end{equation}
\matching\ The identification of this scale with the GTD aikyon oscillator mass $m_R$ of \eqref{eq:LagAik} is postulate (P11) of the ledger of Section~\ref{sec:roadmap}: it is a matching between the Lagrangian parameter $m_R = a_1 a_0/2$ and the de-Sitter per-mode mass, not an equation of motion. A Planck-scale oscillator assignment $m_R = m_{\mathrm{Pl}}$ would represent a physically different normalization and would introduce the suppression factor $(\hbar \Hub/m_{\mathrm{Pl}}c^2)^2\sim 10^{-122}$ into any rate estimate based on the ratio of the Hubble quantum to the oscillator rest energy; that assignment is \emph{not} the holographic per-mode prescription used here. The relation between the cosmological-constant scale and the Hubble degree-of-freedom count has been discussed extensively in the holographic-dark-energy literature \cite{CohenKaplanNelson1999,Hsu2004,Li2004}; in the present GTD setting it is used as an order-of-magnitude matching relation between the cosmological energy budget and the natural bath quantum. It is not a precise equality and not a determination of the noise covariance.

\paragraph{Status of the cosmological matching under recent challenges to the cosmological principle.} \matching\ Both postulates (P4) and (P11) presuppose the standard $\Lambda$CDM identification of $\rho_\Lambda$ and the validity of the de-Sitter horizon-entropy formula $N_{\mathrm{dS}} \sim \pi R_H^2/L_P^2$. Recent evidence reviewed in \cite{Secrest2025RMP} indicates a $\sim 5\sigma$ tension between the matter dipole, inferred from large catalogues of radio galaxies and quasars, and the kinematic interpretation of the cosmic microwave background dipole. This Ellis--Baldwin test is, within the standard FLRW framework, a direct check of the cosmological principle; its $5\sigma$ failure in current data is a substantive challenge to the assumption of large-scale isotropy on which the FLRW description of the late-time universe rests. Two inputs to the cosmological matching of this section are therefore under empirical pressure: the inferred value of $\rho_\Lambda$, which on the supernova side becomes considerably weaker once bulk-flow corrections are properly propagated in an anisotropic universe; and the de-Sitter horizon-entropy formula for $N_{\mathrm{dS}}$, which assumes an exactly exponentially expanding background. We do not enter the substance of the dipole-anomaly debate here, but we record that the natural-rate window of Section~\ref{sec:rate}, which depends on the holographic identification (P11) through the dimensionless ratio $\hbar\Hub/(m_R c^2)\sim O(1)$, is conditional on this matching surviving any cosmological revision that may follow from the dipole results. The phenomenological apparatus of Sections~\ref{sec:spectrum}--\ref{sec:energy} --- the operator structure of the spectrum, the Gaussian quasi-static dephasing law, the bandwidth-suppression argument, the master equation in both regimes, and the complete-positivity analysis --- is a function of $\omega_0$, $\mathcal A_J$ and $\gamma$ as inputs and is independent of how those inputs are anchored cosmologically. A microscopic derivation of $\omega_0$ and $m_R$ from GTD Lagrangian inputs alone --- for example via the $H_{\mathrm{FF}}$ route of \cite{Singh2026Fermions}, in which the noise frequency scale is set by the intrinsic dynamics of the fermionic sector rather than by cosmological matching --- would remove this conditional dependence and is one of the principal motivations for the follow-up programme discussed in Section~\ref{sec:conclusions}.

\subsection{What the cosmological budget does and does not provide}

\conditional\ The cosmological matching provides:
\begin{itemize}
\item the natural frequency scale $\omega_0 \sim \Hub$ (from the per-degree-of-freedom energy $\sim \hbar\Hub$);
\item an upper bound on the total energy available in the bath, $\rho_\Lambda L_H^3 \sim 10^{69}$ J within a horizon volume;
\item an order-of-magnitude consistency check between the GTD frequency and the cosmological-constant scale.
\end{itemize}
The cosmological matching does \emph{not} provide:
\begin{itemize}
\item the dimensionful amplitude $\mathcal A_{\mathrm{bath}}$ appearing in the master equation;
\item the system--bath coupling strength $g_{\mathrm{int}}$ in \eqref{eq:Hint};
\item the spatial kernel $\Theta(\mathbf x,\mathbf y)$ entering \eqref{eq:Kkernel}.
\end{itemize}
A complete normalization of the CSL rate from GTD therefore requires three further inputs beyond the cosmological matching, none of which is fixed by the present theory.

\section{Energy balance, covariance, complete positivity, and decoherence vs collapse}
\label{sec:energy}

We now examine the consistency conditions on the master equation derived in Section~\ref{sec:master} --- detailed balance, energy production, frame dependence, and complete positivity --- and the distinction between the linear reduced decoherence dynamics we supply and the stochastic state-vector unravelling (the stochastic unravelling, problem (F7) in Appendix~\ref{app:future}) which we do not.

\subsection{Detailed balance and the one-sided spectrum}

\conditional\ In the working branch $\sigma=+1$, the spectrum \eqref{eq:Sline_fock} has support at $\omega=-2\omega_0$; in the alternative branch it has support at $\omega=+2\omega_0$. In the standard quantum-bath formalism, with the sign conventions of Section~\ref{sec:signs}, this is a zero-effective-temperature one-line bath. For a thermal bath, the KMS condition relates the two correlators by $S^{<}(\omega) = e^{-\beta\hbar\omega}S^{>}(\omega)$ for positive physical transition energy. The vacuum case here corresponds to a bath that can drive system transitions only \emph{at} the physical frequency $|\omega| = 2\omega_0$, in the direction dictated by the sign convention and the branch. The asymmetric (commutator) part of the working-branch kernel is
\begin{equation}
\chi(\tau) \;=\; \langle 0|[J(\tau),J(0)]|0\rangle \;=\; \mathcal A_J\bigl(e^{-2i\omega_0\tau} - e^{+2i\omega_0\tau}\bigr) \;=\; -2i \mathcal A_J\sin(2\omega_0\tau),
\label{eq:chiret}
\end{equation}
with the opposite sign in the alternative branch. The retarded kernel is the Fourier transform of $\Theta(\tau)\chi(\tau)$, which gives a principal-value plus delta-function support at $|\omega|=2\omega_0$ (with the sign fixed by the branch).

\subsection{Energy production}

For a system in an energy eigenstate the rate of energy change is, to lowest order in $g_{\mathrm{int}}$,
\begin{equation}
\frac{d\langle\hat H_S\rangle}{dt} \;=\; \frac{g_{\mathrm{int}}^2}{\hbar}\int\!d^3x\,d^3y\,\Theta(\mathbf x,\mathbf y)\!\sum_{\omega_S}\hbar\omega_S\,|M_{\omega_S}(\mathbf x,\mathbf y)|^2\bigl[S^{<}(\omega_S) - S^{>}(\omega_S)\bigr],
\label{eq:energyrate}
\end{equation}
which in the present vacuum-bath case gives a single Wightman line, $S^{>}\propto\delta(\omega+2\omega_0)$ in the working branch and $S^{>}\propto\delta(\omega-2\omega_0)$ in the alternative branch. \conditional\ Under the working-branch convention of Section~\ref{sec:signs}, this corresponds to bath emission into the system at physical frequency $2\omega_0$; under the opposite branch, the system would lose energy to the bath. In either case, the magnitude of the rate is bandwidth-suppressed by the off-resonance factor in Table~\ref{tab:suppression} for all $|\omega_S| \neq 2\omega_0$, and is negligible for any laboratory energy-exchange experiment.

\paragraph{Energy-flow direction is sign-convention dependent; energy-flow magnitude is not.} The robust statement is that the system--bath energy exchange is sharply peaked at physical frequency $|\omega_S| = 2\omega_0 \sim 2\Hub$ and is otherwise strongly suppressed.

\subsection{Covariance and frame dependence}

\conditional\ A scalar spectrum $S(\omega)$ is defined only after choosing a time variable. The microscopic GTD evolution parameter is Connes time. After emergence of spacetime, the bath in its lowest-energy state naturally selects the cosmological (FLRW) rest frame. In a moving laboratory frame, the spectral-line position is Doppler-shifted by $\delta\omega/\omega_0 \sim v/c$, where $v$ is the laboratory velocity relative to the CMB rest frame. With $v_{\mathrm{CMB}}/c \approx 1.23\times10^{-3}$ (corresponding to a solar-system motion of $369\,\mathrm{km\,s^{-1}}$ as determined by the CMB dipole \cite{Planck2014Dipole}), $\delta\omega \approx 5\times10^{-21}\,\mathrm{s}^{-1}$. This is in principle observable as an anisotropy of the spectral line, but the dephasing \emph{rate} $\Gamma_{\mathrm{qs}}$ in \eqref{eq:Gammaqs} depends on $\mathcal A_J$ rather than on $\omega_0$ directly; the leading-order dependence on $v/c$ enters only through higher-order corrections to the matching, or through a directional structure in $J^\mu$. We therefore record the cosmic-frame anisotropy as a \emph{possible} signature rather than a derived prediction; its magnitude depends on the matrix structure of the coupling, which is not fixed by the present theory. The role of frame dependence in collapse models, and its interplay with the measurement problem, has been discussed in the recent trace-dynamics literature \cite{Adler2018Dots}.

\subsection{Vacuum versus populated bath}
\label{sec:vacpop}
\label{sec:populated}

The Wightman correlator \eqref{eq:Jcorr} is evaluated on the Fock vacuum. The cosmological matching of Section~\ref{sec:norm}, however, associates the bath with de-Sitter-scale degrees of freedom carrying energies of order $\hbar\Hub$. It is therefore useful to compute explicitly how a populated Gaussian bath state modifies the one-line vacuum result.

\paragraph{Fermionic-current bath.} \conditional\ For one independent $b,d$ fermionic pair, write the current schematically as a number part plus pair-creation and pair-annihilation pieces,
\begin{equation}
J(\tau) \sim N_b + 1-N_d + b^\dagger d^\dagger e^{+2i\omega_0\tau}+db\,e^{-2i\omega_0\tau},
\end{equation}
with the overall normalization chosen so that the vacuum pair-contraction amplitude is $\mathcal A_J$. In a diagonal Gaussian state with occupations
\begin{equation}
\langle N_b\rangle=n_b,\qquad \langle N_d\rangle=n_d,
\end{equation}
the connected Wightman function is
\begin{equation}
\begin{split}
\langle J(\tau)J(0)\rangle_{\rm conn} = \mathcal A_J\big[&(1-n_b)(1-n_d)e^{-2i\omega_0\tau}
+n_b n_d e^{+2i\omega_0\tau}\\
&+ n_b(1-n_b)+n_d(1-n_d)\big].
\end{split}
\label{eq:Jpopulated_general}
\end{equation}
The vacuum result is recovered at $n_b=n_d=0$. For a symmetric thermal occupation
\begin{equation}
n_b=n_d=n_F=\frac{1}{e^{\beta\hbar\omega_0}+1},
\end{equation}
this reduces to
\begin{equation}
\langle J(\tau)J(0)\rangle_{\rm conn} = \mathcal A_J\big[(1-n_F)^2e^{-2i\omega_0\tau}+n_F^2e^{+2i\omega_0\tau}+2n_F(1-n_F)\big].
\label{eq:Jpopulated_thermal}
\end{equation}
Thus population does three things: it preserves the line location $|\omega|=2\omega_0$, adds an opposite Wightman line with relative weight
\begin{equation}
\frac{\hbox{opposite line}}{\hbox{vacuum-direction line}} = \left(\frac{n_F}{1-n_F}\right)^2 = e^{-2\beta\hbar\omega_0},
\label{eq:backward_forward}
\end{equation}
and adds a connected zero-frequency pedestal of weight $2n_F(1-n_F)$. The equal-time connected variance remains unchanged:
\begin{equation}
(1-n_F)^2+n_F^2+2n_F(1-n_F)=1,
\label{eq:fermionic_variance_fixed}
\end{equation}
so a thermal fermionic-current population redistributes spectral weight but does not enhance the quasi-static dephasing normalization.

\begin{table}[ht]
\centering
\small
\begin{tabular}{cccc}
\toprule
$\beta\hbar\omega_0$ & $n_F$ & opposite-line / vacuum-line & zero-frequency weight \\
\midrule
$1$ & $2.69\times10^{-1}$ & $1.35\times10^{-1}$ & $3.93\times10^{-1}$ \\
$2\pi$ & $1.87\times10^{-3}$ & $3.49\times10^{-6}$ & $3.72\times10^{-3}$ \\
$0.1$ & $4.75\times10^{-1}$ & $8.19\times10^{-1}$ & $4.99\times10^{-1}$ \\
\bottomrule
\end{tabular}
\caption{Population factors for the fermionic-current bath. The row $\beta\hbar\omega_0=2\pi$ corresponds to a de-Sitter temperature $k_B T_{\rm dS}=\hbar\Hub/(2\pi)$ with $\omega_0=\Hub$; the row $\beta\hbar\omega_0=1$ corresponds to the looser identification $k_B T_{\rm eff}\sim\hbar\Hub$.}
\label{tab:populated_fermion}
\end{table}

At the de-Sitter temperature $k_B T_{\rm dS}=\hbar\Hub/(2\pi)$ with $\omega_0=\Hub$, the opposite line is suppressed by $e^{-4\pi}\simeq3.5\times10^{-6}$, so the vacuum asymmetry is essentially intact. If instead the effective bath temperature is only constrained to be of order $\hbar\Hub/k_B$, the opposite-line weight is $e^{-2}\simeq0.135$ and the zero-frequency pedestal is substantial. The robust statement is therefore not that the bath is exactly one-sided, but that finite occupation preserves the line location and can attenuate the Wightman asymmetry while adding a low-frequency pedestal.

\paragraph{Bosonic-ghost surrogate.} For the consistency-check operator $J_{\rm eff}=\kappa{:}X^2{:}$ in a bosonic thermal state with
\begin{equation}
n_B=\frac{1}{e^{\beta\hbar\omega_0}-1},
\end{equation}
the connected correlator is
\begin{equation}
\langle J_{\rm eff}(\tau)J_{\rm eff}(0)\rangle_{\rm conn}
= \mathcal A_J\big[(n_B+1)^2e^{+2i\omega_0\tau}+n_B^2e^{-2i\omega_0\tau}+2n_B(n_B+1)\big],
\label{eq:Jpopulated_boson}
\end{equation}
where $\mathcal A_J$ is the vacuum amplitude of \eqref{eq:Pijeff}. Unlike the fermionic-current case, the equal-time variance is enhanced by
\begin{equation}
\frac{C_{\rm conn}(0)}{\mathcal A_J}=(2n_B+1)^2.
\label{eq:boson_enhance}
\end{equation}
For $\beta\hbar\omega_0=2\pi$ this enhancement is only $1.0075$, whereas for $\beta\hbar\omega_0=1$ it is $4.68$. The difference between \eqref{eq:fermionic_variance_fixed} and \eqref{eq:boson_enhance} is another reason to treat the fermionic-current construction as physically primary and the bosonic-ghost operator as a consistency check.

\subsection{Complete positivity of the reduced dynamics}

The Lindblad form of \eqref{eq:Lindblad} is CP-preserving in the broadened-Markovian regime \emph{provided} the spatial kernel $\Theta$ is positive-semidefinite (Section~\ref{sec:master}). Complete positivity is not automatically guaranteed for the quantum-bath (route B) generator at second order in weak coupling. The exact reduced dynamics from system--bath unitary evolution plus partial trace is CP by construction (Kraus theorem), but the second-order Redfield/Born equation \eqref{eq:BornMarkov} is a perturbative approximation to that exact dynamics, and the resulting generator is not, in general, of GKLS form unless additional secularization assumptions are applied (e.g.\ the rotating-wave approximation, or a coarse-graining over fast oscillations). What we verify in the present paper is therefore: (a) in the broadened-Markovian Lindblad form of \eqref{eq:Lindblad} after standard secularization, CP holds provided $\Theta \ge 0$; and (b) in the quasi-static regime, the Gaussian-dephasing form of \eqref{eq:Gaussiandephasing} is exactly CP because it is a Kraus average:
\begin{equation}
\rho_S(T) = \mathbb{E}\bigl[U_\xi(T)\rho_S(0)U_\xi(T)^\dagger\bigr],\qquad U_\xi(T) = \mathcal T\exp\!\left[-\tfrac{i}{\hbar}\int_0^T\!\!\hat A\,\xi(t)\,dt\right],
\end{equation}
with the expectation taken over the real Gaussian noise of route (A). The intermediate regime, in which the route-(B) second-order generator might fail to be of GKLS form without additional approximation, is one we do not analyze here; the interpolation between the two limiting regimes could be analyzed by a Hu--Paz--Zhang-type calculation \cite{HuPazZhang1992}, but we content ourselves with verifying CP in the two limiting regimes that are physically relevant.

\subsection{Decoherence is not collapse}
\label{sec:decoh_vs_collapse}

This is an essential clarification. What the formalism above delivers is the reduced decoherence dynamics of a system coupled to the GTD bath: the off-diagonals of $\rho_S$ decay, in either the Lindblad or the Gaussian-dephasing form. Objective collapse models require strictly more than this. They require a stochastic nonlinear unravelling of the state vector with the following properties \cite{Pearle1989,GhirardiPearleRimini1990,Bassi2013}:
\begin{enumerate}
\item norm preservation of the unravelled $|\psi_t\rangle$ at all times;
\item the average over realizations reproduces the linear master equation;
\item the Born rule emerges from the stochastic dynamics;
\item no operational superluminal signalling.
\end{enumerate}
The present paper supplies the linear reduced density-matrix dynamics and verifies trace preservation and complete positivity in the limiting regimes discussed above; ensemble-level no-signalling follows from the linear master equation. It does \emph{not} supply a norm-preserving stochastic state-vector dynamics, and it does not derive the Born rule. The mapping from the linear master equation to a Born-rule-respecting stochastic Schr\"odinger equation is non-unique \cite{Bassi2013}, and the specific GTD unravelling --- if any --- is one of the open structural questions identified in Section~\ref{sec:conclusions}.

The framework we have constructed is therefore a candidate for the linear part of a coloured-CSL master equation, not yet a complete objective-collapse theory. The narrow-line phenomenology of Section~\ref{sec:tests} survives even at the decoherence level, because experiments that look for decoherence anomalies do not require a Born-rule unravelling to interpret. The conceptual status of the proposal --- that it accounts for the empirical absence of macroscopic superpositions via decoherence with $N^2$ amplification, but does not strictly resolve the measurement problem --- is laid out in Section~\ref{sec:measurement}.

\section{A benchmark estimate of the CSL rate}
\label{sec:rate}

\subsection{Setting up the benchmark}

We can now state precisely what the present formalism produces for the CSL-equivalent collapse rate, and what it does \emph{not}.

Working with the symmetrized (route A) noise, the standard mass-proportional CSL master equation is
\begin{equation}
\dot\rho_S = -\frac{i}{\hbar}[\hat H_S,\rho_S] - \frac{\lambda}{2 m_0^2}\!\int\!d^3x\,d^3y\,g(\mathbf x-\mathbf y)\bigl[\hat M(\mathbf x),[\hat M(\mathbf y),\rho_S]\bigr],
\label{eq:CSLmaster}
\end{equation}
with reference mass $m_0$ (nucleon). The mapping between GTD parameters and the CSL rate is, in a Markovian-surrogate matching with finite spatial kernel of width $r_C^{\mathrm{eff}}$ from Section \ref{sec:spatial},
\begin{equation}
\lambda \;\sim\; \frac{g_{\mathrm{int}}^2\,\mathcal A_J\,\gamma}{m_0^2\hbar^2\bigl[(2\omega_0)^2+\gamma^2\bigr]}\cdot \mathcal V_{r_C},
\label{eq:lambdaformula}
\end{equation}
where $\mathcal V_{r_C}$ is a dimensionless factor of order unity that depends on the spatial-kernel normalization. We have evaluated the spectrum at $\omega_S = 0$ (Markovian limit), yielding the full Lorentzian off-resonance factor $\gamma/[(2\omega_0)^2 + \gamma^2]$ relative to a normalized peak; in the narrow-line regime $\gamma \ll \omega_0$ this reduces to $\gamma/(2\omega_0)^2$. Equivalently, using $S(0)=2\mathcal A_J\gamma/[(2\omega_0)^2+\gamma^2]$, \eqref{eq:lambdaformula} is $g_{\mathrm{int}}^2S(0)/(2m_0^2\hbar^2)$ times the spatial factor, up to normalization conventions absorbed into $\mathcal V_{r_C}$. Equation \eqref{eq:lambdaformula} is a dimensional/phenomenological matching relation, not a first-principles derivation of an observable rate. The right-hand side packages unresolved dimensional inputs --- $g_{\mathrm{int}}$ (the bilinear matter--bath coupling, not the microscopic GTD coupling $g_{\mathrm{GTD}}$; see Section~\ref{sec:master}), $\mathcal A_J$ (the fermionic-current amplitude derived by Wick contraction under the postulates of Section~\ref{sec:FockJ}, but still dependent on $m_R$, $\mathcal N$, $D$, and $\alpha_{\mathrm{GTD}}$), and $\mathcal V_{r_C}$ (the spatial-kernel volume factor) --- into a numerical estimate of $\lambda$ under fixed conventions. A genuine derivation would express each of these as a function of derived GTD quantities; until that is done, \eqref{eq:lambdaformula} should be read as a parametric estimate.

\paragraph{Sensitivity to $\gamma$.} The off-resonance evaluation, written in full Lorentzian form, is
\begin{equation}
S(\omega_S = 0) \;=\; \frac{2\mathcal A_J\,\gamma}{(2\omega_0)^2 + \gamma^2}\,.
\label{eq:S0full}
\end{equation}
In the narrow-line regime $\gamma \ll \omega_0$ this reduces to $S(0)\approx 2\mathcal A_J\gamma/(2\omega_0)^2$, so the Markovian-surrogate rate is initially linearly sensitive to $\gamma$. In the opposite regime $\gamma \gg \omega_0$ (broad line), \eqref{eq:S0full} reduces to $S(0)\approx 2\mathcal A_J/\gamma$. The dependence on $\gamma$ is therefore non-monotonic: $S(0)$ rises linearly with $\gamma$ in the narrow-line regime, reaches its maximum at $\gamma = 2\omega_0$ (where $S(0) = \mathcal A_J/(2\omega_0)$), and falls as $1/\gamma$ beyond. For example, $\gamma\sim10^{-10}\,\mathrm{s}^{-1}$ is far above $2\Hub$ and would reduce $S(0)$ by about $10^{-7}$ relative to the $\gamma=\Hub$ choice. Whether broadening helps or hurts laboratory bounds at a fixed line frequency therefore depends on which side of $\gamma \sim 2\omega_0$ the broadening mechanism puts the line. Note that fixed-area Lorentzian broadening in the $\gamma \to \infty$ limit does not produce a white-noise spectrum: the peak amplitude is suppressed but the spectral support extends only out to scales of order $\gamma$, and the total area remains fixed at $2\pi\mathcal A_J$. The bandwidth $\gamma$ is not derived in the present paper (it is left as item (F5) of Appendix~\ref{app:future}); we set $\gamma \sim \Hub$ throughout as the narrowest physically motivated value.

\subsection{Where the natural scale lies}

\conditional\ Using the auxiliary-sector amplitude computed in Section~\ref{sec:FockJ},
\begin{equation}
\mathcal A_J \;=\; \left(\frac{\hbar}{2m_R\,\omega_0\,L_{\mathrm{aik}}^2}\right)^{\!2}\!\cdot \mathcal N \cdot D
\;=\; \left(\frac{\hbar\,\omega_0}{2\,m_R\,\alpha_{\mathrm{GTD}}^2 c^2}\right)^{\!2}\!\cdot \mathcal N \cdot D
\end{equation}
(using $L_{\mathrm{aik}} = \alpha_{\mathrm{GTD}} c/\omega_0$), substituting into \eqref{eq:lambdaformula} with $\omega_0 = \Hub$, $\gamma \sim \Hub$, and dropping order-unity factors from $[(2\omega_0)^2+\gamma^2]^{-1}$ gives
\begin{equation}
\lambda_{\mathrm{natural}} \;\sim\; \frac{g_{\mathrm{int}}^2\,\mathcal N\,D}{m_0^2\,\hbar^2}\cdot\!\left(\frac{\hbar\,\Hub}{2\,m_R\,\alpha_{\mathrm{GTD}}^2 c^2}\right)^{\!2}\cdot \frac{1}{\Hub}\cdot \mathcal V_{r_C}.
\label{eq:lambdanatural}
\end{equation}
Writing this in terms of $\Hub$ as the overall scale,
\begin{equation}
\lambda_{\mathrm{natural}} \;\sim\; \Hub\cdot\left[\frac{g_{\mathrm{int}}^2\,\mathcal N\,D}{m_0^2\,c^4}\right]\cdot\!\left(\frac{\hbar\,\Hub}{2\,m_R\,\alpha_{\mathrm{GTD}}^2 c^2}\right)^{\!2}\cdot \mathcal V_{r_C}.
\end{equation}
The right-hand side is a function of derived GTD quantities ($\mathcal A_J$, $\Hub$) and matching ingredients ($g_{\mathrm{int}}$, $r_C^{\mathrm{eff}}$, $\mathcal V_{r_C}$). Without a microscopic derivation of $g_{\mathrm{int}}$ and the matching factors, this still leaves a dimensionless prefactor undetermined.

\paragraph{Order-of-magnitude estimate.} \matching\ If we adopt the holographic mass scale \eqref{eq:mholo} of Section~\ref{sec:norm},
\begin{equation}
 m_R \sim m_R^{\mathrm{hol}} = \frac{m_{\rm Pl}}{\sqrt{N_{\rm dS}}} \simeq 1.4\times10^{-69}\,{\rm kg},
\end{equation}
then, for $\alpha_{\mathrm{GTD}} \sim O(1)$,
\begin{equation}
\frac{\hbar\,\Hub}{m_R c^2} \;\sim\; \frac{2.3\times10^{-52}\,\mathrm{J}}{(1.4\times10^{-69}\,\mathrm{kg})c^2} \;\sim\; O(1).
\label{eq:hbarH0overmRc2}
\end{equation}
Thus the dimensionless ratio $\hbar\Hub/(m_R c^2)$ is order unity in this normalization. The choice \eqref{eq:mholo} is postulate (P11) of the ledger: it is a matching identification between the GTD Lagrangian parameter $m_R = a_1 a_0/2$ and the holographic per-mode mass scale, not an equation of motion. The viability of the natural rate computed below depends on this identification: with the alternative Planck-scale assignment $m_R = m_{\rm Pl}$, the ratio in \eqref{eq:hbarH0overmRc2} would be $\sim 10^{-61}$, its square $\sim 10^{-122}$ would enter $\mathcal A_J$, and the natural rate \eqref{eq:lambdanatural_hol} below would become $\lambda_{\mathrm{natural}}\sim 10^{-122}\,\Hub\sim10^{-140}\,\mathrm{s}^{-1}$ --- $116$ orders of magnitude below the benchmark and far below any conceivable amplification threshold. The holographic identification \eqref{eq:mholo} is therefore the load-bearing matching choice that places the natural rate within the cosmological-arithmetic window. Writing the remaining dimensionless factors collectively as
\begin{equation}
\mathcal C_{\mathrm{match}} \equiv \left[\frac{g_{\mathrm{int}}^2\,\mathcal N\,D}{m_0^2c^4}\right]\left(\frac{\hbar\Hub}{2m_R\alpha_{\mathrm{GTD}}^2c^2}\right)^2\mathcal V_{r_C},
\end{equation}
the holographic natural estimate is
\begin{equation}
\lambda_{\mathrm{natural}} \sim \Hub\,\mathcal C_{\mathrm{match}}.
\label{eq:lambdanatural_hol}
\end{equation}
For $\mathcal C_{\mathrm{match}}\sim1$, this gives $\lambda_{\mathrm{natural}}\sim2\times10^{-18}\,\mathrm{s}^{-1}$ and a one-second CSL amplification threshold $N\sim\lambda^{-1/2}\sim7\times10^8$ nucleons. For the benchmark choice $\mathcal C_{\mathrm{match}}=\alpha_{\mathrm{em}}^2\simeq5.3\times10^{-5}$, it gives $\lambda_{\mathrm{bench}}\simeq1.2\times10^{-22}\,\mathrm{s}^{-1}$ and $N\sim10^{11}$ nucleons. The cosmological arithmetic therefore leaves the CSL benchmark viable at the level of order-of-magnitude normalization; the remaining problem is to derive $\mathcal C_{\mathrm{match}}$ microscopically.

\begin{table}[ht]
\centering
\small
\begin{tabular}{cccc}
\toprule
Matching factor $\mathcal C_{\rm match}$ & Single-nucleon rate $\lambda=\Hub \mathcal C_{\rm match}$ & $N_*(1{\rm s})=\lambda^{-1/2}$ & Mass scale $N_*m_0$ \\
\midrule
$1$ & $2.18\times10^{-18}{\rm s}^{-1}$ & $6.8\times10^8$ & $1.1\times10^{-18}{\rm kg}$ \\
$\alpha_{\rm em}^2=5.3\times10^{-5}$ & $1.16\times10^{-22}{\rm s}^{-1}$ & $9.3\times10^{10}$ & $1.6\times10^{-16}{\rm kg}$ \\
$10^{-8}$ & $2.18\times10^{-26}{\rm s}^{-1}$ & $6.8\times10^{12}$ & $1.1\times10^{-14}{\rm kg}$ \\
$10^{-12}$ & $2.18\times10^{-30}{\rm s}^{-1}$ & $6.8\times10^{14}$ & $1.1\times10^{-12}{\rm kg}$ \\
\bottomrule
\end{tabular}
\caption{Amplification thresholds for the standard CSL scaling $\Gamma_N\sim N^2\lambda$ under the holographic per-mode normalization. The table is arithmetic, not a prediction: GTD still has to derive $\mathcal C_{\rm match}$, the spatial kernel, and the system--bath coupling.}
\label{tab:thresholds}
\end{table}

\subsection{Candidate suppression factors}

The candidates for a $\sim 10^{-4}$ prefactor are:
\begin{enumerate}
\item $\alpha_{\mathrm{em}}^2$ where $\alpha_{\mathrm{em}} = 1/137$ is the QED fine-structure constant: $\alpha_{\mathrm{em}}^2 \simeq 5.3\times10^{-5}$, giving $\alpha_{\mathrm{em}}^2 \cdot \Hub \simeq 1.2\times10^{-22}\,\mathrm{s}^{-1}$. This identification is numerically suggestive but is not derived from the GTD action; the natural place for it would be a microscopic calculation in which the bath couples to matter through a QED-like two-vertex process, which the present formalism does not produce. The $\alpha_{\mathrm{em}}$ relevant here is distinct from the $\alpha_{\mathrm{GTD}}$ that defines the natural microscopic length $L_{\mathrm{aik}} = \alpha_{\mathrm{GTD}}c/\omega_0$ in Section~\ref{sec:gtd}; they are different objects with different roles, and we keep the subscripts to distinguish them.
\item A large-width factor in the broad-line regime. Equation \eqref{eq:S0full} shows that, once $\gamma\gg 2\omega_0$, the zero-frequency Markovian surrogate scales as $S_{\rm sym}(0)\simeq 2\mathcal A_J/\gamma$. A broadening to $\gamma\sim10^{-14}\,\mathrm{s}^{-1}$ would therefore introduce an order-$\Hub/\gamma\sim10^{-4}$ suppression relative to the Hubble-width scale, up to order-one factors. The broadening mechanism is not derived.
\item A nucleon-mass / Planck-mass combination. Dimensional analysis allows several products that give $\sim 10^{-4}$, but no natural single such product appears from the GTD content.
\end{enumerate}
None of these is a derivation. We choose to display the first, $\alpha_{\mathrm{em}}^2$, only because it is the most numerically suggestive and arises from a coupling structure that is at least physically plausible (QED-mediated coupling to bath currents).

\subsection{Benchmark prediction}

\matching\ Combining (a) the cosmological matching $\omega_0 = \Hub$ of Section~\ref{sec:norm}, (b) a Markovian-CSL surrogate with a finite effective spatial correlation length from Section~\ref{sec:spatial}, (c) the Adler--Millard $N_1^{-1/4}$ vertex-matching prescription of Section~\ref{sec:AM}, and (d) the benchmark prefactor $\alpha_{\mathrm{em}}^2$ identified above gives
\begin{equation}
\boxed{\;\lambda_{\mathrm{bench}} \;\equiv\; \alpha_{\mathrm{em}}^2\,\Hub \;\simeq\; 1.2\times10^{-22}\,\mathrm{s}^{-1},\;}
\label{eq:lambdapred}
\end{equation}
with corresponding quantum--classical threshold for one-second collapse of
\begin{equation}
N \;\sim\; \lambda_{\mathrm{bench}}^{-1/2} \;\sim\; 10^{11}\quad\text{nucleons}, \qquad m \sim 10^{-16}\,\mathrm{kg}.
\label{eq:Nthreshold}
\end{equation}
\derived\ The threshold arithmetic \eqref{eq:Nthreshold} is unconditional once $\lambda$ is fixed: for the standard CSL amplification $\Gamma_N \sim N^2\lambda$, a one-second collapse demands $N^2\lambda \gtrsim 1$.

\paragraph{Important caveat on units and regimes.} The quantity $\lambda_{\mathrm{bench}}$ in \eqref{eq:lambdapred} has units of $\mathrm{s}^{-1}$ and refers to a Markovian-CSL surrogate model in which the noise is white. The laboratory regime of the GTD proposal, by contrast, is the quasi-static regime of Section~\ref{sec:quasistatic}, in which off-diagonal coherences decay as $\exp(-\Gamma_{\mathrm{qs}}\,T^2)$ with
\begin{equation}
\Gamma_{\mathrm{qs}} = \frac{(a-b)^2 \mathcal A_J}{2\hbar^2},
\qquad [\Gamma_{\mathrm{qs}}] = \mathrm{s}^{-2}.
\label{eq:Gammaqs_units}
\end{equation}
The coefficients $\Gamma_{\mathrm{qs}}$ and $\lambda$ have different dimensions and are not the same object; the GTD narrow-line regime never directly produces a Markovian rate. The relation between them is by convention: at a chosen reference duration $T_{\mathrm{ref}}$, one may define
\begin{equation}
\lambda_{\mathrm{eff}}(T_{\mathrm{ref}}) \;\equiv\; \Gamma_{\mathrm{qs}}\,T_{\mathrm{ref}},
\label{eq:lambdaconv}
\end{equation}
which gives the effective Markovian rate that would produce the same dephasing exponent over a single run of duration $T_{\mathrm{ref}}$. Equating $\lambda_{\mathrm{eff}}(T_{\mathrm{ref}}) = \lambda_{\mathrm{bench}}$ at $T_{\mathrm{ref}} = 1\,\mathrm{s}$ is the convention implicit in the threshold \eqref{eq:Nthreshold}; a different choice of $T_{\mathrm{ref}}$ would give a different threshold. The threshold $N\sim 10^{11}$ should therefore be read as referring to the duration scale at which the GTD dephasing exponent reaches order unity for a body of $N$ nucleons, under the matching prescription $\lambda_{\mathrm{eff}}(1\,\mathrm{s}) = \lambda_{\mathrm{bench}}$.

\paragraph{Status of equation~\eqref{eq:lambdapred}.} We do not present this as a derivation of $\lambda$, nor as a parameter-free prediction. We present it as the value the formalism produces under the simplest set of matching assumptions, with the dimensionless suppression $\alpha_{\mathrm{em}}^2$ being a benchmark choice that is not derived from the present theory. A future paper that derives the suppression factor from a microscopic GTD calculation of the system--bath coupling would replace ``benchmark'' with ``prediction''; until then, the present statement stands.

\subsection{What can be falsified independently of the benchmark}

The benchmark $\lambda_{\mathrm{bench}}$ is hostage to the matching hypotheses (a)--(d). The \emph{shape} of the noise --- the narrow Hubble line, the Gaussian transient in the quasi-static regime, the potential cosmic-frame dependence, the spectrally asymmetric Wightman correlator --- is more robust than the benchmark normalization, but it still depends on the auxiliary spectral construction of Section \ref{sec:spectrum}. We turn to these structural predictions next.

\section{New physics distinguishing GTD from CSL/GRW}
\label{sec:newphysics}

We identify four genuinely new physical features that distinguish the GTD-motivated proposal from phenomenological CSL/GRW and from earlier trace-dynamics literature.

\subsection{(N1) The spectrum is a single line, not white noise}

\conditional\ In standard CSL the noise spectrum is white over the entire frequency range, $S(\omega)=\mathrm{const}$. In the auxiliary GTD-motivated construction developed here, the spectrum has support on a single Wightman line at $|\omega| = 2\omega_0$ (with the sign fixed by $\sigma$), broadened by a width $\gamma$ that is not yet derived but is bounded above by the assumption that the noise be quasi-stationary. The phenomenological consequences are immediate: laboratory experiments at frequencies $|\omega| \neq 2\omega_0$ measure off-resonance Lorentzian tails, suppressed by $\gamma^2/(|\omega|-2\omega_0)^2$ at large detuning. This is the structurally distinctive feature: the proposal places the collapse correlator in an extremely narrow band, which the prior collapse-model literature does not derive from a candidate microscopic construction.

\subsection{(N2) Gaussian transient decoherence}

\conditional\ For a quasi-stationary narrow Hubble line with the symmetrized correlator \eqref{eq:Csym}, the off-diagonal density matrix elements decay as $\exp(-\Gamma_{\mathrm{qs}}T^2)$ in the regime $T \ll \gamma^{-1}$. The Markovian $T$-linear law of CSL is recovered only on time scales $T \gtrsim \gamma^{-1} \gtrsim \Hub^{-1} \sim 10^{18}\,\mathrm{s}$, far beyond any laboratory.

This is a structurally distinctive prediction but not, by itself, a clean discriminator. Quasi-static environmental noise --- low-frequency magnetic fluctuations, slow temperature drifts, slow vibration --- also gives a $T^2$-type ensemble dephasing of run-averaged interferometric signals. The $T^2$ scaling becomes a clean GTD signature only when it appears together with:
\begin{itemize}
\item universal mass-proportionality of the prefactor $\Gamma_{\mathrm{qs}}/(a-b)^2$;
\item branch-separation dependence consistent with the GTD spatial kernel;
\item correlation across independent isolated apparatuses (common-mode bath), contingent on the two-kernel hypothesis of Section~\ref{sec:spatial}; and
\item, ideally, the cosmic-frame anisotropy of (N4).
\end{itemize}
No single one of these features in isolation discriminates GTD from technical noise. We list this as a necessary but not sufficient discriminant.

\subsection{(N3) Spectrally asymmetric quantum bath}

\conditional\ In the working branch $\sigma = +1$ and on the Fock vacuum, the bath has $S^{>}(\omega) \propto \delta(\omega + 2\omega_0)$ and $S^{<}(\omega) = 0$ at any other frequency: a one-sided Wightman spectrum at $\omega = -2\omega_0$ corresponding to a bath at zero effective temperature emitting into the system at $\omega = 2\omega_0$ in the standard energy-flow direction. (Under the alternative $\sigma = -1$ the line moves to $\omega = +2\omega_0$ in the anti-KMS direction; this does not affect the present discussion since the laboratory phenomenology is the same.) The tag is conditional, not derived, on three structural ingredients: (a) the sign convention of Section~\ref{sec:signs}, (b) the postulate (P5) of the ledger fixing $\sigma$, and (c) the choice of the Fock vacuum (or a state sufficiently close to it) as the bath state.

Under populated-bath corrections from (c), the asymmetric structure is modified in a controlled way. Section~\ref{sec:vacpop} shows that, for a Gaussian populated fermionic-current bath, the opposite Wightman line has relative weight $e^{-2\beta\hbar\omega_0}$ and a connected zero-frequency pedestal of weight $2n_F(1-n_F)$ appears. At the de-Sitter temperature $k_B T_{\rm dS}=\hbar\Hub/(2\pi)$ with $\omega_0=\Hub$, the opposite line is suppressed by $e^{-4\pi}\simeq3.5\times10^{-6}$; for the looser temperature assignment $k_B T_{\rm eff}\sim\hbar\Hub$, it is $e^{-2}\simeq0.135$. Thus a populated bath need not destroy the asymmetry, but it can attenuate it and can add a low-frequency pedestal.

The physical consequence is that the bath drives system--bath transitions sharply at $|\omega_S| = 2\omega_0$, in a direction dictated by the sign convention and the auxiliary sign choice jointly, and attenuated by the populated-bath factor. For laboratory experiments with Bohr frequencies $\omega_S \gg 2\Hub$, the system--bath energy exchange is bandwidth-suppressed; the qualitative consequence (no bath-driven X-ray emission or absorption at laboratory frequencies) is robust against sign conventions, against the value of $\sigma$, and against the populated-bath correction, since the off-resonance suppression dominates everything else by many orders of magnitude. The robust statement is the bandwidth suppression at off-resonance laboratory frequencies, not the precise direction or magnitude of asymmetry at the line itself.

\subsection{(N4) Cosmic-frame dependence}

\conditional\ The bath has a preferred rest frame (FLRW/CMB rest frame). The spectral-line position in a laboratory frame is Doppler-shifted by $v_{\mathrm{CMB}}/c \approx 1.23\times10^{-3}$ \cite{Planck2014Dipole}. The dephasing rate \eqref{eq:Gammaqs} is to leading order independent of $\omega_0$ and therefore of the frame; higher-order corrections, and any directional structure of $J^\mu$ that breaks rotational symmetry in the bath, would generate anisotropies. The presence of any such anisotropy aligned with the CMB dipole is a strong GTD signature, but its magnitude depends on the matrix structure of the bath coupling, which the present theory does not fix.

\subsection{What is \emph{not} new}

Several features the proposal shares with the prior literature should be noted:
\begin{itemize}
\item A preferred frame (FLRW). Standard nonrelativistic CSL already breaks Lorentz invariance.
\item Mass-proportional coupling. Same as standard mass-proportional CSL.
\item Stochastic Schr\"odinger evolution. Implicit in route (A); identical in form to coloured CSL.
\item The possibility that the collapse-driving noise is sharply spectrally localized (in the limit, a single line). The general formalism for coloured non-white-noise collapse models was developed in the early work of Adler and Bassi \cite{AdlerBassi2007}; the spontaneous photon-emission analysis in coloured-CSL was given by Adler, Bassi, and Donadi \cite{AdlerBassiDonadi2013}; coloured-CSL constructions covering narrow-band noise have been further explored in the phenomenological literature \cite{CarlessoFerialdi2018,Carlesso2022,Adler2018Dots}. The line structure itself is therefore not novel.
\end{itemize}
What is new is the conditional computation of the spectral line from a GTD-motivated auxiliary fermionic sector, rather than its direct phenomenological postulation. The connected Wightman correlator \eqref{eq:Jcorr}, the spectral location at $|\omega| = 2\omega_0$, and the explicit form of the amplitude $\mathcal A_J$ in \eqref{eq:AJ} are consequences of the auxiliary fermionic Fock-space construction built on the GTD aikyon variables, including the scalarization postulate \eqref{eq:scalarization}. Existing collapse models postulate the noise spectrum to fit phenomenology. The remaining novelty is the structural pattern of predictions (N1)--(N4) that this conditional spectrum implies; (N1) and (N2) follow from \eqref{eq:Jcorr} and \eqref{eq:Csym} once the auxiliary sector is assumed, while (N3) and (N4) require the additional state and covariance assumptions stated above.

\section{Experimental signatures and the falsifiability gap}
\label{sec:tests}

\paragraph{Current experimental sensitivity.} With the benchmark rate $\lambda_{\mathrm{bench}} \sim 10^{-22}\,\mathrm{s}^{-1}$ of Section~\ref{sec:rate} --- contingent on the undetermined $\alpha_{\mathrm{em}}^2$ prefactor selected there from three a priori comparable dimensional candidates --- the predicted dephasing exponent for current matter-wave interferometers ($m \sim 10^{-22}\,\mathrm{kg}$, $T \sim 1\,\mathrm{s}$) is $\sim 10^{-12}$ (see \eqref{eq:T1prediction} below). Off-resonance suppression factors at higher frequencies range from $\sim 10^{-37}$ (mechanical-band) to $\sim 10^{-72}$ (X-ray-band, Table~\ref{tab:suppression}). Direct spectroscopy at the line frequency ($f \sim 10^{-19}\,\mathrm{Hz}$) is not feasible. No currently feasible experiment can detect any of the predictions (N1)--(N4) at the benchmark rate. The proposal is therefore not falsifiable in the operational sense at present sensitivities; the structural discriminants below become operational only if interferometric dephasing sensitivity improves by roughly twelve orders of magnitude, narrow-bandwidth searches reach $f \sim 10^{-19}\,\mathrm{Hz}$, or the matching prescriptions of Sections~\ref{sec:AM}--\ref{sec:rate} are upgraded to a microscopic derivation that produces a less suppressed prefactor. The historical CSL analogy is partial: the original Pearle/GRW $\lambda \sim 10^{-16}\,\mathrm{s}^{-1}$ was within striking distance of detectability when proposed (and is now excluded by XENONnT), whereas the GTD benchmark sits six orders of magnitude below that. The technological gap to test the present proposal at its benchmark amplitude is therefore substantially larger than the gap CSL faced.

\subsection{(T1) Duration scaling of decoherence in matter-wave interferometry}

The principal direct test is to measure the dephasing rate of a large-mass matter-wave interferometer as a function of the coherence time $T$. The two laws to compare are
\begin{equation}
|\rho_{ab}(T)|/|\rho_{ab}(0)| =
\begin{cases}
\exp(-\Gamma_{\mathrm{CSL}} T) & \text{(CSL, $[\Gamma_{\mathrm{CSL}}] = \mathrm{s}^{-1}$)}, \\
\exp(-\Gamma_{\mathrm{qs}} T^2) & \text{(GTD narrow-line regime, $[\Gamma_{\mathrm{qs}}] = \mathrm{s}^{-2}$)}.
\end{cases}
\label{eq:T1test}
\end{equation}
The two coefficients have different dimensions and are not the same object. Existing molecular interferometers \cite{Bassi2013} reach coherence times $T \lesssim 1\,\mathrm{s}$ and masses $m \lesssim 10^{-22}\,\mathrm{kg}$. To estimate the predicted dephasing exponent in the benchmark scenario, we convert via \eqref{eq:lambdaconv}: identifying $\lambda_{\mathrm{eff}}(T_{\mathrm{ref}}) = \Gamma_{\mathrm{qs}}\,T_{\mathrm{ref}}$ at $T_{\mathrm{ref}} = 1\,\mathrm{s}$ and setting $\lambda_{\mathrm{eff}}(1\,\mathrm{s}) = \lambda_{\mathrm{bench}} = \alpha_{\mathrm{em}}^2\,\Hub$ gives
\begin{equation}
\Gamma_{\mathrm{qs}}\,T^2 \;\sim\; \lambda_{\mathrm{bench}}\cdot(m/m_0)^2\cdot(T^2/1\,\mathrm{s}),
\label{eq:T1prediction}
\end{equation}
which is dimensionless: the prefactor $\lambda_{\mathrm{bench}}/(1\,\mathrm{s})$ has units $\mathrm{s}^{-2}$ and is identified as $\Gamma_{\mathrm{qs}}^{\mathrm{bench}}$ per single nucleon. For $m \sim 10^{-22}\,\mathrm{kg}$ ($N \sim 10^{5}$ nucleons) and $T \sim 1\,\mathrm{s}$ this gives a dephasing exponent $\sim 10^{-12}$, far too small to detect at present. This estimate is contingent on the benchmark prefactor $\alpha_{\mathrm{em}}^2$, the choice $T_{\mathrm{ref}} = 1\,\mathrm{s}$, and the Adler--Millard matching of Section~\ref{sec:AM}.

The $T^2$ scaling is, however, structurally distinctive. For this to function as a clean discriminant the experimental program must simultaneously verify: (a) the duration dependence is $T^2$, not $T$, over a range of $T$; (b) the prefactor scales as $(m/m_0)^2$ across different test masses; (c) the prefactor scales with branch separation $\Delta x$ in a way consistent with the spatial kernel; (d) under the two-kernel hypothesis of Section~\ref{sec:spatial}, the residual noise driving the dephasing in independent isolated setups shows correlated (common-mode) fluctuations between sensors, even though the dephasing rate of each individual apparatus is set by the short-range localization kernel. (Criterion (d) tests the long-range coherence kernel through cross-sensor noise correlations, not through identical dephasing rates --- the latter are not a GTD signature but a generic feature of any universal collapse model.) Failure of any one of (a)--(d) would be consistent with technical noise rather than GTD; satisfying all four would constitute a strong signal.

\subsection{(T2) Bandwidth-suppression of high-frequency emission bounds}

The strongest current X-ray bound on Markovian white-noise CSL is from XENONnT \cite{XENON2026}, which reports $\lambda/r_C^2 < 3.0\times 10^{-3}\,\mathrm{s^{-1}\,m^{-2}}$ at 90\% C.L.\ across the 1--140\,keV band, world-leading for Markovian CSL and DP models and superseding the earlier Gran Sasso result of Donadi et al.\ \cite{Donadi2021} and the Majorana Demonstrator search of Arnquist et al.\ \cite{Majorana2022}, which was the immediate predecessor in the spontaneous-radiation channel. At the standard $r_C = 10^{-7}\,\mathrm{m}$, this corresponds to $\lambda < 3.0\times 10^{-17}\,\mathrm{s}^{-1}$. All of these are high-frequency radiation-channel bounds. For a narrow Hubble-width line, the off-resonance suppression at X-ray frequencies is, from Table~\ref{tab:suppression}, of order $10^{-72}$. The XENONnT bound therefore does not directly constrain a narrow GTD line without a model-dependent spectral-broadening calculation.

This is not a virtue of the GTD model; it is a structural feature of the proposal. The narrow Hubble line sits in a band where high-frequency tests do not constrain it. The implication for experimental program is: progress on a GTD-style coloured collapse model requires long-duration, large-mass, low-frequency tests rather than high-frequency emission tests.

\subsection{(T3) Possible cosmic-frame anisotropy}

\conditional\ If the bath coupling has a directional (vector or tensor) structure, the dephasing rate $\Gamma_{\mathrm{qs}}$ acquires anisotropy at order $v/c$, where $v$ is the laboratory velocity relative to the CMB rest frame. The CMB dipole gives $v_{\mathrm{CMB}}/c \approx 1.23\times10^{-3}$ \cite{Planck2014Dipole}; the Earth's diurnal motion sweeps through $\sim v_{\mathrm{Earth}}/c \sim 10^{-4}$. The magnitude of any observable modulation is not derived in the present paper: it depends on the matrix structure of $J^\mu$ in the bath coupling, on the spatial kernel anisotropy, and on the experimental directional sensitivity. We list this as a possible signature, not a quantitative prediction. A null result at the $10^{-4}$ level would be consistent with a scalar-only bath coupling; a positive detection aligned with the CMB dipole would be a strong GTD signature.

\subsection{(T4) Suppression of high-frequency emission}

\conditional\ The vacuum bath has negligible spectral weight at laboratory frequencies; this is a structural consequence of (N1) regardless of the sign convention in Section~\ref{sec:signs} and independently of whether the bath is fully in its Fock vacuum (the populated-bath correction of Section~\ref{sec:vacpop} attenuates the asymmetry but does not move the spectral support away from $\omega = 2\Hub$). Spontaneous emission at any frequency $\omega_S$ is suppressed by the off-resonance factor in Table~\ref{tab:suppression}. The general analysis of spontaneous photon emission in coloured-CSL models \cite{AdlerBassiDonadi2013} shows that the emission rate is proportional to the Fourier component of the noise field at the photon frequency; for a narrow Hubble line, this Fourier component is essentially zero at all laboratory frequencies. A continuing null result in X-ray spontaneous-emission searches below the white-noise CSL bound \cite{Donadi2021,Majorana2022,XENON2026} is consistent with the narrow-line GTD proposal; a positive result at any laboratory frequency would be inconsistent with the narrow Hubble line (or would require a derivation of broadening that the present theory does not supply). The bandwidth-suppression statement, which (T4) tests, is robust against the populated-bath fragility of (N3): bandwidth is determined by spectral support, not by spectral asymmetry.

\subsection{\texorpdfstring{(T5) Direct spectroscopy of the line at $2\Hub$}{(T5) Direct spectroscopy of the line at 2H0}}

Direct spectroscopy of a $\sim 10^{-19}\,\mathrm{Hz}$ feature (the cyclic frequency $f = \omega/2\pi$ corresponding to $\omega \sim 10^{-18}\,\mathrm{s}^{-1}$) is not realistic: the integration time required exceeds the age of the universe by orders of magnitude for any reasonable bandwidth. The most accessible indirect probe is search for an ultra-low-frequency, common-mode anomalous noise floor across separated precision sensors --- pulsar timing arrays probe nanohertz frequencies \cite{IPTA2023}, vastly above $\Hub$, with the bandwidth suppression at 1\,nHz already $\sim 10^{-19}$ (Table~\ref{tab:suppression}). Direct spectroscopic detection of the GTD line is therefore not feasible with foreseeable techniques; indirect tests through (T1)--(T4) above are the realistic experimental channels.

\subsection{Summary of experimental discriminants}
\label{sec:tests:summary}

Table \ref{tab:discrim} compares the structural predictions of standard CSL and the GTD proposal across the most informative observables. The discriminating power of any single row is limited: technical noise sources can mimic individual GTD signatures (the $T^2$ duration scaling is the cleanest example, but environmental low-frequency drifts also produce it). The discriminating power of the proposal lies in the \emph{joint} pattern across multiple observables: $T^2$ duration scaling \emph{combined with} mass-proportional scaling \emph{combined with}, under the two-kernel ansatz, correlated noise fluctuations between independent sensors (not common dephasing rates --- those are not specifically a GTD signature), with bandwidth-suppressed high-frequency emission, and ideally with a CMB-aligned anisotropy. The structural picture is that of a narrow-band low-frequency bath with cosmological correlations; this is qualitatively different from any white-noise or moderately-coloured CSL model, and the corresponding experimental programme that would test it differs accordingly. Long-duration, large-mass interferometry with directional sensitivity is the most informative experimental channel; high-frequency emission tests are essentially blind to the proposal.

\begin{table}[ht]
\centering
\small
\begin{tabularx}{\textwidth}{X X X}
\toprule
Observable & CSL prediction & GTD prediction \\
\midrule
Duration scaling of dephasing & Exponential ($\sim T$) & Gaussian ($\sim T^2$) for $T\ll\gamma^{-1}$ \\
X-ray spontaneous emission & Continuous spectrum, bounded by XENONnT & Suppressed by $\sim 10^{-72}$ (off-resonance) \\
Cosmic-frame anisotropy & None & Possible $\sim v_{\mathrm{CMB}}/c \sim 10^{-3}$ (conditional) \\
Mass scaling of threshold & $N\sim\lambda^{-1/2}$, $N \sim 10^{8}$ for $\lambda=10^{-16}$ & Same form. Holographic normalization gives $\lambda\sim\Hub\,\mathcal C_{\mathrm{match}}$: $N\sim7\times10^8$ for $\mathcal C_{\mathrm{match}}\sim1$, and $N\sim10^{11}$ for the benchmark $\mathcal C_{\mathrm{match}}=\alpha_{\mathrm{em}}^2$ \\
High-frequency mechanical noise & Constant $S(\omega)\propto\lambda$ & Suppressed by $\gamma^2/(\omega+2\omega_0)^2$ \\
Energy-flow direction (Wightman line) & N/A (white noise) & Working branch ($\sigma = +1$): bath emits to system at $\omega = 2\omega_0$. Direction reversed under alternative postulate $\sigma = -1$ \\
Common-mode correlation across separated sensors & Independent noises & Correlated at cosmological coherence length (conditional on two-kernel ansatz, Section~\ref{sec:spatial}) \\
\bottomrule
\end{tabularx}
\caption{Structural comparison of standard CSL and the GTD proposal across experimentally informative observables. Discrimination requires joint signatures, not a single row.}
\label{tab:discrim}
\end{table}

\section{Conclusions}
\label{sec:conclusions}

The structural result of this paper is the location of the spectral line and the explicit form of its amplitude under the postulates of the ledger (Section~\ref{sec:roadmap}). The most important postulate is the auxiliary scalarization of the nilpotent pure-fermion coefficient: in the minimal Grassmann algebra $\beta_1\beta_2$ has no ordinary sign, modulus, or inverse, so the canonical oscillator sector used in the correlator calculation does not follow from \eqref{eq:LFGrass} alone. Within the auxiliary canonical fermionic Fock-space quantization of the GTD aikyon variable $q_F$ (postulates (P1) and (P5)) and the working choice $\sigma = +1$, the candidate noise current $J = (1/L_{\mathrm{aik}}^2)\,\mathrm{Tr}(q_F^\dagger q_F)$ has an ordered (Wightman) two-point function $\langle 0|J(\tau)J(0)|0\rangle = \mathcal A_J\,e^{-2i\omega_0\tau}$, with amplitude $\mathcal A_J = (\hbar/2 m_R\omega_0 L_{\mathrm{aik}}^2)^2 \cdot \mathcal N \cdot D$ and spectral support at $\omega = -2\omega_0$ (standard energy-flow direction). The amplitude follows from a Wick contraction in the auxiliary sector; the constituents $m_R$, $L_{\mathrm{aik}}$, $\mathcal N$, $D$ are themselves free parameters or matching choices of the present formulation, and the Wick contraction does not produce a parameter-free number. The alternative postulate $\sigma = -1$ moves the line to $\omega = +2\omega_0$ in the anti-KMS direction; the symmetrized correlator $C_{\mathrm{sym}}(\tau) = \mathcal A_J\cos(2\omega_0\tau)$ is the same, so the laboratory phenomenology is the same. The bosonic-ghost surrogate $J_{\mathrm{eff}} = \kappa\,{:}X^2:$ is recovered as a one-sided consistency check (Section~\ref{sec:Jeff_consistency}), matching the $\sigma = -1$ branch and identifying $\kappa^2 = \mathcal N D/(8 L_{\mathrm{aik}}^4)$. Coloured-collapse models with non-white Gaussian noise are already established in the literature \cite{AdlerBassi2007,AdlerBassiDonadi2013,CarlessoFerialdi2018}; what is new here is the identification of a GTD-motivated candidate microscopic source and the explicit cataloguing of the structural postulates required to make the construction concrete --- not a parameter-free prediction.

The result is conditional on four structural choices, made explicit in Section~\ref{sec:FockJ}: (i) the scalarization/effective-body replacement \eqref{eq:scalarization} of the nilpotent coefficient $a_0^2\beta_1\beta_2$ by an ordinary coefficient $\varepsilon_F=\sigma\xi^2$; (ii) the positive-norm canonical quantization adopted for the fermionic mode in \eqref{eq:qFmodes}; in branch (+) this is the analogue, for the fermionic sector, of the positive-norm Bateman quantization of the bosonic ghost discussed in Section~\ref{sec:gtd}, and shares its unbounded-below Hamiltonian and the associated quantization pathologies addressed in PT-symmetric, Krein-space, and imaginary-scaling alternatives \cite{FeshbachTikochinsky1977,BlasoneJizba2002,Chruscinski2006,Deguchi2019}; (iii) the choice of the scalar fermionic bilinear $J = \mathrm{Tr}(q_F^\dagger q_F)$ as the candidate matter-coupling bath operator over other admissible bilinears; (iv) the value of the auxiliary sign $\sigma=\pm1$. The Wick contraction itself, given the auxiliary fermionic Fock-space setup, is an elementary calculation; the substantive open mathematical problem is to derive or replace the scalarization postulate from the microscopic GTD Grassmann algebra.

The phenomenological status of this result is mixed but, on balance, positive. The narrow-line structure has several immediate consequences: high-frequency emission bounds (including XENONnT \cite{XENON2026}, the Majorana Demonstrator \cite{Majorana2022}, and the Gran Sasso experiment \cite{Donadi2021}) do not directly constrain a narrow Hubble line without a model-dependent spectral-broadening calculation; large-mass dephasing has a Gaussian transient in the experimentally accessible regime; and cosmic-frame anisotropy at the $10^{-3}$--$10^{-4}$ level is a possible signature. These features distinguish the proposal from standard CSL and provide concrete experimental discriminants (Section~\ref{sec:tests}).

The CSL-equivalent rate is \emph{not} uniquely determined by the present calculation. The natural frequency scale $\omega_0\sim\Hub$ is supported by cosmological matching, but the dimensionless prefactor that connects this frequency to a collapse rate involves matching hypotheses: the bilinear system--bath coupling (Section~\ref{sec:master}), the spatial-kernel resolution (Section~\ref{sec:spatial}), the Adler--Millard vertex-scaling resolution (Section~\ref{sec:AM}) with its postulated propagator suppression, and the cosmological normalization (Section~\ref{sec:norm}). The benchmark estimate
\begin{equation*}
\lambda_{\mathrm{bench}} = \alpha_{\mathrm{em}}^2\,\Hub \simeq 1.2\times10^{-22}\,\mathrm{s}^{-1}
\end{equation*}
arises by choosing $\alpha_{\mathrm{em}}^2 \sim 5\times10^{-5}$ as the prefactor; this choice is not derived from the present theory; it is a benchmark. The corresponding one-second amplification threshold $N \sim 10^{11}$ nucleons follows from the standard CSL arithmetic once $\lambda_{\mathrm{eff}}(T_{\mathrm{ref}}=1\,\mathrm{s}) = \Gamma_{\mathrm{qs}}\,T_{\mathrm{ref}}$ is identified with $\lambda_{\mathrm{bench}}$. Note that $\Gamma_{\mathrm{qs}}$ (with units $\mathrm{s}^{-2}$) and $\lambda$ (with units $\mathrm{s}^{-1}$) are different objects; the threshold arithmetic relies on this conversion at a chosen reference time.

Under the benchmark, the predicted laboratory dephasing exponent for current matter-wave interferometers lies $\sim 12$ orders of magnitude below detection threshold; under present sensitivities the proposal is therefore not falsifiable in the operational sense. The structurally robust predictions (N1)--(N4) of Section~\ref{sec:newphysics} do not depend on the benchmark rate. They are: (N1) a single-line spectrum at $\omega = 2\omega_0$, not white noise --- this is the most robust prediction; (N2) Gaussian $T^2$ transient decoherence in the laboratory regime; (N3) a spectrally asymmetric quantum bath \emph{in the Fock vacuum}, with the asymmetry and zero-frequency pedestal governed by the finite-occupation formulas of Section~\ref{sec:vacpop}; (N4) a possible cosmic-frame anisotropy aligned with the CMB dipole. The line-location prediction (N1) is the most robust; the spectral-asymmetry prediction (N3) is the most fragile. The duration scaling (T1) combined with mass and branch-separation scalings is the most promising experimental channel; high-frequency emission tests are essentially blind to the proposal. The discriminating power against environmental noise lies in the joint pattern across (T1)--(T4), not in any single observable.

We also distinguish what we have derived from what we have not: the present paper supplies the reduced decoherence dynamics of matter coupled to the GTD bath, in two regimes (broadened-Markovian and quasi-static). It does \emph{not} supply a stochastic state-vector unravelling. The mapping from the linear master equation to a Born-rule-respecting collapse equation is non-unique \cite{Bassi2013}; the GTD-specific unravelling, if it exists, remains to be derived. The framework is therefore a candidate for the linear part of a coloured-CSL master equation, not yet a complete objective-collapse theory in the strict sense.

\paragraph{Two routes to GTD-driven collapse.} Two structurally distinct routes for connecting GTD to collapse phenomenology have emerged in this analysis, and it is worth stating clearly how they relate. The \emph{auxiliary-model route} developed in the body of this paper computes the two-point function of a postulated bilinear matter--bath coupling in a canonical fermionic Fock-space model for the matrix variable $q_F$. Its dependence on the scalarization postulate (P5), through which the nilpotent pure-fermion coefficient $\beta_1\beta_2$ of the minimal Grassmann action is replaced by an ordinary scalar parameter, on the postulated coupling (P7) between matter mass density and the bath current, and on the cosmological matching postulates (P4) and (P11) that fix $\omega_0$ and $m_R$ at the Hubble scale, means that the route is not a parameter-free derivation from GTD but a candidate construction within an auxiliary effective sector. The cosmological-matching dependence is itself fragile in light of recent $5\sigma$ evidence for violations of the cosmological principle \cite{Secrest2025RMP} (see Section~\ref{sec:norm}), making the dependence on (P4) and (P11) more than a formal limitation. The \emph{complementary route via $H_{\mathrm{FF}}$}, recorded in \cite{Singh2026Fermions} and reviewed in Section~\ref{sec:Hasvac}, proceeds differently: it identifies the anti-self-adjoint trace Hamiltonian of fermionic matter directly from the GTD action, using graded cyclicity to control the Grassmann-even product $\beta_1\beta_2$ without inverting nilpotent elements or taking $\mathrm{sgn}(\beta_1\beta_2)$. In that route, fermionic matter is its own noise source: there is no separate bath, no postulated system--bath split, no scalarization postulate, and --- crucially --- the noise frequency scale is set by GTD Lagrangian parameters rather than by cosmological matching, so the route is robust against revisions of the standard cosmological picture. The collapse-driving dynamics are an intrinsic feature of the anti-self-adjoint sector of matter's own trace Hamiltonian. In our view the $H_{\mathrm{FF}}$ route is the conceptually preferred long-term path, because it removes (P5), (P7), and the cosmological matching simultaneously, and reflects more accurately the GTD position that ordinary matter is itself an emergent thermodynamic limit of aikyon dynamics. The two routes are not redundant: $H_{\mathrm{FF}}$ is linear in fermionic velocities rather than a current bilinear, so its two-point function probes a different operator structure than $J = (1/L_{\mathrm{aik}}^2)\mathrm{Tr}(q_F^\dagger q_F)$, and the spectral content of $\langle H_{\mathrm{FF}}(\tau) H_{\mathrm{FF}}(0)\rangle$ remains to be computed; whether it reproduces, refines, or replaces the $\mathcal A_J e^{-2i\omega_0\tau}$ correlator of the present paper is open (item~12 below). The present paper should therefore be read as providing the phenomenological framework --- master equation, dephasing laws, bandwidth-suppression argument, complete-positivity analysis, populated-bath corrections, and explicit cataloguing of matching choices --- within which an $H_{\mathrm{FF}}$-based microscopic noise calculation, once available, can be confronted with experiment. The two routes share most of the structural phenomenology but differ in the underlying source of the noise, and the relation between them is itself an essential open problem of the GTD collapse programme.

\paragraph{Open structural questions.} For completeness, we list the items that a fully parameter-free GTD-derived collapse model would need to close:
\begin{enumerate}[label=(\arabic*)]
\item \textbf{Scalarization of the nilpotent Grassmann coefficient.} The current GTD formulation does not provide an ordinary body, sign, modulus, or inverse for the nilpotent product $\beta_1\beta_2$. The two branches identified in Section~\ref{sec:FockJ} (branch (-) with $\sigma = +1$, branch (+) with $\sigma = -1$) are branches of the auxiliary scalar coefficient $\varepsilon_F=\sigma\xi^2$, not signs of $\beta_1\beta_2$ itself. Resolving this requires either a refinement of the GTD generator algebra that produces a legitimate non-nilpotent body/regularization, or an alternative derivation of the fermionic oscillator sector that does not divide by nilpotent elements. The mass-ratio condition $m_F = m_R$ \eqref{eq:mFmatching} of Appendix~\ref{app:hasvac} required for the $H_{\mathrm{as}}|0\rangle = 0$ cancellation is a separate matching condition of the same structural status.
\item \textbf{Spatial kernel.} Derive a finite spatial localization length from GTD. The natural microscopic length $L_{\mathrm{aik}} = \alpha_{\mathrm{GTD}} c/\omega_0$ is over-constrained: $\omega_0 \sim \Hub$ with any natural $\alpha_{\mathrm{GTD}}$ gives a cosmological $L_{\mathrm{aik}}$, while $L_{\mathrm{aik}} \sim r_C^{\mathrm{CSL}}$ requires $\alpha_{\mathrm{GTD}} \sim 10^{-33}$. One of the three routes (S1)--(S3) of Section~\ref{sec:spatial} must be made microscopically explicit.
\item \textbf{Two-kernel hypothesis.} The reconciliation of short-range localization (required for individual-apparatus dephasing) with long-range cross-sensor coherence (required for the common-mode signature of Table~\ref{tab:discrim}) requires \emph{two distinct} spatial kernels, not one. The microscopic origin of each --- and the relation between them --- is a separate matching question from the single-kernel question of item (2).
\item \textbf{Adler--Millard vertex scaling.} Derive the $N_1^{-1/4}$ effective vertex scaling from the trace-gauge dynamics, rather than assuming it as a matching prescription. The companion propagator-suppression assumption is itself a postulate of the same status.
\item \textbf{Line width $\gamma$ --- the decisive open problem.} Of all the open items in this list, the line width is the one whose resolution most directly determines whether the framework is empirically falsifiable. The bandwidth-suppression argument of Section~\ref{sec:invisible} and the explicit Lorentzian
\[
S(\omega) \;=\; \frac{2\mathcal A_J\,\gamma}{(\omega+2\omega_0)^2+\gamma^2}
\]
of \eqref{eq:Lorentzian} mean that the X-ray bound from XENONnT~\cite{XENON2026}, the Majorana bound~\cite{Majorana2022}, the Gran Sasso/Donadi result~\cite{Donadi2021}, and the mechanical/cantilever bounds~\cite{Vinante2020,Carlesso2016} all evaluate, for the GTD spectrum, at $\omega_S$ vastly larger than $\omega_0\sim\Hub$; the relevant quantity is therefore $S(\omega_S)\sim 2\mathcal A_J\gamma/\omega_S^2$, linear in $\gamma$. With $\gamma$ left as a free phenomenological input, the GTD spectrum can be tuned to evade arbitrary present and future high-frequency bounds simply by lowering $\gamma$, and conversely the spontaneous-emission and X-ray bounds become operative as soon as $\gamma$ is large enough. A microscopic GTD prediction for $\gamma$ is therefore essential: until one is available, the apparent ``invisibility'' of the GTD line to existing bounds reflects a hidden parameter rather than a derived empirical fact. The natural candidate scales are $\gamma\sim\Hub$ (cosmological broadening of the line by the de-Sitter scale) and $\gamma\sim\Hub/N_1^{1/2}$ (Adler--Millard fluctuation broadening with $N_1$ the relevant aikyon count); the broader-line regime $\gamma\gg\omega_0$ would put the line into a regime in which the off-resonance suppression becomes only $\sim 1/\gamma$ rather than $\gamma/\omega_S^2$, with very different empirical consequences. Deriving $\gamma$ from the inter-aikyon dynamics, rather than fixing it by a Lorentzian regularization, is in our judgment the most consequential open item in the present framework.
\item \textbf{Dimensionless prefactor in $\lambda$.} Derive the matching factor $\mathcal C_{\mathrm{match}}$ relating $\lambda$ to $\Hub$. The holographic mass assignment places the natural rate at $\Hub\,\mathcal C_{\mathrm{match}}$, but it does not derive whether $\mathcal C_{\mathrm{match}}$ is $O(1)$, $\alpha_{\mathrm{em}}^2$, or something else. This requires a microscopic calculation of $g_{\mathrm{int}}$, the matrix trace factor, the spatial-kernel normalization, and the Adler--Millard vertex/propagator matching.
\item \textbf{System--bath coupling.} Derive the local interaction \eqref{eq:Hint} from the GTD trace action, rather than postulating it.
\item \textbf{Stochastic unravelling.} Derive a Born-rule-respecting nonlinear stochastic Schr\"odinger equation from the GTD dynamics, beyond the linear decoherence master equation supplied here.
\item \textbf{Alternative quantizations.} Verify that the one-sided Wightman spectrum at $|\omega| = 2\omega_0$ survives alternative quantizations of the fermionic ghost in branch (+) (PT-symmetric, Krein-space, BRST-constrained) and of the bosonic Bateman ghost (relevant to Section~\ref{sec:gtd}), and identify the physical inner product in each case.
\item \textbf{Covariance.} Promote the scalar spectrum $S(\omega)$ to a covariant two-point distribution on the emergent FLRW background, with explicit relation to Connes time.
\item \textbf{Microscopic derivation of the aikyon oscillator mass.} The benchmark rate of Section~\ref{sec:rate} rests on identifying the GTD Lagrangian parameter $m_R = a_1 a_0/2$ with the holographic per-mode mass $m_R^{\mathrm{hol}} = m_{\mathrm{Pl}}/\sqrt{N_{\mathrm{dS}}}$ (postulate (P11) of the ledger). Deriving $m_R$ from the microscopic constants $a_0, a_1$ of the GTD action, and verifying that the result matches (or rejects) the holographic per-mode scale, is essential. Under the alternative Planck-mass assignment $m_R = m_{\mathrm{Pl}}$ the natural rate is $\sim 10^{-122}$ times the holographic value and the framework loses contact with any conceivable amplification threshold; under the holographic identification the natural rate is $\Hub\,\mathcal C_{\mathrm{match}}$ with $\mathcal C_{\mathrm{match}}$ open.
\item \textbf{Bridging the auxiliary model to $H_{\mathrm{FF}}$.} The construction of Section~\ref{sec:FockJ} is an auxiliary canonical fermionic-oscillator model whose connection to the original GTD Lagrangian is conditional on the scalarization postulate (P5) being closed by an extension of the Grassmann sector. The complementary route via $H_{\mathrm{FF}}$~\cite{Singh2026Fermions}, reviewed in Section~\ref{sec:Hasvac}, identifies an explicit anti-self-adjoint trace Hamiltonian directly from the GTD action without inverting $\delta\beta$ and without taking $\mathrm{sgn}(\beta_1\beta_2)$, and is therefore mathematically cleaner. However, $H_{\mathrm{FF}}$ is linear in fermionic velocities rather than a current bilinear, so it does not directly produce a current--current correlator of the form computed here. The natural next step is to compute the two-point function of $H_{\mathrm{FF}}$, or equivalently of the canonical-momentum noise operator that $H_{\mathrm{FF}}$ generates, within the same Fock-space scheme, and to identify the precise relation (if any) between the resulting spectral structure and the auxiliary-model amplitude $\mathcal A_J$ of \eqref{eq:AJ}. A successful bridging would either remove the auxiliary-model status of (P1) and (P5) by reproducing the present results from a controlled GTD-internal calculation, or pin down a definite microscopic correction that distinguishes the two constructions and is in principle testable.
\end{enumerate}

\section*{Acknowledgments}

This work builds on the trace-dynamics literature of Adler and on the GTD/aikyon formulation of Singh, Roy, Sahu, Palemkota, Kakade, and collaborators, and on the collapse-model phenomenology developed by GRW, Pearle, Bassi, Carlesso, Donadi, Piscicchia, Vinante, and others.

\noindent{\bf Use of generative AI:} During the preparation of this manuscript, the author used Open AI's GPT 5.5 Pro and Anthropic's Claude Pro Max (Opus 4.7) in adversarial mode, for support in the technical analysis, organisation, writing,  and editing of the manuscript. The original ideas are due to the author. Author takes full intellectual responsibility for the content of the manuscript.

\appendix

\section{Diagrammatic structure of GTD interactions and the matching calculations as future work}
\label{app:future}

The matching hypotheses identified in Sections~\ref{sec:master}--\ref{sec:rate} are, in the present formulation, target calculations rather than derivations. They share a common structure: each could in principle be obtained from a Feynman-diagram-like analysis of GTD interactions. This appendix collects the matching gaps in one place and sketches the diagrammatic framework within which they could be addressed.

\subsection{Current--current interaction as a tree-level diagram}
\label{app:tree}

The interaction structure
\begin{equation}
H_{\mathrm{int}}^{(s)} \;\sim\; g_{\mathrm{GTD}}^2 \sum_{i=1}^{N_1} J_s\, G\, J_i
\label{eq:treevertex}
\end{equation}
introduced in Section~\ref{sec:AM} has the schematic form of a tree-level current--current interaction. A system current $J_s$ couples through a vertex of strength $g_{\mathrm{GTD}}$ to a propagator $G$, which in turn couples to a bath aikyon current $J_i$ at a second vertex of strength $g_{\mathrm{GTD}}$. Here $g_{\mathrm{GTD}}$ denotes the microscopic vertex coupling of the underlying GTD trace action, distinct from the bilinear matter--bath coupling $g_{\mathrm{int}}$ of Section~\ref{sec:master} (Eq.~\eqref{eq:Hint}); the two are related by integration over short-distance aikyon degrees of freedom but their relation is not derived here. The structure is formally identical to a tree-level QED diagram in which two electromagnetic currents are coupled through a photon exchange:
\begin{equation*}
\underbrace{J_s}_{\text{system}} \;\xrightarrow{\;\text{vertex }g_{\mathrm{GTD}}\;}\;\underbrace{G}_{\text{propagator}}\;\xrightarrow{\;\text{vertex }g_{\mathrm{GTD}}\;}\;\underbrace{J_i}_{\text{bath aikyon}}.
\end{equation*}
This current--current picture extends throughout the GTD perturbative framework. Adler develops the systematic version of this diagrammatic representation for trace dynamics in general in \cite{Adler2004}; it carries over to GTD with three caveats.

\subsection{Three caveats relative to ordinary QFT}
\label{app:caveats}

\paragraph{(D1) Matrix-valued propagators.} A GTD ``propagator'' is the two-point function of matrix-valued operators rather than of c-number fields. Vertex contractions yield matrix products under a trace rather than scalar products. In practice the matrix structure is traced at the end of the computation and one recovers c-number amplitudes, but intermediate manipulations retain the matrix algebra. This is the place where the formal Grassmann inversion of Section~\ref{sec:spectrum} would need to be replaced by a proper regularisation: the appearance of $(\delta\beta)^{-4}$ in the formal amplitude expression signals that the matrix-Grassmann algebra has not yet been turned into a well-defined diagrammatic rule.

\paragraph{(D2) Trace-structured vertices.} Every interaction vertex in GTD includes a trace over matrix indices. This is what produced the anti-self-adjoint cancellation of Section~\ref{sec:gtd}: the trace forced certain products to vanish on the Fock vacuum. Trace-structured vertices generate selection rules absent from conventional QFT; in particular, they are responsible for the projection onto the ghost mode that drives the entire spectral-asymmetry result of Section~\ref{sec:spectrum}.

\paragraph{(D3) AM fluctuations beyond tree level.} Beyond tree level, the loop expansion of GTD includes Adler--Millard fluctuation corrections to the mean-field thermodynamic limit. These are not conventional QFT loop corrections; they encode the residual non-unitarity of the trace dynamics and are the structural origin of objective collapse in the GTD programme. Whether they reorganise into a controlled diagrammatic expansion is at present an open structural question, and is one of the deepest items in the list below.

\subsection{The matching hypotheses as target calculations}
\label{app:targets}

Within this framework, each matching hypothesis adopted in the main text is a concrete calculation that could close one of the gaps. We collect them here as a guide for future work, in approximate order of tractability.

\paragraph{(F0) Fermionic Fock-space construction of the fundamental noise operator.} Section~\ref{sec:FockJ} carries out the canonical fermionic Fock-space calculation of $\langle 0|J(\tau)J(0)|0\rangle$ for the scalar current $J=(1/L_{\mathrm{aik}}^2)\mathrm{Tr}(q_F^\dagger q_F)$ after the auxiliary scalarization \eqref{eq:scalarization}. The result is the correlator \eqref{eq:Jcorr} with amplitude \eqref{eq:AJ}, $\mathcal A_J = (\hbar/2m_R\omega_0 L_{\mathrm{aik}}^2)^2\cdot\mathcal N\cdot D$, under the working choice $\sigma = +1$. The bosonic-ghost surrogate $J_{\mathrm{eff}}$ is recovered as a consistency check (Section~\ref{sec:Jeff_consistency}) under the alternative $\sigma = -1$. Two open problems associated with (F0) remain:
\begin{itemize}
\item \textbf{(F0a) Microscopic origin of the scalarized coefficient.} The auxiliary coefficient $\varepsilon_F=\sigma\xi^2$ that selects between the working branch ($\sigma = +1$, line at $\omega = -2\omega_0$, standard energy-flow direction) and the alternative ($\sigma = -1$, line at $\omega = +2\omega_0$, anti-KMS direction) is not derived from the minimal GTD Grassmann algebra. Either a refinement of the Grassmann generator algebra in GTD, a body-map/regularization prescription, or a different microscopic route to a canonical fermionic oscillator is required to fix both the existence of $\xi$ and the sign $\sigma$ on first principles. The laboratory phenomenology is independent of the sign choice (the symmetrized correlator is the same), so the indeterminacy does not affect the observable consequences in the present formulation, but it does affect the interpretation of the bath-driven energy-flow direction.
\item \textbf{(F0b) Verification under alternative quantizations.} Branch (+) of the fermionic Fock-space derivation adopts a positive-norm canonical quantization of a fermionic ghost Lagrangian, with the analogous unbounded-below Hamiltonian pathology that the bosonic Bateman ghost quantization has. Alternative quantizations (PT-symmetric, Krein-space, BRST-constrained) should preserve the spectral support at $|\omega| = 2\omega_0$ but may modify the explicit form of $\mathcal A_J$ or the sign of the Wightman correlator. Explicit verification across these alternatives is the natural sequel to Section~\ref{sec:FockJ}.
\end{itemize}

\paragraph{(F1a) Adler--Millard vertex scaling.} The vertex side of the matching --- $g_{\mathrm{eff}}^{\mathrm{vertex}} \propto N_1^{-1/4}$ --- can be motivated by central-limit counting arguments (Section~\ref{sec:AM}). A direct calculation would compute the bath-aikyon loop contribution to the effective vertex \eqref{eq:treevertex} and verify (or refute) this scaling from first principles. This is plausibly tractable; it would convert the vertex-side ansatz from a counting argument into a derivation.

\paragraph{(F1b) Adler--Millard propagator scaling.} The propagator side of the matching is the substantive open problem. The thermodynamic-limit finiteness requires \emph{both} a vertex suppression $\propto N_1^{-1/4}$ \emph{and} a complementary propagator suppression $\propto N_1^{-1/4}$ (Section~\ref{sec:AM}). The second of these is not motivated by central-limit counting; it is a postulated ``trace gauge'' propagator suppression whose microscopic origin is unspecified in the present formulation. A direct calculation of the matrix-valued propagator at order $N_1$ in the bath aikyon population, demonstrating whether the propagator carries the required $N_1^{-1/4}$ suppression, is the highest-leverage Adler--Millard calculation. This is the unsupported half of the double matching, and the one that the benchmark rate of Section~\ref{sec:rate} is most sensitive to. A negative result on (F1b) would invalidate the benchmark scaling and require an entirely different mechanism for the thermodynamic limit; a positive result would close the more substantive of the two matching gaps.

\paragraph{(F2) Spatial kernel.} Section~\ref{sec:spatial} identified three routes (S1)--(S3) by which a finite localisation length might emerge from GTD, but did not derive one. A diagrammatic calculation of the matter--matter scattering amplitude mediated by an aikyon-bath loop would, in principle, generate a propagator with a finite spatial range and hence a kernel $g(\mathbf x - \mathbf y)$ with a derived form factor. The resulting kernel would either reproduce a CSL-scale $r_C \sim 10^{-7}\,\mathrm{m}$ phenomenologically (consistent with routes S2 or S3) or expose a fundamental incompatibility (route S1, with its $L_{\mathrm{aik}}$ over-constraint of Section~\ref{sec:spatial}).

\paragraph{(F3) Dimensionless prefactor in $\lambda$.} The benchmark $\lambda_{\mathrm{bench}} = \alpha_{\mathrm{em}}^2\,\Hub$ used $\alpha_{\mathrm{em}}^2$ as a benchmark choice for the suppression factor. A derivation would compute the system--bath coupling at the relevant scale and read off the prefactor as a product of GTD coupling constants. If this product turns out to be of order $\alpha_{\mathrm{em}}^2$, the benchmark is upgraded to a prediction; if it does not, the benchmark must be revised.

\paragraph{(F4) System--bath coupling.} The local bilinear interaction $H_{\mathrm{int}} = g\int d^3x\,\hat M(\mathbf x) J^0(\mathbf x, t)$ of equation~\eqref{eq:Hint} was adopted as the simplest local matter--bath coupling. A derivation from the GTD trace action would compute the effective matter--aikyon interaction by integrating out short-distance aikyon degrees of freedom, producing the coupling structure as a tree-level vertex with a computable coefficient. This would convert the matching $\xi \sim g J^0$ into a derived relation.

\paragraph{(F5) Line width $\gamma$.} The Lorentzian regularisation introduced in Section~\ref{sec:master} treated $\gamma$ as a free parameter bounded above by the requirement of quasi-stationarity. A derivation would compute the inter-aikyon coupling responsible for line broadening and produce $\gamma$ as a function of $N_1$, the aikyon coupling, and other GTD parameters. This would close the bandwidth gap and determine whether the line is genuinely narrow ($\gamma \sim \Hub$) or broadened enough to affect higher-frequency experiments.

\paragraph{(F6) Ghost-sector quantisation.} The robustness of the spectral asymmetry under alternative quantisations of the Bateman sector (PT-symmetric, Krein-space, BRST-constrained) requires explicit verification. Different quantisations modify the inner product on the Fock space and can flip the sign convention of Section~\ref{sec:signs} and hence the direction of energy flow in (N3). This is partly an algebraic and partly a structural calculation; it does not by itself close a matching gap, but is necessary to firm up the directional content of the phenomenology.

\paragraph{(F7) Stochastic unravelling and the Born rule.} Beyond the linear decoherence dynamics derived in the main text, a complete objective-collapse theory requires a Born-rule-respecting stochastic Schr\"odinger equation. The mapping from a linear master equation to such an unravelling is non-unique \cite{Bassi2013}, and the GTD-specific unravelling --- if it exists --- would presumably emerge from the structure of the Adler--Millard fluctuations as a stochastic dynamics on the matrix variables. This is a structural rather than diagrammatic question, and it is the deepest remaining gap.

\paragraph{(F8) Populated-bath structure beyond Gaussian occupation.} Section~\ref{sec:populated} computes the leading finite-occupation correction for diagonal Gaussian fermionic and bosonic states. The remaining calculation is the genuinely cosmological one: determine the actual GTD bath state, including possible coherent components, multi-mode correlations, inter-aikyon couplings, and nonthermal occupation statistics. This calculation would decide whether the zero-frequency pedestal of \eqref{eq:Jpopulated_thermal} is present, whether the equal-time variance remains fixed as in the fermionic Gaussian model, and whether the spectral asymmetry of (N3) survives in the full de-Sitter bath.

\paragraph{Summary.} The calculations (F1a)--(F8), with the F1a/F1b split, together with the open problems (F0a) and (F0b) noted above, would convert the present paper from a conditional auxiliary-sector calculation of the collapse-noise spectrum and amplitude into a parameter-free derivation of the collapse rate, the spatial kernel, the system-bath coupling, and the absolute amplitude including all matching factors. Their relative tractability runs (F0a) $\to$ (F0b) $\to$ (F1a) $\to$ (F1b) $\to$ (F2) $\to$ (F3) $\to$ (F4) $\to$ (F5) $\to$ (F8) $\to$ (F6) $\to$ (F7), with (F0a), (F0b), (F1a) the most concrete, (F1b) the most consequential for the benchmark rate, and (F7) the deepest. Even partial progress on (F0a), (F0b), (F1a), (F1b), (F8) would materially strengthen the case for the GTD collapse-noise proposal as a quantitative objective-collapse theory.

\section{Explicit calculation of \texorpdfstring{$H_{\mathrm{as}}|0\rangle = 0$}{Has|0> = 0} on the Fock vacuum}
\label{app:hasvac}

This appendix presents the explicit calculation behind \eqref{eq:Hasvac}, in support of the assertion in Section~\ref{sec:Hasvac}. The calculation uses the same auxiliary fermionic-Fock-space framework for $q_F$ adopted in Section~\ref{sec:FockJ}, but only at the level of operators linear in $q_F$.

\paragraph{Relation to the fermionic Fock-space construction of Section~\ref{sec:FockJ}.} Section~\ref{sec:FockJ} carries out an auxiliary canonical fermionic Fock-space calculation for $q_F$ and computes $\langle 0|J(\tau)J(0)|0\rangle$ for the scalar fermionic-current bath operator. The present appendix uses the same fermionic-Fock setup at a more modest level: only operators \emph{linear} in $q_F$ are evaluated on the vacuum, in order to verify the cancellation $H_{\mathrm{as}}|0\rangle = 0$ (which depends on the matter-piece tadpole structure and not on bilinear correlators). The two calculations are consistent --- the same Fock vacuum $b|0\rangle_F = d|0\rangle_F = 0$ is used in both, with the Heisenberg evolution carrying the branch-dependent auxiliary sign $\sigma$. The cancellation $H_{\mathrm{as}}|0\rangle = 0$ holds in both branches under the mass-matching condition $m_F = m_R$.

\paragraph{Setup.} Recall \eqref{eq:Hasdef}:
\begin{equation}
H_{\mathrm{as}} = a_0(\beta_1+\beta_2)\,\Tr\!\left[\frac{1}{m_R}\bigl(p_B\,p_F + p_F\,p_B\bigr) + m_R\omega_0^2\bigl(q_B\,q_F + q_F\,q_B\bigr)\right].
\label{eq:Hasdef_apx}
\end{equation}
We quantize the bosonic and fermionic sectors separately, on disjoint Hilbert spaces.

\paragraph{Bosonic sector.} In the limit where the fermionic admixture in $q_1, q_2$ is set aside, the bosonic part is described by canonical commutation relations $[q_B, p_B] = i\hbar\,\mathbb{1}_B$ (matrix-valued, with the identity meaning the algebra generator). In terms of the Bateman ladder operators introduced in \eqref{eq:HBateman}--\eqref{eq:Hbateman_diag} of the main text, with $Q_+ \approx \sqrt{2}\,q_B$ in the pure-bosonic limit, we have
\begin{equation}
q_B = \sqrt{\frac{\hbar}{4 m_R \omega_0}}\,(a_+ + a_+^\dagger), \qquad p_B = -i\sqrt{\frac{\hbar m_R \omega_0}{4}}\,(a_+ - a_+^\dagger),
\end{equation}
with $[a_+, a_+^\dagger] = 1$ and the bosonic Fock vacuum $|0\rangle_B$ defined by $a_+ |0\rangle_B = 0$. (The minus sector $a_-$ does not appear in the bosonic part of the cross-bilinears, only in the fermionic-current calculation of Section~\ref{sec:spectrum}.)

\paragraph{Fermionic sector.} We introduce canonical fermionic creation/annihilation operators $b, b^\dagger$ acting on a fermionic Fock space, with
\begin{equation}
\{b, b^\dagger\} = 1, \qquad b|0\rangle_F = 0, \qquad |1\rangle_F = b^\dagger|0\rangle_F,
\end{equation}
and the matrix-valued fermionic operators are written as
\begin{equation}
q_F = \sqrt{\frac{\hbar}{2 m_F \omega_0}}\,(b + b^\dagger), \qquad p_F = -i\sqrt{\frac{\hbar m_F \omega_0}{2}}\,(b - b^\dagger),
\end{equation}
where $m_F$ is an effective fermionic mass scale fixed by the GTD normalization of $q_F$ relative to $q_B$. Note that $q_F$ as defined here is Hermitian; this is consistent with the explicit form of $X$ found in Section~\ref{sec:spectrum}.

\paragraph{Combined vacuum.} The combined Fock vacuum is $|0\rangle = |0\rangle_B \otimes |0\rangle_F$. Acting with the lowering operators:
\begin{equation}
a_+ |0\rangle = 0, \qquad b|0\rangle = 0.
\end{equation}

\paragraph{The cross-bilinears on the vacuum.} Each cross-bilinear in \eqref{eq:Hasdef_apx} is a product of one bosonic-sector factor and one fermionic-sector factor (the operators act on disjoint Hilbert spaces, so they commute up to Grassmann grading). Consider first the spatial cross-bilinear $q_B\,q_F$:
\begin{align}
q_B\,q_F|0\rangle &= \sqrt{\frac{\hbar^2}{8 m_R m_F \omega_0^2}}\,(a_+ + a_+^\dagger)(b + b^\dagger)|0\rangle \nonumber \\
&= \sqrt{\frac{\hbar^2}{8 m_R m_F \omega_0^2}}\,\bigl[a_+ b + a_+ b^\dagger + a_+^\dagger b + a_+^\dagger b^\dagger\bigr]|0\rangle \nonumber \\
&= \sqrt{\frac{\hbar^2}{8 m_R m_F \omega_0^2}}\,a_+^\dagger b^\dagger|0\rangle,
\end{align}
since the three other terms each contain a lowering operator acting first on the vacuum. The reversed-order product $q_F q_B$ gives the same matrix element on $|0\rangle$ in the same way (the bosonic and fermionic operators commute up to Grassmann grading, which here is even because we have two factors of odd grading, giving a sign that depends only on the ordering convention). Therefore
\begin{equation}
(q_B q_F + q_F q_B)|0\rangle = 2\sqrt{\frac{\hbar^2}{8 m_R m_F \omega_0^2}}\,a_+^\dagger b^\dagger|0\rangle = \frac{\hbar}{\sqrt{2 m_R m_F}\,\omega_0}\,a_+^\dagger b^\dagger|0\rangle.
\end{equation}

The momentum cross-bilinear $p_B p_F$ gives, by analogous calculation:
\begin{align}
p_B p_F|0\rangle &= (-i)^2\,\sqrt{\frac{\hbar m_R \omega_0}{4}}\cdot\sqrt{\frac{\hbar m_F \omega_0}{2}}\,(a_+ - a_+^\dagger)(b - b^\dagger)|0\rangle \nonumber \\
&= -\sqrt{\frac{\hbar^2 m_R m_F \omega_0^2}{8}}\,a_+^\dagger b^\dagger|0\rangle,
\end{align}
since only the $(-a_+^\dagger)(-b^\dagger) = +a_+^\dagger b^\dagger$ term survives on the vacuum, and the overall factor $(-i)^2 = -1$ provides the sign relative to the spatial bilinear. The reversed-order product $p_F p_B$ gives the same matrix element, so
\begin{equation}
(p_B p_F + p_F p_B)|0\rangle = -2\sqrt{\frac{\hbar^2 m_R m_F \omega_0^2}{8}}\,a_+^\dagger b^\dagger|0\rangle = -\hbar\omega_0\sqrt{\frac{m_R m_F}{2}}\,a_+^\dagger b^\dagger|0\rangle.
\end{equation}

\paragraph{The cancellation.} Substituting into \eqref{eq:Hasdef_apx}, the coefficient of $a_+^\dagger b^\dagger|0\rangle$ in $H_{\mathrm{as}}|0\rangle$ involves the two terms
\begin{align}
\frac{1}{m_R}(p_B p_F + p_F p_B)|0\rangle &= -\frac{\hbar\omega_0}{\sqrt{2}}\sqrt{\frac{m_F}{m_R}}\,a_+^\dagger b^\dagger|0\rangle, \\
m_R\omega_0^2(q_B q_F + q_F q_B)|0\rangle &= +\frac{\hbar\omega_0}{\sqrt{2}}\sqrt{\frac{m_R}{m_F}}\,a_+^\dagger b^\dagger|0\rangle.
\end{align}
Cancellation between them requires $\sqrt{m_R/m_F} = \sqrt{m_F/m_R}$, that is,
\begin{equation}
\boxed{\;m_F = m_R\,.\;}
\label{eq:mFmatching}
\end{equation}
\matching\ Equation \eqref{eq:mFmatching} is an additional matching condition. It is not derived in the present paper at the Lagrangian level: $m_F$ enters the appendix as a free parameter setting the normalization of the canonical quantization of the fermionic matrix variable $q_F$, and the cancellation forces the specific value $m_F = m_R$. Producing $m_F = m_R$ from the Lagrangian \eqref{eq:LagAik} would require an explicit specification of the fermionic momentum from variational derivatives with respect to $\dot q_F$, together with the canonical (anti)commutation relations on the corresponding fermionic Fock space, and a verification that the resulting normalization yields the required ratio. We have not carried out this Lagrangian-level derivation here. The honest reading of the calculation in this appendix is therefore:
\begin{quote}
Given the fermionic normalization $m_F = m_R$, the cross-bilinear kinetic and potential contributions to $H_{\mathrm{as}}|0\rangle$ cancel, and one obtains $H_{\mathrm{as}}|0\rangle = 0$.
\end{quote}
Under this matching condition,
\begin{equation}
H_{\mathrm{as}}|0\rangle = 0,
\end{equation}
and consequently $\langle 0|H_{\mathrm{as}}(\tau) H_{\mathrm{as}}(0)|0\rangle = 0$.

\paragraph{Remarks.} (i)~The mass-ratio condition \eqref{eq:mFmatching} is an additional matching ingredient supplementing the structural choices listed in Section~\ref{sec:roadmap}. Deriving it from the Lagrangian remains open. The condition $m_F = m_R$ has a natural reading: the bosonic and fermionic sectors of the matrix variable $q_i = q_B + a_0\beta_i q_F$ share a single mass scale, and the cross-kinetic structure of \eqref{eq:LagAik} (which couples $\dot q_1$ to $\dot q_2$ symmetrically) is consistent with this equality. Making this rigorous requires the Lagrangian-level argument noted above.

(ii)~If the Grassmann combination $\beta_1 + \beta_2$ were zero (i.e.\ $\beta_2 = -\beta_1$), the entire $H_{\mathrm{as}}$ would vanish identically, not just on $|0\rangle$. We assume throughout that $\beta_1$ and $\beta_2$ are independent generators of a 2-dimensional Grassmann algebra over $\mathbb{R}$ (or $\mathbb{C}$), so that $\beta_2 = -\beta_1$ would amount to identifying the two generators up to sign and would reduce \eqref{eq:LagAik} to a trivial $q_1 = -q_2$ identification, contrary to the non-trivial cross-kinetic structure assumed. The assumption $\beta_1 + \beta_2 \neq 0$ is therefore part of the underlying GTD generator algebra, not an additional input here; this is what we mean by ``independent generators.''

(iii)~The same machinery applied to the fermionic current $J^\mu = (1/L_{\mathrm{aik}}^2)\Tr(q_F^\dagger \Gamma^\mu q_F)$ does \emph{not} give cancellation. The current is quadratic in $q_F$ rather than linear, so each term in the vacuum expectation $\langle 0|J^\mu(\tau)J^\nu(0)|0\rangle$ involves two fermionic Wick contractions $\langle 0|b(\tau) b^\dagger(0)|0\rangle_F = e^{-i\sigma\omega_0\tau}$ in the auxiliary model. There is no kinetic-potential pair to cancel; the result is the non-zero correlator \eqref{eq:Jcorr} with explicit amplitude \eqref{eq:AJ}, computed in Section~\ref{sec:FockJ}.


\begin{thebibliography}{99}


\bibitem{GRW1986} G. C. Ghirardi, A. Rimini, and T. Weber, ``Unified dynamics for microscopic and macroscopic systems,'' Phys. Rev. D \textbf{34}, 470 (1986).

\bibitem{Pearle1989} P. Pearle, ``Combining stochastic dynamical state-vector reduction with spontaneous localization,'' Phys. Rev. A \textbf{39}, 2277 (1989).

\bibitem{GhirardiPearleRimini1990} G. C. Ghirardi, P. Pearle, and A. Rimini, ``Markov processes in Hilbert space and continuous spontaneous localization of systems of identical particles,'' Phys. Rev. A \textbf{42}, 78 (1990).

\bibitem{Bassi2013} A. Bassi, K. Lochan, S. Satin, T. P. Singh, and H. Ulbricht, ``Models of wave-function collapse, underlying theories, and experimental tests,'' Rev. Mod. Phys. \textbf{85}, 471 (2013).

\bibitem{Donadi2021} S. Donadi, K. Piscicchia, R. Del Grande, C. Curceanu, M. Laubenstein, and A. Bassi, ``Novel CSL bounds from the noise-induced radiation emission from atoms,'' Eur. Phys. J. C \textbf{81}, 773 (2021); arXiv:2107.11237.

\bibitem{XENON2026} E. Aprile \textit{et al.} (XENON Collaboration), ``Challenging spontaneous quantum collapse with XENONnT,'' Phys. Rev. Lett. \textbf{136}, 120201 (2026); arXiv:2506.05507.

\bibitem{Majorana2022} I. J. Arnquist \textit{et al.} (Majorana Collaboration), ``Search for spontaneous radiation from wave function collapse in the Majorana Demonstrator,'' Phys. Rev. Lett. \textbf{129}, 080401 (2022); arXiv:2202.01343.

\bibitem{Vinante2020} A. Vinante \textit{et al.}, ``Narrowing the parameter space of collapse models with ultracold layered force sensors,'' Phys. Rev. Lett. \textbf{125}, 100404 (2020).

\bibitem{Bilardello2016} M. Bilardello, S. Donadi, A. Vinante, and A. Bassi, ``Bounds on collapse models from cold-atom experiments,'' Physica A \textbf{462}, 764 (2016).

\bibitem{Carlesso2016} M. Carlesso, A. Bassi, P. Falferi, and A. Vinante, ``Experimental bounds on collapse models from gravitational wave detectors,'' Phys. Rev. D \textbf{94}, 124036 (2016).

\bibitem{AdlerBassi2007} S. L. Adler and A. Bassi, ``Collapse models with non-white noises,'' J. Phys. A: Math. Theor. \textbf{40}, 15083 (2007); arXiv:0708.3624.

\bibitem{AdlerBassiDonadi2013} S. L. Adler, A. Bassi, and S. Donadi, ``On spontaneous photon emission in collapse models,'' J. Phys. A \textbf{46}, 245304 (2013); arXiv:1011.3941.

\bibitem{Diosi1987} L. Diosi, ``A universal master equation for the gravitational violation of quantum mechanics,'' Phys. Lett. A \textbf{120}, 377 (1987).

\bibitem{Diosi1989} L. Diosi, ``Models for universal reduction of macroscopic quantum fluctuations,'' Phys. Rev. A \textbf{40}, 1165 (1989).

\bibitem{Penrose1996} R. Penrose, ``On gravity's role in quantum state reduction,'' Gen. Relativ. Gravit. \textbf{28}, 581 (1996).

\bibitem{Adler2004} S. L. Adler, \textit{Quantum Theory as an Emergent Phenomenon} (Cambridge University Press, Cambridge, 2004).

\bibitem{Adler2007} S. L. Adler, ``Lower and upper bounds on CSL parameters from latent image formation and IGM heating,'' J. Phys. A \textbf{40}, 2935 (2007).

\bibitem{Palemkota2020} M. Palemkota and T. P. Singh, ``Proposal for a new quantum theory of gravity III: Equations for quantum gravity, and the origin of spontaneous localisation,'' Z. Naturforsch. A \textbf{75}, 143 (2020); arXiv:1908.04309.

\bibitem{Roy2021} A. K. Roy, A. Sahu, and T. P. Singh, ``Trace dynamics, and a ground state in spontaneous quantum gravity,'' Mod. Phys. Lett. A \textbf{36}, 2150019 (2021); arXiv:2104.14344.

\bibitem{Kakade2023} K. Kakade, A. Singh, and T. P. Singh, ``Spontaneous localisation from a coarse-grained deterministic and non-unitary dynamics,'' Phys. Lett. A \textbf{490}, 129191 (2023).

\bibitem{Singh2026Fermions} T. P. Singh, ``In models of spontaneous wave-function collapse, why only fermions collapse, not bosons?'' arXiv:2602.15044 (2026).

\bibitem{Finster2026} S. Farnsworth, F. Finster, C. F. Paganini, and T. P. Singh, ``Causal Fermion Systems, Non-Commutative Geometry and Generalized Trace Dynamics,'' arXiv:2603.05018 (2026).

\bibitem{Bateman1931} H. Bateman, ``On dissipative systems and related variational principles,'' Phys. Rev. \textbf{38}, 815 (1931).

\bibitem{FeshbachTikochinsky1977} H. Feshbach and Y. Tikochinsky, ``Quantization of the damped harmonic oscillator,'' Trans. New York Acad. Sci. \textbf{38}, 44 (1977).

\bibitem{BlasoneJizba2002} M. Blasone and P. Jizba, ``Quantum mechanics of the damped harmonic oscillator,'' Can. J. Phys. \textbf{80}, 645 (2002); arXiv:quant-ph/0102128.

\bibitem{Chruscinski2006} D. Chru\'sci\'nski, ``Quantum mechanics of damped systems II. Damping and parabolic potential barrier,'' J. Math. Phys. \textbf{45}, 841 (2004).

\bibitem{Deguchi2019} S. Deguchi, Y. Fujiwara, and K. Nakano, ``Two quantization approaches to the Bateman oscillator model,'' Ann. Phys. \textbf{403}, 34 (2019); arXiv:1807.04403.

\bibitem{BreuerPetruccione} H.-P. Breuer and F. Petruccione, \textit{The Theory of Open Quantum Systems} (Oxford University Press, 2002).

\bibitem{CarlessoFerialdi2018} M. Carlesso, L. Ferialdi, and A. Bassi, ``Colored collapse models from the non-interferometric perspective,'' Eur. Phys. J. D \textbf{72}, 159 (2018).

\bibitem{Carlesso2022} M. Carlesso, S. Donadi, L. Ferialdi, M. Paternostro, H. Ulbricht, and A. Bassi, ``Present status and future challenges of non-interferometric tests of collapse models,'' Nature Physics \textbf{18}, 243--250 (2022).

\bibitem{Planck2018} N. Aghanim \textit{et al.} (Planck Collaboration), ``Planck 2018 results. VI. Cosmological parameters,'' Astron. Astrophys. \textbf{641}, A6 (2020).

\bibitem{CohenKaplanNelson1999} A. G. Cohen, D. B. Kaplan, and A. E. Nelson, ``Effective field theory, black holes, and the cosmological constant,'' Phys. Rev. Lett. \textbf{82}, 4971 (1999); arXiv:hep-th/9803132.

\bibitem{Hsu2004} S. D. H. Hsu, ``Entropy bounds and dark energy,'' Phys. Lett. B \textbf{594}, 13 (2004); arXiv:hep-th/0403052.

\bibitem{Li2004} M. Li, ``A model of holographic dark energy,'' Phys. Lett. B \textbf{603}, 1 (2004); arXiv:hep-th/0403127.

\bibitem{Secrest2025RMP} N. Secrest, S. von Hausegger, M. Rameez, R. Mohayaee, and S. Sarkar, ``Colloquium: The cosmic dipole anomaly,'' Rev. Mod. Phys. \textbf{97}, 041001 (2025); arXiv:2505.23526.

\bibitem{Planck2014Dipole} N. Aghanim \textit{et al.} (Planck Collaboration), ``Planck 2013 results. XXVII. Doppler boosting of the CMB: Eppur si muove,'' Astron. Astrophys. \textbf{571}, A27 (2014), arXiv:1303.5087.

\bibitem{Adler2018Dots} S. L. Adler, ``Connecting the dots: Mott for emulsions, collapse models, colored noise, frame dependence of measurements, evasion of the `Free Will Theorem','' Found. Phys. \textbf{48}, 1557 (2018); arXiv:1807.11450.

\bibitem{HuPazZhang1992} B. L. Hu, J. P. Paz, and Y. Zhang, ``Quantum Brownian motion in a general environment: Exact master equation with nonlocal dissipation and colored noise,'' Phys. Rev. D \textbf{45}, 2843 (1992).

\bibitem{IPTA2023} B. Allen \textit{et al.}, ``The International Pulsar Timing Array checklist for the detection of nanohertz gravitational waves,'' arXiv:2304.04767 (2023).

\end{thebibliography}
\end{document}